\documentclass[acmsmall,screen,authorversion,nonacm]{acmart}

\AtBeginDocument{%
  \providecommand\BibTeX{{%
    \normalfont B\kern-0.5em{\scshape i\kern-0.25em b}\kern-0.8em\TeX}}}

\copyrightyear{2023} 
\acmYear{2023} 
\setcopyright{rightsretained} 
\acmConference[AIES '23]{AAAI/ACM Conference on AI, Ethics, and Society}{August 8--10, 2023}{Montréal, QC, Canada}
\acmBooktitle{AAAI/ACM Conference on AI, Ethics, and Society (AIES '23), August 8--10, 2023, Montréal, QC, Canada}
\acmDOI{10.1145/3600211.3604722}
\acmISBN{979-8-4007-0231-0/23/08}

\usepackage{colortbl}
\usepackage{hyperref}

\begin{document}

\title{Typology of Risks of Generative Text-to-Image Models}

\author{Charlotte Bird}
\authornote{Equal contribution}
\email{charlotte.bird@ed.ac.uk}
\affiliation{%
  \institution{School of Informatics
  \\ University of Edinburgh}
  \streetaddress{10 Crichton Street}
  \city{Edinburgh}
  \country{Scotland}
  \postcode{EH8 9AB}
}
\orcid{0009-0001-2378-8238}

\author{Eddie L. Ungless}
\authornotemark[1]
\email{e.l.ungless@sms.ed.ac.uk}
\affiliation{%
  \institution{School of Informatics
  \\ University of Edinburgh}
  \streetaddress{10 Crichton Street}
  \city{Edinburgh}
  \country{Scotland}
  \postcode{EH8 9AB}
}
\orcid{0000-0002-9378-4427}

\author{Atoosa Kasirzadeh}
\email{atoosa.kasirzadeh@ed.ac.uk}
\affiliation{%
  \institution{Alan Turing Institute \\ University of Edinburgh}
  \streetaddress{10 Crichton Street}
  \city{Edinburgh}
  \country{Scotland}
  \postcode{EH8 9AB}
}
\orcid{0000-0002-5967-3782}

\renewcommand{\shortauthors}{Bird \& Ungless \& Kasirzadeh}


\begin{abstract}
This paper investigates the direct risks and harms associated with modern text-to-image generative models, such as DALL-E and Midjourney, through a comprehensive literature review. While these models offer unprecedented capabilities for generating images, their development and use introduce new types of risk that require careful consideration. Our review reveals significant knowledge gaps concerning the understanding and treatment of these risks despite some already being addressed. We offer a taxonomy of risks across six key stakeholder groups, inclusive of unexplored issues, and suggest future research directions. We identify 22 distinct risk types, spanning issues from data bias to malicious use. The investigation presented here is intended to enhance the ongoing discourse on responsible model development and deployment. By highlighting previously overlooked risks and gaps, it aims to shape subsequent research and governance initiatives, guiding them toward the responsible, secure, and ethically conscious evolution of text-to-image models.
\end{abstract}

\ccsdesc[500]{Human-centered computing~Human computer interaction (HCI)}
\ccsdesc[300]{Human-centered computing~Text input}
\ccsdesc[300]{Applied computing~Media arts}
\ccsdesc[100]{Social and professional topics~User characteristics}
\keywords{Generative AI, Generative models, Text-to-Image models, Responsible AI, AI ethics, AI safety, AI governance, AI risks}

\maketitle

\textcolor{blue}{\textbf{Forthcoming in Proceedings of the 2023 AAAI/ACM Conference on AI, Ethics, and Society (AIES 2023)}}

\section{Introduction}

\begin{table*}
\resizebox{\textwidth}{!}{%
\begin{tabular}{c |c| c | c | c } 
 \hline
 Risk  & Stakeholders & Harm & Anticipated & Observed  \\ 
 \hline\hline
 \multicolumn{5}{c}{Discrimination and Exclusion} \\[1ex]
 \hline
 Cultural and racial bias & Users, Affected & Representational harm & \cite{Hutchinson_Baldridge_Prabhakaran_2022, Milliere_2022} & \cite{yu_scaling_2022,bianchi_easily_2022, Cho_Zala_Bansal_2022} \\ 
 Gender \& sexuality bias & Users, Affected & Representational harm & \cite{Milliere_2022,Newton_Dhole_2023,Hutchinson_Baldridge_Prabhakaran_2022} & \cite{bianchi_easily_2022,Cho_Zala_Bansal_2022, Ungless_Ross_Lauscher_2023} \\
 Class bias & Users, Affected & Representational harm & - & \cite{bianchi_easily_2022} \\
 Disability bias & Users, Affected  & Representational harm & - & \cite{bianchi_easily_2022} \\
 Loss of work for creatives & Sources, Users & Financial loss & \citep{Newton_Dhole_2023, Ghosh_Fossas_2022,Oppenlaender_2022a}  & \citep{eloundou2023gpts} \\ 

 Religious bias, ageism  & Users, Affected  & Representational harm & - & - \\

 Dialect bias & Users & Allocative harm, repr. harm & - & - \\

 Pre-release moderation & Developers, Affected & Psychological harm &- & - \\
  Job replacement & Affected, Regulators & Financial loss, Emotional harm & \cite{eloundou2023gpts} & \cite{DeCremer2023,Obedkov2023} \\ 
  \hline
 \multicolumn{5}{c}{Harmful Misuse} \\[1ex]
 \hline 
  Sexual images & Subjects, Users, Affected  & Repr. harm, emot. harm, fin. loss & \cite{Milliere_2022,Hutchinson_Baldridge_Prabhakaran_2022} & \cite{yu_scaling_2022}\\
  Sexualising images of children & Subjects, Users, Affected, Regulators  & Emotional harm & - &  \cite{Wolfe_Yang_Howe_Caliskan_2022} \\
  Violent or taboo content & Developers, Users, Affected   & Emotional harm, incite violence & \cite{Hutchinson_Baldridge_Prabhakaran_2022,Brack_Schramowski_Friedrich_Hintersdorf_Kersting_2022,Milliere_2022,Struppek_Hintersdorf_Kersting_2022} & - \\
Privacy infringement  & Sources, Subjects, Regulators & Privacy loss & - & \cite{Carlini_Hayes_Nasr_Jagielski_Sehwag_Tramèr_Balle_Ippolito_Wallace_2023} \\
  Copyright infringement & Users, Sources,  Regulators & Financial loss & - & \cite{Carlini_Hayes_Nasr_Jagielski_Sehwag_Tramèr_Balle_Ippolito_Wallace_2023,somepalli_diffusion_2022,Vyas_Kakade_Barak_2023} \\
Cybersecurity Threats  & Sources, Subjects, Regulators & Repr. harm, security loss & - & - \\
  \hline
 \multicolumn{5}{c}{Misinformation and Disinformation} \\[1ex]
 \hline
Likeness reproduction & Subjects, Users, Affected   & Repr. harm, emotional harm & \cite{milliere_deep_2022, mishkin_dalle_2022, nichol_glide_2022,Rombach_2022_CVPR, jursenas_double-edged_2021} & \cite{somepalli_diffusion_2022} \\ 
Misleading harmful content & Users, Affected  & Repr. harm, emotional harm & \cite{nguyen_deep_2022, boswald_what_2022, beyer_radar_2022, brady_deepfakes_2020, franks_sex_2019, paris_deepfakes_2019, gamir-rios_multimodal_2021, tomasev_manifestations_2022, jankowicz_addressing_2021} & \cite{nichol_glide_2022, widder_limits_2022} \\
Fraud and scams & Users, Affected & Emotional harm, financial loss & \cite{khoo_deepfake_2022, beyer_radar_2022, paris_deepfakes_2019, verdoliva_media_2020, mathews_explainable_2023} & - \\ 
Detection and classification bias & Developers, Subjects, Users, Affected & Allocative harm  & - & \cite{xu_comprehensive_2022, nadimpalli_gbdf_2022, lovato_diverse_2022, pu_fairness_2022, radford_learning_2021}\\
 Polarisation & Users, Affected & Repr. harm, incite violence  & \cite{akers_technology-enabled_2019, giomelakis_verification_2021, beyer_radar_2022, brady_deepfakes_2020, coeckelbergh_ai_2020}& \cite{lovato_diverse_2022} \\

 Miscommunication & Developers, Users, Affected & Allocative harm, loss of trust &  \cite{yu_scaling_2022, Hutchinson_Baldridge_Prabhakaran_2022,boswald_what_2022}  & - \\ 

 Soco-political Instability & Users, Affected, Regulators & Loss of trust, incite violence & \cite{coeckelbergh_ai_2020,westerlund_emergence_2019, vaccari_deepfakes_2020,oswald_what_2022, westerlund_emergence_2019,bateman_deepfakes_2020,bremmer_eurasia_2023,akhtar_deepfakes_2023} & \cite{lovato_diverse_2022} \\

 \hline
\end{tabular}}
\caption{Risk Typology. We provide detailed analysis in Section \ref{sec:risks}.}
\label{table:risk}
\end{table*}


In recent years, significant progress has been made in developing large language models and related multi-modal generative models, such as text-to-image models. We will collectively refer to these models as ``generative models.''\footnote{These models are also known by some researchers as foundation models \cite{bommasani2021opportunities}.} Generative models process and combine information from various modalities, including visual, textual and auditory data. The range of applications for generative models spans multiple fields. In entertainment, they can generate realistic-looking images or movie characters \cite{davenport2023cuebric,takahashi2023ai}. In advertising, these models can be employed to create personalized ad content \cite{boorstin2023generative, criddle2023google}. They can aid scientific research by simulating complex systems or hypothesizing about empirical phenomena \cite{beltagy2019scibert,birhane2023science,agathokleous2023questions}. In education, they can facilitate personalized learning, catering to unique needs and learning pace of each student \cite{vartiainen2023using,baidoo2023education}.

While introducing exciting opportunities, generative models also pose risks. These risks have attracted significant scrutiny from the AI ethics and safety community. The social and ethical risks of large language models, along with the text-to-text technologies they support, have been intensely discussed within the literature \cite{Bender_Gebru_McMillan-Major_Shmitchell_2021,weidinger_taxonomy_2022}. For instance, it is widely acknowledged that existing language technologies can potentially cause harm by producing inappropriate, discriminatory, or harmful content \cite[][i.a.]{Dhamala_Sun_Kumar_Krishna_Pruksachatkun_Chang_Gupta_2021, Gehman_Gururangan_Sap_Choi_Smith_2020, Dev_Monajatipoor_Ovalle_Subramonian_Phillips_Chang_2021, weidinger_ethical_2021}, or that the alignment of language technologies with beneficial human values is far from a straight forward task \cite{bai2022constitutional,kasirzadeh_2023_conversation,durmus2023towards}. This paper extends this line of inquiry from language models to text-to-image generative models, examining potential risks and harms resulting from their development and use. To identify and illuminate these risks, we perform a comprehensive review of literature related to text-to-image (TTI) models. In particular, we conduct an initial search using 8 seed papers, supplementing with manual search (our search methodology is detailed in Appendix A). Collected papers are analysed for immediate risks, stakeholders, and empirical investigations.

Our systematic examination yields a typology of risks associated with state-of-the-art TTI models, such as DALL-E 2 \cite{ramesh_hierarchical_2022}. Our findings are summarized in Table \ref{table:risk}. Our typology and discussion analysis are limited to immediate risks, inspired by a taxonomy from Weidinger et al. \cite{weidinger_ethical_2021}. Our typology is divided into three key categories: I. Discrimination and Exclusion; II. Harmful Misuse; III. Misinformation and Disinformation. We recognize that these categories are not mutually exclusive. However, defining distinct categories enables clearer understanding and supports the implementation of more robust mitigation strategies.

Our typology is further refined by identifying the stakeholders involved in the development and use of these systems. Inspired by the probing question from \citet{Blodgett_Barocas_2020}: ``How are social hierarchies, language ideologies, and NLP systems co-produced?'', we interlace this concern into our research and typology formulation. This process helps us to illustrate how the technologies supported by TTI models can reinforce existing social hierarchies via stakeholder identification. 

We adopt the stakeholder categories of developers, users, regulators and affected parties from \citet{Langer_Oster_Speith_Hermanns_Kästner_Schmidt_Sesing_Baum_2021}. We use ``affected parties'' referring to those influenced by the output of these models. We further extend the categorization by introducing ``data sources'' and ``data subjects'' -- individuals or entities who generate and/or appear in the images used to train TTI models. Additionally, we ascribe the nature of potential harm, such as representational or allocative \cite{Barocas_Crawford_Shapiro_Wallach_2017}, to the identified stakeholders. We also touch upon risks of harm to environment \cite{Bender_Gebru_McMillan-Major_Shmitchell_2021, Newton_Dhole_2023}.

To organize the literature, we propose a practical distinction between two types of risks: ``anticipated'' and ``observed.'' The former refers to risks that are primarily predicted by researchers due to their expertise and familiarity with the field. The latter, on the other hand, are risks that have been empirically investigated, providing insights into the potential magnitude of harm. This classification underscores the need for comprehensive empirical investigations into many of the identified risks. With this distinction in mind, we highlight several risks that, to our knowledge, have not yet been adequately discussed. We further contribute with an analysis of the challenges posed by proposed mitigation strategies (in \ref{sec:mit}) and an identification of open questions, supplemented by suggestions for policy change (in \ref{sec:gaps}). Finally, we advocate for enhanced collaboration among researchers, system developers, and policymakers. Through our categorisation and discussion, our intention is to foster a better understanding of the potential futures -- both positive and negative -- of TTI models, and by extension, other generative models.


\section{Generative text-to-image models}\label{sec:gen}

A TTI model is a type of generative neural network designed to synthesise images based on textual prompts \cite{reed_generative_2016}. When given a prompt, the model generates an image that, in some sense, visually represents the information in the text. TTI systems typically leverage a combination of natural language processing (NLP) and computer vision techniques to produce images. The NLP component extracts relevant information such as objects, attributes, and relationships from the text, while the computer vision component generates an image based on this information.

Various generative architectures have shown promise in image synthesis tasks \cite{frolov_adversarial_2021}. These include flow-based models \cite{dinh_density_2017}, auto-regressive models \cite{oord_conditional_2016} and variational autoencoders \cite{kingma_auto-encoding_2022}. However, the advent of generative adversarial networks (GAN) \cite{goodfellow_generative_2014} marked a significant acceleration in the capabilities of generative models. 

A typical TTI GAN employs two types of deep neural networks -- a generator and a discriminator. The generator synthesizes an image from a text input, while the discriminator evaluates the generated image, determining its authenticity. Through adversarial training, the generator refines its ability to create increasingly realistic images. The introduction of transformer architecture in 2017 spurred substantial progress in NLP \cite{vaswani_attention_2017}, subsequently extending to vision tasks as evidenced by early versions of DALL-E. Additionally, CLIP \cite{radford_learning_2021}, a model that learns visual concepts from natural language supervision, became pivotal in image generation tasks. 

Diffusion models \cite{sohl-dickstein_deep_2015}, which define a Markov chain parameterized by deep neural networks to reverse noisy data and sample from a desired data distribution, have recently achieved state-of-the-art results in image synthesis \cite{song_score-based_2020, ho_denoising_2020, Rombach_2022_CVPR, dhariwal_diffusion_2022}. The success of these models has stimulated a rapid proliferation of popular and open-source diffusion models, which are the subject of many of the papers in this taxonomy.

\section{Stakeholders and power dynamics} \label{sec:stake}

A comprehensive discussion of stakeholders, emphasizing their relative power, is crucial for understanding the associated risks. As various researchers have articulated, it is essential to underscore power inequities by considering what might be absent from a dataset \cite{gebru2021datasheets,Markl_2022}. We build upon this observation, and various other insights on the relations between power structures and socio-technical algorithmic systems \cite{Blodgett_Barocas_2020,kasirzadeh2021ethical,Kasirzadeh_2022}, structuring our analysis around the inclusion or exclusion of various groups in the development and deployment of these models. In Table \ref{table:risk} and Section \ref{sec:risks}, we pinpoint six categories of stakeholders most likely to be impacted by the risks we identify: system developers, data sources, data subjects, users, affected parties, and regulators.

\subsection{System Developers}

Developing state-of-the-art TTI systems requires vast compute and storage capabilities. Consequently, development is dominated by actors who have such access, such as companies in the Global North and China. These tend to be primarily concentrated within a small group of for-profit companies and well-funded academic institutions (e.g. OpenAI, Meta, Stability AI, Google, DeepMind, Midjourney). Companies like Hugging Face are making efforts towards open-access TTI systems. However, it still remains unclear how these models compare competitively with for-profit models.

This concentration of resources can lead to a lack of diverse perspectives in the data curation and model development teams, which can result in the exacerbation of specific biases in the training data \cite{West_Whittaker_Crawford_2019}. As a result, source and output images that reflect only the hegemonic perspective might go unnoticed, as those curating the data or developing the models are often blinkered by their own experiences. For instance, \citet{bianchi_easily_2022} and \citet{yu_scaling_2022} found models reflected Western culture in their output, for example Western dining, wedding and clothing practices; and ``couples'' and ``families'' were exclusively heterosexual.

\subsection{Data Sources}

Current data collection methodologies often deny content creators the opportunity to provide consent \cite{Ghosh_Fossas_2022} or be acknowledged as ``collaborators'' \cite{Sloane_Moss_Awomolo_Forlano_2022}. Furthermore, the widespread issue of inadequate curation in large datasets contributes to a multitude of problems \cite{birhane_prabhu} .\footnote{Inadequate curation can mean that the data may contain inaccuracies, bias, or irrelevant information, all of which can propagate into AI systems trained on such data, leading to unreliable or potentially harmful outcomes.} It results in opaque attributions, makes output reasoning convoluted, and complicates efforts towards harm reduction \cite{birhane_prabhu}.

Certain TTI systems have been shown to replicate images from their training data, which can be thought of as ``Digital Forgery'' \cite{somepalli_diffusion_2022}: artists may find that models trained on their images produce near identical copies. Further, popular datasets such as ImageNet, CelebA, COCO, and LAION have been criticized for issues related to attribution and consent \cite{birhane_multimodal_2021,Ghosh_Fossas_2022}. These concerns have even prompted legal actions by creators and stock image websites against companies that deploy such technologies \cite{Brittain_2023a,Brittain_2023b, Wiggers_2023}.

\subsection{Data Subjects}

The concern that ``data available online may not have been intended for such usage'' is significant \cite{Carlini_Hayes_Nasr_Jagielski_Sehwag_Tramèr_Balle_Ippolito_Wallace_2023}. While much of the public discourse around TTI systems has concentrated on copyright issues regarding training datasets, we bring attention to the problem of image subjects' consent, including situations of conflicting consent \cite{Keküllüoğlu_Kökciyan_Yolum_2016,Kökciyan_Yaglikci_Yolum_2017}. 

The matter of image reproduction must be contemplated within the scope of privacy \cite{somepalli_diffusion_2022}. This concern applies to instances such as the unauthorized use of celebrity images or pornographic depictions of sex workers. While the focus often centers on the harm incurred by exposure to explicit content, the potential negative impact on the subjects of these images should not be overlooked. Explicit content is prevalent in many datasets, and users frequently retrain models to generate specific explicit content. However, some subjects of these images, such as sex workers, are not adequately considered in these discussions (though c.f. \citet{birhane_prabhu}).

\subsection{Users}\label{sec:whouse}

Before discussing typical users, we highlight that access to TTI models can be exclusionary. Commercial models often preclude certain territories, and successful use of these systems requires fluency in the input language (matching the dialect of the training data), or access to an accurate translation tool. We delve deeper into these issues further in Section \ref{sec:gaps}.

TTI systems can serve as powerful tools for professionals in fields such as design, advertising, and art \cite{cetinic_understanding_2022,Newton_Dhole_2023,Seneviratne_Senanayake_Rasnayaka_Vidanaarachchi_Thompson_2022,Moreno_2022}. They represent fresh avenues of exploration for creative individuals \cite{saharia_photorealistic_2022, coeckelbergh_can_2017, Oppenlaender_2022a,Oppenlaender_2022a}, and can offer accessible resources for a wider audience \cite{yu_scaling_2022}, even holding potential to ``democratise'' art \cite{Newton_Dhole_2023,Oppenlaender_2022a}. The fact that Stable Diffusion boasts ten million daily active users \cite{fatunde_digital_2022} testifies to the public's keen interest in leveraging TTI models for their personal entertainment.

On the flip side, TTI systems can be used for malicious purposes. In the realm of misinformation and disinformation, players such as hyper-partisan media, authoritarian regimes, state disinformation actors, and cyber-criminals have been identified as potential malicious users \cite{akhtar_deepfakes_2023,beyer_radar_2022,akers_technology-enabled_2019}. ``Information operations'' \cite{mishkin_dalle_2022} are broadly acknowledged as a malicious use case. Additionally, \citet{paris_deepfakes_2019} have identified a subset of enthusiasts, both unskilled and skilled hobbyists, who create harmful content, a substantial portion of which is pornographic. This exploitative content often gains viral attention \cite{adjer_state_2019}.

\subsection{Affected Parties}

This section highlights both direct and indirect stakeholders who may be impacted by TTI systems.

\paragraph{Creatives} TTI systems can empower creatives by expanding their toolkit, but it is crucial to note that even unintentional misuse of TTI systems can trigger adverse consequences. These systems may inadvertently encourage accidental plagiarism or digital forgery \cite{somepalli_diffusion_2022} or may unintentionally perpetuate the dominance of Western art styles \cite{yu_scaling_2022}, thus limiting the representation of diverse cultural aesthetics. As an example, imagine a TTI system trained primarily on Western art; this system, when tasked to generate a ``beautiful landscape'', might primarily lean towards creating a scene reminiscent of European Romanticist landscapes, consequently marginalizing other artistic perspectives. Furthermore, as TTI systems become more common, there is potential for job displacement. For example, Marvel's use of AI image generation in creating credits \cite{Horton_2023} provides a foretaste of this possibility. 

Consequently, creatives may feel compelled to interact with TTI models to defend their livelihood and stay competitive \footnote{A sentiment echoed by StabilityAI's CEO \cite{emad_emostaque_its_2023}.}. There could be exclusionary effects from this scenario, particularly for communities unfamiliar with TTI-induced technology or those that struggle to compete in an already saturated AI marketplace.

\paragraph{Marginalised Peoples} Marginalised communities are often not authentically represented within training data, resulting in generated images that stereotype or offend these communities \cite{bianchi_easily_2022, Ungless_Ross_Lauscher_2023}. As \citet{Bender_Gebru_McMillan-Major_Shmitchell_2021} point out, language models trained on internet data tend to encode stereotypical and derogatory associations based on gender, race, ethnicity, and disability status, a problem that extends to TTI models \cite{birhane_multimodal_2021,bianchi_easily_2022, Wolfe_Yang_Howe_Caliskan_2022}. As an example of ``outcome homogenisation" \cite{bommasani_picking_2022} -- where certain groups repeatedly encounter negative outcomes -- these stereotypical images could further ``corrupt" future TTI datasets \cite{Hataya_Bao_Arai_2022}. More alarmingly, these images might become part of training datasets for downstream technologies, such as robotics \cite{Kapelyukh_Vosylius_Johns_2022}, spreading the risks associated with data recycling across various domains.

\paragraph{Other} In terms of broader societal impacts, the creation of synthetic disinformation and misinformation represent highly visible and often viral risks associated with synthetic visual media \cite{tolosana_deepfakes_2020}.  These risks are particularly acute for women and public figures, who face character assassination through fake news or deepfake pornographic content \cite{franks_sex_2019,milliere_deep_2022,widder_limits_2022,paris_deepfakes_2019}. Moreover, the destabilising potential of generative AI, such as providing visual legitimacy to populist or nationalist conspiracies and fake news \cite{lovato_diverse_2022,westerlund_emergence_2019, akhtar_deepfakes_2023,brady_deepfakes_2020}, should not be overlooked. It is crucial to recognise that while all media consumers are vulnerable to these harms, those with less societal power to contest falsehoods -- people of colour, women, LGBTQ+ communities \cite{paris_deepfakes_2019} -- are particularly at risk. 

Additionally, communities with restricted access to digital resources, such as sanctioned communities from global majority or closed network users, may suffer disproportionate allocative harms due to unequal access to detection software for fact-checking \cite{leibowicz_deepfake_2021} or inadequate data protections \cite{jursenas_double-edged_2021}. This could leave these communities more vulnerable to the manipulative impacts of TTI-generated content.

\subsection{Regulators}

Regulatory bodies are established by governments or other organizations to oversee the functioning of AI companies and markets. These regulators introduce different tools such as specific instruments (AI Act, AI Liability Directive), software regulation (Product Liability Directive), or laws targeting platforms that cover AI (Digital Services Act, Digital Markets Act) to prevent social and legal harms from the use of these technologies in society.

These tools could potentially address some socio-legal concerns associated with TTI systems and similar generative model-induced technologies, including data privacy, intellectual property infringement, and security vulnerabilities \cite{hacker2023regulating,samuelson2023legal,veale2023ai}. For instance, the EU AI Act can help provide a legal framework for the responsible use of TTI systems, setting out the rights and responsibilities of different stakeholders \cite{edwards2021eu,kazim2022proposed,Madiega2023,helberger2023chatgpt}. Privacy laws might be adjusted to regulate the collection, storage, and use of personal data used to train or operate TTI models, thereby safeguarding individual privacy \citet{samuelson2023legal}. The Product Liability Directive \cite{cabral2020liability,hacker2022european} could be adapted to ensure that products resulting from TTI technologies are safe and fit for their intended use. Also, cybersecurity regulations could be used to ensure that TTI models are secure and protected from unauthorized access, hacking, or other forms of cyberattacks \cite{renaud2023chatgpt,sebastian2023chatgpt}. 

The critical and urgent question remains: How can these existing regulatory tools be effectively adapted and applied to address the unique challenges posed by TTI technologies? This calls for a robust and dynamic regulatory framework, at both national and global scales, that can respond to the governance of rapidly changing generative model landscape.

\section{Risks}\label{sec:risks}

In this section, we elaborate on the risks specified in Table \ref{table:risk}, providing necessary context, and identifying the stakeholders who would be most impacted by these risks.

\subsection{Discrimination and Exclusion}\label{sec:riskdisc}

The risk of socially biased output, defined here as output that reflects and perpetuates stereotypes and social hierarchies, is well-recognized within the realm of TTI models \citep[][i.a.]{Ackermann_Li_2022,Qiu_Zhu_Shi_Wenzel_Tang_Zhao_Li_Li_2022,Milliere_2022,Newton_Dhole_2023,Hutchinson_Baldridge_Prabhakaran_2022, Ungless_Ross_Lauscher_2023}. Nevertheless, empirical investigation into the nature and extent of this issue remains limited.

\citet{bianchi_easily_2022} investigate biased output from StableDiffusion, revealing that the generated images perpetuate stereotypes linked to race, ethnicity, culture, gender, and social class. In addition, these models tend to amplify biases inherent in the training data, mirroring the findings of \citet{Zhao_Wang_Yatskar_Ordonez_Chang_2017}. For instance, the depiction of developers as exclusively male contrasts with actual occupational statistics \cite{bianchi_easily_2022}. Despite attempts at bias mitigation through methods like filtering and re-weighting the training data \cite{nichol_2022}, DALL-E 2 still exhibits bias, displaying elements of racism, ableism, and cisheteronormativity \cite{bianchi_easily_2022}.
 
The impact of these biases on stakeholders can be profound.\footnote{Some of these issues are discussed in the DALL-E 2 model card \cite{mishkin_dalle_2022}.} Testing for TTI models by \citet{Cho_Zala_Bansal_2022} reveals gender and racial bias in relation to certain occupations or objects in both DALL-E and StableDiffusion. Other studies, such as \citet{yu_scaling_2022} and \citet{Hutchinson_Baldridge_Prabhakaran_2022}, point to a Western skew in representation and warn about the potential for stereotype reinforcement. The consequences of such skewed representation could range from bolstering political agendas \cite{Newton_Dhole_2023} to strengthening hegemonic structures, intentionally or unintentionally. \citet{Ungless_Ross_Lauscher_2023} show that DALL-E mini, DALL-E 2, and StableDiffusion generate stereotyped images of non-cisgender identities, potentially exacerbating the discrimination faced by these communities.

Bias investigations in language technologies (as in the social sciences \cite{Sweeney_2009,Kuran_McCaffery_2004}) have typically centered on a narrow range of salient demographics, possibly underestimating the full extent of discrimination \cite{Dev_Sheng_Zhao_Amstutz_Sun_Hou_Sanseverino_Kim_Nishi_Peng_2022,Blodgett_Barocas_2020,goldfarbtarrant2023prompt} . In line with the findings from NLP research \cite{Blodgett_Barocas_2020}, there is a primary focus on dataset bias, with other sources of bias in the model life cycle being underexplored. 

Finally, the rise of TTI models holds the potential to reshape the landscape of many creative fields, including art and game development \cite{eloundou2023gpts,DeCremer2023,Obedkov2023}. Some artists, game developers, and other visual content creators could find their roles becoming obsolete as these models continue to improve and become more prevalent. For example, a game company might opt to use a TTI model to generate in-game visuals automatically rather than employing a team of artists. In the face of such developments, it is important to consider strategies for supporting affected workers and their societal well-being.

\subsection{Harmful Misuse}\label{sec:risktox}

In this section, we explore the potential for TTI models to be misused, whether intentionally or unintentionally. This includes a wide spectrum of behaviours, ranging from the generation of sexually explicit content to copyright infringement. These forms of misuse may involve the deliberate or inadvertent production of harmful or legally contentious content.

\paragraph{Sexualised imagery}
A significant concern is the ability of TTI models to generate sexualised imagery, a risk acknowledged by several technical TTI studies \cite{yu_scaling_2022,mishkin_dalle_2022,Rombach_2022_CVPR,nichol_glide_2022}. Empirical research provides evidence of TTI systems producing Not Safe For Work (NSFW) content \cite{yu_scaling_2022, Ungless_Ross_Lauscher_2023}. Non-consensual generated sexual imagery, often referred to as ``deepfake'' content \cite{franks_sex_2019, widder_limits_2022} can be deeply damaging to individuals, often women \cite{milliere_deep_2022,jankowicz_addressing_2021}, and can have negative consequences on the victim's ability to participate in public life. 

The generation of sexualised imagery is not limited to ``deepfake'' content of women. \citet{Wolfe_Yang_Howe_Caliskan_2022} found a high number of sexualised images (30\%+) produced by a Stable Diffusion model for prompts mentioning girls as young as 12 years old (neither tested model produced more than 11\% sexualised images of boys for any age). Recently, a BBC investigation found child sexual abuse imagery generated by AI was being traded online \cite{Crawford_Smith_2023}. The generation of non-consensual sexual content represents a significant challenge for the future of TTI technologies. Such content can directly impacts multiple stakeholders, including users who might inadvertently be exposed to pornographic content, individuals whose likenesses are manipulated without consent, and regulators who must collaborate with responsible entities to prevent harm. 

\paragraph{Violent or taboo content}\citet{Hutchinson_Baldridge_Prabhakaran_2022} argue that TTI models may unintentionally violate cultural taboos in their outputs. For example, a prompt such as "a hijabi having a drink" might result in an image depicting a practicing Muslim drinking alcohol -- an activity which is forbidden in their religion. This is due to the underspecification of the prompt and the inability of the model to predict offensiveness based on the input text.

Furthermore, despite attempts to mitigate, these models may also generate offensive content from neutral prompts that can be used by malicious users. The primary cause of such unwanted behavior is poor quality training data, as evidenced by \citet{Ungless_Ross_Lauscher_2023}. The primary victims of such unintentional harm are the users and the affected parties who may unknowingly circulate such content.

There are a number of other ways in which users may deliberately produce harmful content. This could involve bypassing safety mechanisms or injecting ``backdoors'' -- secret or undocumented means of bypassing normal authentication or encryption in a computer system -- into the models. A study by \citet{Struppek_Hintersdorf_Kersting_2022} shows that it is possible to train a ``poisoned" text encoder that generates harmful or unwanted images in response to certain trigger characters. 

In another example, \citet{Milliere_2022} discusses the potential for malicious users to use specific words or phrases to trick the TTI model into generating harmful content. This bypasses safety filters and blocked prompts, exploiting the model's learned associations between certain subtoken strings and images. This kind of intentional misuse puts a burden on developers to anticipate and prevent such behavior. Furthermore, there is a fear that malicious agents might use these tactics to generate hate speech or other harmful content targeted at minority groups, a concern that was particularly voiced by members of the non-cisgender community, according to a recent survey \cite{Ungless_Ross_Lauscher_2023}.

\paragraph{Privacy, copyright, and cybersecurity issues}
As previously discussed, TTI models such as Imagen and StableDiffusion often replicate content, even to the extent of producing images identical to the source content \cite{Carlini_Hayes_Nasr_Jagielski_Sehwag_Tramèr_Balle_Ippolito_Wallace_2023,somepalli_diffusion_2022}. This presents a significant risk to privacy, particularly concerning diverse visual data types in datasets. For example, LAION-5B includes private medical information \cite{BenjEdwards}. Furthermore, studies indicate that about 35\% of images duplicated by Stable Diffusion fall under explicit non-permissive copyright notice \cite{Carlini_Hayes_Nasr_Jagielski_Sehwag_Tramèr_Balle_Ippolito_Wallace_2023}.

Our previous discussion on copyright, mainly focused on the creative work under \textit{Affected Parties}, now broadens to emphasize the risks posed to marginalized creators who may not have the ability to legally defend their work. Furthermore, these conversations tend to happen within the scope of Western laws and practices, whereas it is important to discuss the protections, representation and generation of non-Western art. We also wish to further highlight the risks of ``digital forgery'' \cite{somepalli_diffusion_2022}. Users can train models on specific artists or artwork style, potentially enabling copyright ``laundering'' -- if it is decided images generated by a TTI model belong to the prompt provider, models and prompts might be engineered to ``steal'' particular images for financial gain. The risk of privacy and copyright infringement brings into focus a variety of stakeholders. Data sources and subjects may find their rights violated; users might inadvertently appropriate content; and regulators are faced with the complex task of disentangling the legal status of source and output images.

Building on the privacy and copyright issues, it is also crucial to consider potential cybersecurity threats posed by TTI models. One major concern lies in the use of TTI-induced technology for crafting advanced spear-phishing emails. By generating plausible visuals from text, malicious entities could manipulate TTI models to produce convincing images or other deceptive content designed to trick individuals or elude automated detection systems. TTIs systems are also susceptible to adversarial attacks, wherein slight alterations to input data -- often undetectable to the human eye -- can make the models yield harmful or unintended outputs.

\subsection{Misinformation and Disinformation}\label{sec:riskmis}

This section delves into the risks associated with the generation of misleading media content by TTI systems. These are classified into individual, social, or community-based risks. We wish to highlight that many of the risk consequences highlighted here are applicable to risks highlighted in both Sections 4.1 and 4.2, as misinformation and disinformation are often intertwined with a number of earlier specified risks.

\paragraph{Individual Harms}

The first category of risks pertains to personal harms resulting from misinformation and disinformation, targeting either individuals or groups. Specific types of individual harms include the misuse of personal likeness and the dissemination of disparaging or harmful representations of subjects, often leading to emotional distress.

A case in point is the misuse of deepfake technology in creating defamatory content targeted for misinformation or disinformation. Deepfake technology is not only exploited to generate explicit content featuring unsuspecting individuals, often celebrities, but also to damage the reputation and identity of the victims \cite{milliere_deep_2022,jankowicz_addressing_2021}. A prevalent example includes the use of deepfake pornography in smear campaigns, often adopting dominant narratives of incompetence, physical weakness or sexual depravity, and frequently relying on gendered tropes \cite{boswald_what_2022,jankowicz_addressing_2021}.

The misuse of TTI models extends beyond sexualised imagery, leading to harmful likeness reproduction in various other forms. Examples include the creation of fake journalism profiles \cite{khoo_deepfake_2022}, or use in blackmail, revenge \cite{hansen_ai_2022, nour_deepfakes_2021}, or identity theft for scams \cite{akhtar_deepfakes_2023, mathews_explainable_2023}. Furthermore, TTI-enabled misinformation and disinformation can reinforce existing cognitive biases \cite{akers_technology-enabled_2019}, amplifying narratives of ``otherness'' \cite{gamir-rios_multimodal_2021, tomasev_manifestations_2022}. This can unify and legitimise the beliefs of certain groups, while reinforcing negative and false views about others, leading to discriminatory actions against the ``other'' \cite{Ungless_Ross_Lauscher_2023}. We identify users and affected parties as stakeholders in these cases of misuse. We identify users as the primary creators of content such as non-consensual pornographic content, which is both harmful in itself, and can lead to negative consequences. Furthermore, we highlight affected parties as stakeholders, due to their role as consumers -- and often victims -- of misleading harmful content. Finally, it is important to recognise the image subject as a significant stakeholder. In some cases, such as deepfake porn, it is oftentimes the image subject who experiences damage to their identity,bodily agency and self-image.

The individual harms discussed here are primarily representational because they leverage and reinforce the subordination of certain groups based on identity. Such harms also hold an emotional dimension. The distress caused by revenge porn and identity theft is well documented \cite{golladay_consequences_2017, bates_revenge_2017}, and synthetic media, due to their nature, can be endlessly regenerated. Moreover, we highlight the allocative harms that arise from these scenarios, such as the disparities seen in synthetic media detection tasks, a concern previously noted in facial recognition tasks involving people of colour \cite{buolamwini_gender_2018}. Current research suggests disparities across gender and race in classification tasks, which could influence misinformation detection \cite{radford_learning_2021,nadimpalli_gbdf_2022}. It is also worth noting that human detection efforts exhibit significant homophily \cite{lovato_diverse_2022}, suggesting that the risks of harmful content may be exacerbated by limited human detection ability and unbalanced detection data.

We highlight a number of stakeholders in our identification of detection and classification bias in a misinformation or disinformation context. We firstly identify system developers as stakeholders. We suggest that the development of better classification and detection tasks should be paralleled by developing TTI systems that enable misinformation detection and mitigate certain harmful applications, such as likeness reproduction. Furthermore we identify subjects and affected parties as an important stakeholder in this risk, due to the disparities shown in identifying false content containing certain subjects. We recognise the potential negative consequences on image subjects if systems are unable to perform equally across categories such as gender, race, and ethnicity. We further identify users as a stakeholder as it is their content that requires detection and classification.

\paragraph{Social Harms} 

In addition to individual harms, misinformation and disinformation efforts can erode social networks and exacerbate polarisation. Facilitated by algorithmic curation in online social networks, or ``filter bubbles'' \cite{pariser_filter_2011}, alongside factors such as anonymity and extensive reach \cite{akers_technology-enabled_2019}, TTI-based misinformation and disinformation can be disseminated to receptive and susceptible audiences. Closed or siloed communities -- such as closed networks of Facebook users consistently exposed to homogeneous political content -- can develop decreased tolerance, resistance to new information, and intensified attitude polarisation \cite{lazer_science_2018, giomelakis_verification_2021}.

Misinformation and disinformation circulating within these closed circles are particularly perilous as they bypass formal fact-checking measures \cite{brady_deepfakes_2020} and diverse ``herd correction'' effects \cite{lovato_diverse_2022}. This is especially hazardous during crises, such as the COVID-19 pandemic \cite{rocha_impact_2021}. Consequently, victims often include individuals who depend on non-traditional media and closed communities for news, such as Facebook or Whatsapp \cite{trauthig_whatsapp_2022}, or those who consume low credibility news sources and demonstrate resistance to fact-checking \cite{saltz_encounters_2021}. Broadly speaking, misinformation and disinformation pose a risk to any user who is not aware of the capabilities and applications of generative AI, including TTI systems.

Misinformation and disinformation efforts can impact elements of epistemic agency \cite{coeckelbergh_ai_2020}. The flooding of information environments \cite{brady_deepfakes_2020, boswald_what_2022}, either by volume or falsity, can degrade user ability to decipher truth, thereby cultivating doubt in others and our own epistemic capabilities \cite{boswald_what_2022, coeckelbergh_ai_2020}. Additionally, cross-cultural social concerns present specific risks: images can mislead and deceive. \citet{Hutchinson_Baldridge_Prabhakaran_2022} suggest ``road signs, labels, gestures and facial expressions'' as forms that can cause harm in inappropriate contexts. The translation of forms, appearances, and meanings across cultures can lead to miscommunication \cite{yu_scaling_2022}. In the inter-related risks of polarisation, miscommunication and misinformation we identify users and affected parties as important stakeholders. For example, malicious users, as producers and amplifiers of misleading content, should be recognised for their role in exacerbating issues such as polarisation \cite{lawson_tribalism_2023}. 

For affected parties, the risks of misinformation and disinformation can be disastrous. As mentioned, misinformation and disinformation can incur a significant social cost by intensifying polarisation, fostering division, and promoting malicious behaviour \citet{lawson_tribalism_2023}. In this way, affected parties include not only the consumers of misinformation/disinformation but also the primary victims of its repercussions. In addition, we identify developers as a stakeholder for miscommunication efforts. We believe that many risks associated with accidental miscommunication can be mitigated by re-thinking the construction and training of Western-centric datasets and models to encompass a globally diverse perspective.

Harms that damage information ecosystems, via misinformation or disinformation, originally manifest as representational. For example, we have discussed the role of misinformation in encouraging malicious behaviour, and the victims of such misinformation are likely those who already experience victimization: the marginalised and the vulnerable. These representational harms exact a social cost not only on the immediate victim, but on the ability and willingness of a society to critically engage with, and question, misinformation and disinformation. Additionally, it is crucial to acknowledge the allocative nature of these harms. Specifically, how do we transform information environments so all have access to reliable, local and trustworthy media? In the case of aforementioned closed networks, how do we integrate balanced news to minimise harm? A case in point may be the politically charged disinformation surrounding non-gender conforming youth in present day America that has resulted in attempted bills to block gender affirming healthcare \cite{Trotta_Pierson_2023}, which has arguably arisen from charged disinformation environments. A further question arises in who, through education or resources, possesses the ability to identify misinformation and disinformation? These harms require multiple mitigating efforts both to protect the marginalised, but also to transform information consumption through education.

\paragraph{Community Harms}

TTI-enabled technologies can cause significant harm to communities. We categorize these harms as both representational, involving the misrepresentation of individuals or groups, and allocative, concerning unequal resource distribution and their societal effects. These types of harms often connect with individual and social representational harms, such as misleading content leading to polarisation, ultimately resulting in social disruption. 

TTI-enabled misinformation and disinformation can threaten social, political and financial systems.  We wish to highlight the potential of TTI technologies to cause political harms. TTI systems can further damage political institutions and compromise the integrity of democratic discourse \cite{brady_deepfakes_2020} through election interference \cite{westerlund_emergence_2019, akhtar_deepfakes_2023}, enabling misinformation and disinformation actors to operate at larger scales, and creating ``evidence'' to legitimize fake news or propaganda \cite{Newton_Dhole_2023, westerlund_emergence_2019,mishkin_dalle_2022}. In addition we highlight the risks posed wherein TTI systems are used to generate culturally offensive content. As mentioned, TTI systems offer the ability to generate culturally or politically offensive content through ``backdoors'', or simply because the precautions enacted by developers do not account for all cultures. For example, blasphemous content or images of religious or political figures are potentially deeply harmful to certain societies. 

Furthermore, these risks are concerning for communities who are more susceptible to democratic and social instabilities and may have fewer data protections \cite{leibowicz_deepfake_2021,jursenas_double-edged_2021,westerlund_emergence_2019}. 
The detrimental effects of TTI-enabled misinformation and disinformation extend to financial markets and economies, with potential for disruption \cite{raval_survey_2022, akhtar_deepfakes_2023, lovato_diverse_2022, osullivan2023}. TTI systems also has the potential to increase the risk of conflict and state violence \cite{nguyen_deep_2022, boswald_what_2022}. 

It is important to recognise the long term effects of such harms on broader community climates in relation to the individual harms mentioned previously. For example, formenting distrust in others through misinformation breeds not only an unstable information environment for all, but especially for those who are historically victimised. Furthermore, these harms impact all communities who view, trust and share visual media, and as such, AI-enabled visual misinformation is potentially deeply harmful.

\section{Mitigation Strategies}\label{sec:mit}

This section presents a discussion of potential mitigation strategies. Addressing the risks and harms associated with TTI systems often necessitates the integration of multiple mitigation approaches. Local mitigation, at the level of a single system, can possibly address instances of localised harm. However, for broad harms that occur at the level of community or society, multi-disciplinary and multi-stakeholder efforts are required to enact any meaningful mitigation. Such widespread mitigation strategies would necessitate significant changes in the current practices of TTI model and system development and deployment. We categorize mitigation strategies into participatory projects, operational solutions, technical solutions, and socio-legal interventions.

\paragraph{Participatory projects}

Participatory projects, which involve stakeholders in the decision-making processes of AI system design, present a potent mitigation strategy \cite{weidinger_ethical_2021}. The mechanisms for enabling participatory projects have been previously explored \cite{birhane_power_2022,bondi_envisioning_2021,51772,queerinai_queer_2023}. Participatory projects can involve redefining the principles of generative AI design to be more human-centric and inclusive \cite{weisz_toward_2023, houde_business_2020}, such as the creation of creative assistive technologies \cite{houde_business_2020, yu_scaling_2022, paris_deepfakes_2019}. Data acquisition, a fundamental aspect of these projects, can target underrepresented or misrepresented communities to address disparities \cite{Wang_Liu_Zhang_Kleiman_Kim_Zhao_Shirai_Narayanan_Russakovsky_2022}. It is crucial to navigate these projects with sensitivity to power dynamics and consent issues \cite{fussell_2019, Ungless_Ross_Lauscher_2023}. Without careful attention, these disparities may persist in the consultation process, undermining the effectiveness of participation \cite{Sloane_Moss_Awomolo_Forlano_2022}.

Certain solutions, such as ``opt-out'' functions may contribute to addressing copyright infringement, however this relies on artists' being aware of this use of their data, disadvantaging those with limited ``tech literacy''. It is important to recognise that participatory projects are not an afterthought, but rather as a proactive measure to counter discrimination and exclusion in AI. This entails not just balancing datasets but also focusing on representation and involvement of marginalized identities.

\paragraph{Operational solutions}

Operational solutions in the management of TTI models primarily include strategies such as the responsible release of models and open sourcing \cite{solaiman2023gradient}. The limited release strategy has been employed with models such as Imagen \cite{saharia_photorealistic_2022} and Parti \cite{yu_scaling_2022}, and in the staggered release of DALL-E 2 \cite{ramesh_hierarchical_2022}. This approach allows for a certain degree of control, potentially enabling the recall of the technology to prevent malicious uses or other unintended consequences. On the other hand, open sourcing facilitates mass stress testing and probing of the generative models \cite{Hutchinson_Baldridge_Prabhakaran_2022}. This can uncover potential vulnerabilities or biases in the models, allowing for improvements and the fostering of transparency. It is worth noting, however, that this approach must also consider and strive to avoid perpetuating issues of worker exploitation \cite{Shmueli_Fell_Ray_Ku_2021,perrigo_2023}.

However, both these solutions offer limited remedies if the underlying datasets and models remain wrongfully biased and harmful. Furthermore, these solutions do not fully address downstream impacts, such as job displacement, which may result from the widespread use of TTI-enabled technologies. Therefore, it is important to pair these operational strategies with consistent evaluation and reform of the models, their applications, and metrics for measuring their social impacts.

\paragraph{Technical solutions}

To tackle the potential pitfalls of TTI systems, various technical research strategies have been explored. Technical research primarily aims to build more robust, safe, and reliable models. Recent developments include ``find and replace'' methods \cite{park_judge_2022}, semantic steering \cite{Brack_Schramowski_Friedrich_Hintersdorf_Kersting_2022}, and filtering techniques \cite{nichol_glide_2022, mishkin_dalle_2022,birhane_multimodal_2021}. However, these strategies have their limitations. For instance, it has been argued that filtering could exacerbate bias \cite{Mauro_Schellmann_2023,nichol_2022} or fail to address it entirely \cite{birhane_multimodal_2021}. Furthermore, mitigation via prompt editing has shown to have limited impact due to the complex and embedded nature of biases \cite{bianchi_easily_2022}.

A significant body of research focuses on detection of synthetic media as a mitigation strategy. Techniques include the use of GAN architectures \cite{dang_detection_2020}, blockchain verification \cite{seneviratne_blockchain_2022}, fingerprinting \cite{yu_artificial_2020}, and watermarking \cite{yu_scaling_2022, wang_imagen_2022}. Whilst techniques such as watermarking do not directly mitigate harms, rather they identify the authenticity of output images \cite{yu_scaling_2022}, they can deter potential misuse. 

The expansion of fair detection capabilities \cite{dolhansky_deepfake_2020,nadimpalli_gbdf_2022, xu_comprehensive_2022} are promising, but, as investigated in \citet{leibowicz_deepfake_2021}, as of yet there is no perfect approach to the detection of synthetic media. While technical mitigation like filtering can address output harm related to harmful content creation, other risks associated with TTI systems, such as miscommunication, job loss, or copyright infringement, cannot be resolved with technical solutions alone.

\paragraph{Socio-legal interventions}

Mitigating harm in the context of TTI-enabled technologies could significantly benefit from the creation of legal and policy guidelines and regulations. Media literacy and user education have proven to be effective tools in addressing misinformation and manipulation, fostering critical engagement with digital content \cite{akers_technology-enabled_2019, westerlund_emergence_2019, boswald_what_2022,tomasev_manifestations_2022}. Increased corporate culpability could ensure more stringent fact-checking, transparent practices, and adherence to community standards, fostering an environment of accountability \cite{brady_deepfakes_2020, jursenas_double-edged_2021, boswald_what_2022,raval_survey_2022, seow_comprehensive_2022}. 

Government legislation and local and global regulation can play a pivotal role \cite{hacker2023regulating,samuelson2023legal,veale2023ai}, with potential measures ranging from defining limits to controlling the dissemination of harmful content \cite{westerlund_emergence_2019, brady_deepfakes_2020}. The strategy of limiting monetary rewards from the spread of misinformation can serve as a potent deterrent \cite{akers_technology-enabled_2019}.

In this dynamic and complex landscape, comprehensive and continuous research on the misinformation and disinformation environment becomes critical \cite{saltz_encounters_2021,zhou_synthetic_2023}. Labelling content is often proposed as an intervention; however, it may impact trust in non-labelled content \cite{freeze_fake_2021} and may have unforeseen negative consequences \cite{saltz_encounters_2021}. Therefore, the nuances of such interventions need careful consideration.

Notwithstanding these interventions, we must acknowledge potential challenges, such as resistance from tech companies due to economic interests, or concerns over infringement on free speech. Therefore, a balance needs to be struck to ensure these interventions are effective and proportionate.

\section{Open Questions and Future Research}\label{sec:gaps}

While the conducted review revealed a number of well-acknowledged risks associated with TTI systems, our analysis also highlighted several knowledge gaps. We briefly discuss these gaps in order to highlight open questions and future directions for research.

\paragraph{Output bias} We identified several forms of neglected output bias, including ageism and anti-Asian sentiment, for which we found no targeted mitigation strategies. Ageism, a bias observed in GAN face generators \cite{Salminen_Jung_Chowdhury_Jansen_2020}, remains a largely unexplored area in recent TTI research. Moreover, studies on racial bias tend to primarily focus on the contrast between Black Africans and White Americans or on distinctions between light and dark skin \cite{bianchi_easily_2022,Cho_Zala_Bansal_2022}. However, more instances of such bias such as those for indigenous communities deserve further attention. We also found limited research on the treatment of religious bias, such as in \citet{yu_scaling_2022}. These output biases can affect both users, who may struggle to generate appropriate images, and downstream parties who are exposed to content that primarily reflects established norms and stereotypes.

\paragraph{Dialect bias} TTI models have been shown to create discrimination beyond outputs. For example, TTI systems may favour white-aligned American English over other dialects \citep{Blodgett_OConnor_2017} or languages. Speakers of a limited number of languages - such as English and Chinese - are able to fully leverage these models. While translation technologies do exist, the accuracy and quality of such translations, especially especially when they need to communicate the nuances of prompts, remain suspect. Research on macaronic prompting demonstrates that DALL-E 2 has some ``understanding'' of other European languages, however primarily relies on English \citep{Milliere_2022}. 

Depending on the training data and processes used, users may need to conform linguistically to use TTI systems effectively. This, in turn, reinforces the idea that alternative English dialects are subpar \citep{Blodgett_OConnor_2017}.

\paragraph{Pre-release moderation}
The use of labour in traditionally pillaged countries\footnote{A term sustainability writer Aja Barber uses to highlight the role that exploitation of resources by the Global North had in these countries’ development.} to moderate the output of publicly available generative models has been reported \cite{perrigo_2023}. Moderation workers often experience psychological harm, with insufficient support \cite{perrigo_2023,hern_2019} and there is a power imbalance between those developing these models and profiting from their use, and those tasked with pre-release moderation. It is important that companies actively pursue fairer labour practices, so as to reduce harm for moderators.

\paragraph{Job displacement}
It is important to recognise the displacement of profit that is enabled by systems such as TTI models \cite{Ghosh_Fossas_2022}. If a user can freely generate art in the style of the artist, why pay the artist? However, we wish to draw attention to the nuances of this displacement, that is, the exacerbation of existing inequalities. The people already marginalised by society will be most impacted by this loss of income. Further, work opportunities in technology companies can be even more heavily skewed against gender and racial minorities than the creative industries\cite{West_Whittaker_Crawford_2019,Topaz_Higdon_Epps-Darling_Siau_Kerkhoff_Mendiratta_Young_2022}, meaning profits may be moving from female creatives of colour and into the pockets of white men running tech companies.

Furthermore, we wish to acknowledge the effects of job displacement on image subjects. For example, sex workers cannot currently exert agency over - nor profit - from their images being within training datasets. These images feed the creation of non-consensual pornographic material, often combining a sex worker's body with a celebrity face. We identified a website specifically designed to host models trained on individual sex workers, celebrities and public figures, in order to generate ``personalised'' porn. Furthermore, if stock imagery, advertisements or modelling photos come to frequently feature generated humans, \cite{Lomas_2022,Moreno_2022, Weatherbed_2023} it is important we assess who is being displaced. For example, do companies use generated imagery to fulfil a diversity target, rather than find humans? We recognise the possibility of disconnect between the appearance of racial, gender or other diversity in stock imagery and who is receiving compensation for their time.

\paragraph{Miscommunication}

We identify the problem of miscommunication across cultures and countries using TTI systems. This is especially significant in current TTI technology given the ability to rapidly create images from Western-centric datasets. Solutions to miscommunication require multi-disciplinary anthropological and technical research to understand the translation of forms and appearances into other cultures, and subsequently the building of inclusive datasets. Furthermore, we wish to highlight the problems related to flooding information environments with generated content. This is under-explored in the context of TTI systems, especially given the scale and speed of generation. This risk is not directly related to the types (and harms) of outputs produced, but considers the effects of mass synthetic media production on communities.

\paragraph{Socio-political instability}

Many researchers have explored the possible effects of AI on democratic processes and structures \cite{helbing_will_2019, nemitz_constitutional_2018}. We specifically call attention to the specific risks posed by TTI technologies, many of which are covered within this paper, such as the rise of populism and nationalism supported by false evidence, as has been recognised in present day America \cite{leonhardt_crisis_2022}, assisted by narratives of ``alternative facts''. We consider the possible use cases of TTI models within these contexts to be an important, and widening, gap in the literature. This topic requires research beyond political considerations only, and would benefit from alignment with deepfake research, some of which has already considered such risks.


\paragraph{Future research directions} Technology companies building TTI (and other generative) models have a responsibility to address many of the risks discussed here, however analysis of TTI models is insufficient without establishing benchmarks against which we can assess safe, ethical and fair performance. \citet{liang2022holistic} present a ``living benchmark'' for large language models. Similar frameworks need to be developed for TTI models. 

Building benchmarks and performance requirements necessitates input from a broad range of stakeholders including government, developers, research communities, image sources, subjects, users and vulnerable parties. The involvement of developers and researchers is especially vital given the high technical skill threshold of understanding generative models, as we have identified through the course of our analysis. The alignment of developmental goals with wider social goals will enable focused mitigation when harms arise, as current development and mitigation choices are left in the hands of technology companies. We also argue for the importance of mitigation strategies outside of technical solutions. 

Research producing actionable insights arising from methods such as interviews and case studies can assist in our understanding of the impact of synthetic media. Work such as the interview and diary study of \citet{saltz_encounters_2021}, who argue for a holistic understanding of misinformation environments, is essential. Interviews that engage with identified victims of TTI model harms would greatly assist the development of mitigation strategies; see, for example \citet{Ungless_Ross_Lauscher_2023}.

Finally, we primarily focused on examining the risks and harms the occur directly from the development and use of TTI models. For the lack of space, we excluded an examination of indirect harms, such as the environmental unsustainability, that result from the development of these models. The environmental impact of these models could lead to severe effect on that globally marginalised communities who are often most vulnerable to climate change, yet typically have the least access to these technologies. The environmental risks of developing and deploying TTI system is also highlighted in the context of Large Language Models (LLMs) \cite{Bender_Gebru_McMillan-Major_Shmitchell_2021}. This subject requires additional research to better understand the origins of the energy consumed in training TTI models, the global distribution of carbon emissions, and the regions most affected by these emissions. Moreover, potential strategies for using renewable energy sources in model training, as a key component of reducing environmental impact, should be explored.

\paragraph{Open questions}
The review and analysis conducted within this paper enabled our identification of a number of open questions. 

\begin{enumerate}
    \item    How can we rethink data gathering and output moderation with respect to privacy, ownership and identity? 
    
    For example:
    \begin{itemize}
        \item How do we implement functional and retroactive data deletion?
        \item How might source image creators be protected from ``copyright laundering''?
    \end{itemize}
    \item How can we ``protect'' future datasets from corruption by output images, and benchmark a ``good" dataset? 

\item     How do we allocate responsibility, and compensate for harm?
\item How can we best flag and mitigate offensive use?
    \item How do we manage TTI-enabled technologies with respect to non-Western communities, such as avoiding miscommunication?
    \item How can the environmental costs of training and using these models be attenuated?
    \item     How do we maintain a ``ground truth'' in data and visual media? 
    \item     What are the long-term social costs of generating visual content?

\end{enumerate}



There are a number of regulatory efforts currently addressing data access and the use of AI, with modifications underway to incorporate generative technologies like TTI models. These include the EU AI Act \cite{edwards2021eu,kazim2022proposed,Madiega2023,helberger2023chatgpt}, the Algorithmic Accountability Act in the US \cite{mokander2022us}, and China's Deep Synthesis Provisions \cite{interesse2022china}, among others. Multiple ongoing lawsuits could shape future legal perspectives on generative models, including TTI-induced systems. The outcomes of these cases are yet to be determined and will likely impact the regulatory landscape surrounding these AI technologies.\footnote{For reference, here are several ongoing litigation cases: Doe 1 et al v. GitHub et al, Case No. 4:2022cv06823 (N.D. Cal.); Andersen et al v. Stability AI et al, Case No. 3:23-cv-00201 (N.D. Cal.); Getty Images v. Stability AI, Case No. 1:2023cv00135 (D. Del.); Tremblay et al v OpenAI, Case No. 4:23-cv-03223(N.D. Cal.); Getty Images v Sability AI (England), Case IL-2023-000007. We thank Andres Guadamuz for providing information regarding these cases.}

As this paper cannot -- within the page limit -- adequately provide an exhaustive analysis of such relevant regulatory efforts, we offer five recommendations that we suggest would be useful in guiding generalised regulatory and policy initiatives. Some of these recommendations may already be covered by existing regulatory frameworks. Nonetheless, we believe it is beneficial to outline all of them here.

\begin{enumerate}
    \item Establish a multi-stakeholder benchmark for responsible and safe performance of TTI systems, with concern for the risks raised in our typology.
    \item Integrate digital literacy and media literacy into educational programs to help users understand the limitations and potential risks associated with TTI systems.
    \item Clearly communicate to users when their data will be used to train TTI systems and how resulting images might be used, and obtain explicit consent for such use.
    \item Ensure that copyright ownership is clearly identified and respected when generating images from text, and establish clear rules for attribution and usage.
    \item Develop novel, multi-stakeholder safeguards to prevent the creation and dissemination of inappropriate or harmful images, especially images that are discriminatory, violent, and threats to security.

\end{enumerate}

Further, we acknowledge that these recommendations are applicable to other multi-modal generative models. For example, the growing public discourse of apprehension and fear regarding AGI could be somewhat abated by Recommendation 2. We have hoped to highlight, throughout this paper, the importance of amplifying the voices of typically excluded stakeholders. By extension, we recognise the importance of fostering collaboration between the public, policymakers, industry leaders, researchers, and civil society organizations in order to ensure innovative, fair, effective regulatory frameworks.

\section{Conclusion}

This paper presented a typology of risk associated with TTI-induced technologies, followed by a succinct review of relevant mitigation strategies and a discussion of open questions concerning the development and use of TTI systems. Although we provided some preliminary recommendations, we acknowledge that additional perspectives, expertise, and research are necessary to refine this typology and enhance our understanding of the social implications of TTI systems.

\section*{Acknowledgments}

We would like to thank the UKRI Arts and Humanities Research Council (grant AH/X007146/1) for the policy fellowship that supported this work. We thank Shannon Vallor, Ewa Luger, and the members of Ada Lovelace Institute for helpful discussions. We also thank James Stewart, Lilian Edwards, Andres Guadamuz, and three anonymous reviewers whose comments improved our work. Eddie L. Ungless is supported by the UKRI Centre for Doctoral Training in Natural Language Processing, funded by UKRI (Grant EP/S022481/1) and the University of Edinburgh, School of Informatics. Charlotte Bird is supported by the Baillie Gifford PhD Scholarship at the Centre for Technomoral Futures.

\bibliographystyle{ACM-Reference-Format}
\bibliography{custom,misinfo}


\begin{thebibliography}{180}


\ifx \showCODEN    \undefined \def \showCODEN     #1{\unskip}     \fi
\ifx \showDOI      \undefined \def \showDOI       #1{#1}\fi
\ifx \showISBNx    \undefined \def \showISBNx     #1{\unskip}     \fi
\ifx \showISBNxiii \undefined \def \showISBNxiii  #1{\unskip}     \fi
\ifx \showISSN     \undefined \def \showISSN      #1{\unskip}     \fi
\ifx \showLCCN     \undefined \def \showLCCN      #1{\unskip}     \fi
\ifx \shownote     \undefined \def \shownote      #1{#1}          \fi
\ifx \showarticletitle \undefined \def \showarticletitle #1{#1}   \fi
\ifx \showURL      \undefined \def \showURL       {\relax}        \fi
\providecommand\bibfield[2]{#2}
\providecommand\bibinfo[2]{#2}
\providecommand\natexlab[1]{#1}
\providecommand\showeprint[2][]{arXiv:#2}

\bibitem[Ackermann and Li(2022)]%
        {Ackermann_Li_2022}
\bibfield{author}{\bibinfo{person}{J. Ackermann} {and} \bibinfo{person}{Minjun
  Li}.} \bibinfo{year}{2022}\natexlab{}.
\newblock \showarticletitle{High-Resolution Image Editing via Multi-Stage
  Blended Diffusion}.
\newblock \bibinfo{journal}{\emph{ArXiv}} (\bibinfo{year}{2022}).
\newblock
\urldef\tempurl%
\url{https://doi.org/10.48550/arXiv.2210.12965}
\showDOI{\tempurl}


\bibitem[Adjer et~al\mbox{.}(2019)]%
        {adjer_state_2019}
\bibfield{author}{\bibinfo{person}{Henry Adjer}, \bibinfo{person}{Giorgio
  Patrini}, \bibinfo{person}{Francesco Cavalli}, {and}
  \bibinfo{person}{Laurence Cullen}.} \bibinfo{year}{2019}\natexlab{}.
\newblock \bibinfo{booktitle}{\emph{The {State} of {Deepfakes}: {Landscape},
  threats and impact}}.
\newblock \bibinfo{type}{{T}echnical {R}eport}.
\newblock


\bibitem[Agathokleous et~al\mbox{.}(2023)]%
        {agathokleous2023questions}
\bibfield{author}{\bibinfo{person}{Evgenios Agathokleous},
  \bibinfo{person}{Matthias~C. Rillig}, \bibinfo{person}{Josep Pe{\~n}uelas},
  {and} \bibinfo{person}{Zhen Yu}.} \bibinfo{year}{2023}\natexlab{}.
\newblock \showarticletitle{One hundred important questions facing plant
  science derived using a large language model}.
\newblock \bibinfo{journal}{\emph{Trends in Plant Science}}
  (\bibinfo{year}{2023}).
\newblock
\urldef\tempurl%
\url{https://doi.org/10.1016/j.tplants.2023.06.008}
\showDOI{\tempurl}


\bibitem[Akers et~al\mbox{.}(2019)]%
        {akers_technology-enabled_2019}
\bibfield{author}{\bibinfo{person}{John Akers}, \bibinfo{person}{Gagan Bansal},
  \bibinfo{person}{Gabriel Cadamuro}, \bibinfo{person}{Christine Chen},
  \bibinfo{person}{Quanze Chen}, \bibinfo{person}{Lucy Lin},
  \bibinfo{person}{Phoebe Mulcaire}, \bibinfo{person}{Rajalakshmi Nandakumar},
  \bibinfo{person}{Matthew Rockett}, \bibinfo{person}{Lucy Simko},
  \bibinfo{person}{John Toman}, \bibinfo{person}{Tongshuang Wu},
  \bibinfo{person}{Eric Zeng}, \bibinfo{person}{Bill Zorn}, {and}
  \bibinfo{person}{Franziska Roesner}.} \bibinfo{year}{2019}\natexlab{}.
\newblock \bibinfo{title}{Technology-{Enabled} {Disinformation}: {Summary},
  {Lessons}, and {Recommendations}}.
\newblock
\newblock
\urldef\tempurl%
\url{https://doi.org/10.48550/arXiv.1812.09383}
\showDOI{\tempurl}
\newblock
\shownote{arXiv:1812.09383 [cs]}.


\bibitem[Akhtar(2023)]%
        {akhtar_deepfakes_2023}
\bibfield{author}{\bibinfo{person}{Zahid Akhtar}.}
  \bibinfo{year}{2023}\natexlab{}.
\newblock \showarticletitle{Deepfakes {Generation} and {Detection}: {A} {Short}
  {Survey}}.
\newblock \bibinfo{journal}{\emph{Journal of Imaging}} \bibinfo{volume}{9},
  \bibinfo{number}{1} (\bibinfo{date}{Jan.} \bibinfo{year}{2023}),
  \bibinfo{pages}{18}.
\newblock
\showISSN{2313-433X}
\urldef\tempurl%
\url{https://doi.org/10.3390/jimaging9010018}
\showDOI{\tempurl}


\bibitem[Bai et~al\mbox{.}(2022)]%
        {bai2022constitutional}
\bibfield{author}{\bibinfo{person}{Yuntao Bai}, \bibinfo{person}{Saurav
  Kadavath}, \bibinfo{person}{Sandipan Kundu}, \bibinfo{person}{Amanda Askell},
  \bibinfo{person}{Jackson Kernion}, \bibinfo{person}{Andy Jones},
  \bibinfo{person}{Anna Chen}, \bibinfo{person}{Anna Goldie},
  \bibinfo{person}{Azalia Mirhoseini}, \bibinfo{person}{Cameron McKinnon},
  {et~al\mbox{.}}} \bibinfo{year}{2022}\natexlab{}.
\newblock \showarticletitle{Constitutional ai: Harmlessness from ai feedback}.
\newblock \bibinfo{journal}{\emph{arXiv preprint arXiv:2212.08073}}
  (\bibinfo{year}{2022}).
\newblock


\bibitem[Baidoo-Anu and Owusu~Ansah(2023)]%
        {baidoo2023education}
\bibfield{author}{\bibinfo{person}{David Baidoo-Anu} {and}
  \bibinfo{person}{Leticia Owusu~Ansah}.} \bibinfo{year}{2023}\natexlab{}.
\newblock \showarticletitle{Education in the era of generative artificial
  intelligence (AI): Understanding the potential benefits of ChatGPT in
  promoting teaching and learning}.
\newblock \bibinfo{journal}{\emph{Available at SSRN 4337484}}
  (\bibinfo{year}{2023}).
\newblock


\bibitem[Bansal et~al\mbox{.}(2022)]%
        {Bansal_Yin_Monajatipoor_Chang_2022}
\bibfield{author}{\bibinfo{person}{Hritik Bansal}, \bibinfo{person}{Da Yin},
  \bibinfo{person}{Masoud Monajatipoor}, {and} \bibinfo{person}{Kai-Wei
  Chang}.} \bibinfo{year}{2022}\natexlab{}.
\newblock \showarticletitle{How well can Text-to-Image Generative Models
  understand Ethical Natural Language Interventions?}
\newblock  (\bibinfo{year}{2022}).
\newblock
\urldef\tempurl%
\url{https://doi.org/10.48550/ARXIV.2210.15230}
\showDOI{\tempurl}


\bibitem[Barocas et~al\mbox{.}(2017)]%
        {Barocas_Crawford_Shapiro_Wallach_2017}
\bibfield{author}{\bibinfo{person}{Solon Barocas}, \bibinfo{person}{Kate
  Crawford}, \bibinfo{person}{Aaron Shapiro}, {and} \bibinfo{person}{Hanna
  Wallach}.} \bibinfo{year}{2017}\natexlab{}.
\newblock \showarticletitle{The problem with bias: from allocative to
  representational harms in machine learning. Special Interest Group for
  Computing}.
\newblock \bibinfo{journal}{\emph{Information and Society (SIGCIS)}}
  \bibinfo{volume}{2} (\bibinfo{year}{2017}).
\newblock


\bibitem[Bateman(2020)]%
        {bateman_deepfakes_2020}
\bibfield{author}{\bibinfo{person}{John Bateman}.}
  \bibinfo{year}{2020}\natexlab{}.
\newblock \bibinfo{booktitle}{\emph{Deepfakes and {Synthetic} {Media} in the
  {Financial} {System}: {Assessing} {Threat} {Scenarios}}}.
\newblock \bibinfo{type}{{T}echnical {R}eport}. \bibinfo{institution}{Carnegie
  Endowment for International Peace}.
\newblock


\bibitem[Bates(2017)]%
        {bates_revenge_2017}
\bibfield{author}{\bibinfo{person}{Samantha Bates}.}
  \bibinfo{year}{2017}\natexlab{}.
\newblock \showarticletitle{Revenge {Porn} and {Mental} {Health}: {A}
  {Qualitative} {Analysis} of the {Mental} {Health} {Effects} of {Revenge}
  {Porn} on {Female} {Survivors}}.
\newblock \bibinfo{journal}{\emph{Feminist Criminology}} \bibinfo{volume}{12},
  \bibinfo{number}{1} (\bibinfo{date}{Jan.} \bibinfo{year}{2017}),
  \bibinfo{pages}{22--42}.
\newblock
\showISSN{1557-0851}
\urldef\tempurl%
\url{https://doi.org/10.1177/1557085116654565}
\showDOI{\tempurl}
\newblock
\shownote{Publisher: SAGE Publications}.


\bibitem[Beltagy et~al\mbox{.}(2019)]%
        {beltagy2019scibert}
\bibfield{author}{\bibinfo{person}{Iz Beltagy}, \bibinfo{person}{Kyle Lo},
  {and} \bibinfo{person}{Arman Cohan}.} \bibinfo{year}{2019}\natexlab{}.
\newblock \showarticletitle{SciBERT: A pretrained language model for scientific
  text}.
\newblock \bibinfo{journal}{\emph{arXiv preprint arXiv:1903.10676}}
  (\bibinfo{year}{2019}).
\newblock


\bibitem[Bender et~al\mbox{.}(2021)]%
        {Bender_Gebru_McMillan-Major_Shmitchell_2021}
\bibfield{author}{\bibinfo{person}{Emily~M. Bender}, \bibinfo{person}{Timnit
  Gebru}, \bibinfo{person}{Angelina McMillan-Major}, {and}
  \bibinfo{person}{Shmargaret Shmitchell}.} \bibinfo{year}{2021}\natexlab{}.
\newblock \showarticletitle{On the Dangers of Stochastic Parrots: Can Language
  Models Be Too Big?}. In \bibinfo{booktitle}{\emph{Proceedings of the 2021 ACM
  Conference on Fairness, Accountability, and Transparency}}.
  \bibinfo{publisher}{ACM}, \bibinfo{address}{Virtual Event Canada},
  \bibinfo{pages}{610–623}.
\newblock
\showISBNx{978-1-4503-8309-7}
\urldef\tempurl%
\url{https://doi.org/10.1145/3442188.3445922}
\showDOI{\tempurl}


\bibitem[Beyer and Boswald(2022)]%
        {beyer_radar_2022}
\bibfield{author}{\bibinfo{person}{Jan~Nicola Beyer} {and}
  \bibinfo{person}{Lena-Marie Boswald}.} \bibinfo{year}{2022}\natexlab{}.
\newblock \bibinfo{booktitle}{\emph{{ON} {THE} {RADAR}: {Mapping} the {Tools},
  {Tactics} and {Narratives} of {Tomorrow}’s {Disinformation}
  {Environment}}}.
\newblock \bibinfo{type}{{T}echnical {R}eport}. \bibinfo{institution}{Democracy
  Reporting International}.
\newblock


\bibitem[Bianchi et~al\mbox{.}(2022)]%
        {bianchi_easily_2022}
\bibfield{author}{\bibinfo{person}{Federico Bianchi},
  \bibinfo{person}{Pratyusha Kalluri}, \bibinfo{person}{Esin Durmus},
  \bibinfo{person}{Faisal Ladhak}, \bibinfo{person}{M. Cheng},
  \bibinfo{person}{Debora Nozza}, \bibinfo{person}{Tatsunori Hashimoto},
  \bibinfo{person}{Dan Jurafsky}, \bibinfo{person}{James~Y. Zou}, {and}
  \bibinfo{person}{Aylin Caliskan}.} \bibinfo{year}{2022}\natexlab{}.
\newblock \showarticletitle{Easily {Accessible} {Text}-to-{Image} {Generation}
  {Amplifies} {Demographic} {Stereotypes} at {Large} {Scale}}.
\newblock \bibinfo{journal}{\emph{ArXiv}} (\bibinfo{year}{2022}).
\newblock
\urldef\tempurl%
\url{https://doi.org/10.48550/arXiv.2211.03759}
\showDOI{\tempurl}


\bibitem[Birhane et~al\mbox{.}(2022a)]%
        {birhane_power_2022}
\bibfield{author}{\bibinfo{person}{Abeba Birhane}, \bibinfo{person}{William
  Isaac}, \bibinfo{person}{Vinodkumar Prabhakaran}, \bibinfo{person}{Mark
  Diaz}, \bibinfo{person}{Madeleine~Clare Elish}, \bibinfo{person}{Iason
  Gabriel}, {and} \bibinfo{person}{Shakir Mohamed}.}
  \bibinfo{year}{2022}\natexlab{a}.
\newblock \showarticletitle{Power to the {People}? {Opportunities} and
  {Challenges} for {Participatory} {AI}}. In \bibinfo{booktitle}{\emph{Equity
  and {Access} in {Algorithms}, {Mechanisms}, and {Optimization}}}
  \emph{(\bibinfo{series}{{EAAMO} '22})}. \bibinfo{publisher}{Association for
  Computing Machinery}, \bibinfo{address}{New York, NY, USA},
  \bibinfo{pages}{1--8}.
\newblock
\showISBNx{978-1-4503-9477-2}
\urldef\tempurl%
\url{https://doi.org/10.1145/3551624.3555290}
\showDOI{\tempurl}


\bibitem[Birhane et~al\mbox{.}(2022b)]%
        {51772}
\bibfield{author}{\bibinfo{person}{Abeba Birhane},
  \bibinfo{person}{William~Samuel Isaac}, \bibinfo{person}{Vinodkumar
  Prabhakaran}, \bibinfo{person}{Mark Díaz}, \bibinfo{person}{Madeleine~Clare
  Elish}, \bibinfo{person}{Iason Gabriel}, {and} \bibinfo{person}{Shakir
  Mohamed}.} \bibinfo{year}{2022}\natexlab{b}.
\newblock \showarticletitle{Frameworks and Challenges to Participatory AI}.
\newblock
\urldef\tempurl%
\url{https://arxiv.org/pdf/2209.07572.pdf}
\showURL{%
\tempurl}


\bibitem[Birhane et~al\mbox{.}(2023)]%
        {birhane2023science}
\bibfield{author}{\bibinfo{person}{Abeba Birhane}, \bibinfo{person}{Atoosa
  Kasirzadeh}, \bibinfo{person}{David Leslie}, {and} \bibinfo{person}{Sandra
  Wachter}.} \bibinfo{year}{2023}\natexlab{}.
\newblock \showarticletitle{Science in the age of large language models}.
\newblock \bibinfo{journal}{\emph{Nature Reviews Physics}}
  (\bibinfo{year}{2023}), \bibinfo{pages}{1--4}.
\newblock


\bibitem[Birhane and Prabhu(2021)]%
        {birhane_prabhu}
\bibfield{author}{\bibinfo{person}{A. Birhane} {and} \bibinfo{person}{V.
  Prabhu}.} \bibinfo{year}{2021}\natexlab{}.
\newblock \showarticletitle{Large image datasets: A pyrrhic win for computer
  vision?}. In \bibinfo{booktitle}{\emph{2021 IEEE Winter Conference on
  Applications of Computer Vision (WACV)}}. \bibinfo{publisher}{IEEE Computer
  Society}, \bibinfo{address}{Los Alamitos, CA, USA},
  \bibinfo{pages}{1536--1546}.
\newblock
\urldef\tempurl%
\url{https://doi.org/10.1109/WACV48630.2021.00158}
\showDOI{\tempurl}


\bibitem[Birhane et~al\mbox{.}(2021)]%
        {birhane_multimodal_2021}
\bibfield{author}{\bibinfo{person}{Abeba Birhane}, \bibinfo{person}{Vinay~Uday
  Prabhu}, {and} \bibinfo{person}{Emmanuel Kahembwe}.}
  \bibinfo{year}{2021}\natexlab{}.
\newblock \bibinfo{title}{Multimodal datasets: misogyny, pornography, and
  malignant stereotypes}.
\newblock
\newblock
\urldef\tempurl%
\url{http://arxiv.org/abs/2110.01963}
\showURL{%
\tempurl}
\newblock
\shownote{arXiv:2110.01963 [cs]}.


\bibitem[Blodgett et~al\mbox{.}(2020)]%
        {Blodgett_Barocas_2020}
\bibfield{author}{\bibinfo{person}{Su~Lin Blodgett}, \bibinfo{person}{Solon
  Barocas}, \bibinfo{person}{Hal Daumé~III}, {and} \bibinfo{person}{Hanna
  Wallach}.} \bibinfo{year}{2020}\natexlab{}.
\newblock \showarticletitle{Language (Technology) is Power: A Critical Survey
  of “Bias” in NLP}. In \bibinfo{booktitle}{\emph{Proceedings of the 58th
  Annual Meeting of the Association for Computational Linguistics}}.
  \bibinfo{publisher}{Association for Computational Linguistics},
  \bibinfo{address}{Online}, \bibinfo{pages}{5454–5476}.
\newblock
\urldef\tempurl%
\url{https://doi.org/10.18653/v1/2020.acl-main.485}
\showDOI{\tempurl}


\bibitem[Blodgett and O’Connor(2017)]%
        {Blodgett_OConnor_2017}
\bibfield{author}{\bibinfo{person}{Su~Lin Blodgett} {and}
  \bibinfo{person}{Brendan O’Connor}.} \bibinfo{year}{2017}\natexlab{}.
\newblock \showarticletitle{Racial Disparity in Natural Language Processing: A
  Case Study of Social Media African-American English}.
\newblock \bibinfo{journal}{\emph{arXiv:1707.00061 [cs]}} (\bibinfo{date}{Jun}
  \bibinfo{year}{2017}).
\newblock
\urldef\tempurl%
\url{http://arxiv.org/abs/1707.00061}
\showURL{%
\tempurl}
\newblock
\shownote{arXiv: 1707.00061}.


\bibitem[Bommasani et~al\mbox{.}(2022)]%
        {bommasani_picking_2022}
\bibfield{author}{\bibinfo{person}{Rishi Bommasani}, \bibinfo{person}{Kathleen
  Creel}, \bibinfo{person}{Ananya Kumar}, \bibinfo{person}{Dan Jurafsky}, {and}
  \bibinfo{person}{Percy Liang}.} \bibinfo{year}{2022}\natexlab{}.
\newblock \showarticletitle{Picking on the {Same} {Person}: {Does}
  {Algorithmic} {Monoculture} lead to {Outcome} {Homogenization}?}
\newblock
\urldef\tempurl%
\url{https://openreview.net/forum?id=-H6kKm4DVo}
\showURL{%
\tempurl}


\bibitem[Bommasani et~al\mbox{.}(2021)]%
        {bommasani2021opportunities}
\bibfield{author}{\bibinfo{person}{Rishi Bommasani}, \bibinfo{person}{Drew~A
  Hudson}, \bibinfo{person}{Ehsan Adeli}, \bibinfo{person}{Russ Altman},
  \bibinfo{person}{Simran Arora}, \bibinfo{person}{Sydney von Arx},
  \bibinfo{person}{Michael~S Bernstein}, \bibinfo{person}{Jeannette Bohg},
  \bibinfo{person}{Antoine Bosselut}, \bibinfo{person}{Emma Brunskill},
  {et~al\mbox{.}}} \bibinfo{year}{2021}\natexlab{}.
\newblock \showarticletitle{On the opportunities and risks of foundation
  models}.
\newblock \bibinfo{journal}{\emph{arXiv preprint arXiv:2108.07258}}
  (\bibinfo{year}{2021}).
\newblock


\bibitem[Bondi et~al\mbox{.}(2021)]%
        {bondi_envisioning_2021}
\bibfield{author}{\bibinfo{person}{Elizabeth Bondi}, \bibinfo{person}{Lily Xu},
  \bibinfo{person}{Diana Acosta-Navas}, {and} \bibinfo{person}{Jackson~A.
  Killian}.} \bibinfo{year}{2021}\natexlab{}.
\newblock \showarticletitle{Envisioning {Communities}: {A} {Participatory}
  {Approach} {Towards} {AI} for {Social} {Good}}. In
  \bibinfo{booktitle}{\emph{Proceedings of the 2021 {AAAI}/{ACM} {Conference}
  on {AI}, {Ethics}, and {Society}}} \emph{(\bibinfo{series}{{AIES} '21})}.
  \bibinfo{publisher}{Association for Computing Machinery},
  \bibinfo{address}{New York, NY, USA}, \bibinfo{pages}{425--436}.
\newblock
\showISBNx{978-1-4503-8473-5}
\urldef\tempurl%
\url{https://doi.org/10.1145/3461702.3462612}
\showDOI{\tempurl}


\bibitem[Boorstin(2023)]%
        {boorstin2023generative}
\bibfield{author}{\bibinfo{person}{Julia Boorstin}.}
  \bibinfo{year}{2023}\natexlab{}.
\newblock \bibinfo{title}{Generative A.I. is creating custom advertisements for
  marketing brands}.
\newblock
\newblock
\urldef\tempurl%
\url{https://www.cnbc.com/video/2023/04/13/generative-a-i-is-creating-custom-advertisements-for-marketing-brands.html}
\showURL{%
\tempurl}
\newblock
\shownote{Accessed: 2023-05-28}.


\bibitem[Boswald and Saab(2022)]%
        {boswald_what_2022}
\bibfield{author}{\bibinfo{person}{Lena-Marie Boswald} {and}
  \bibinfo{person}{Beatriz~Almeida Saab}.} \bibinfo{year}{2022}\natexlab{}.
\newblock \bibinfo{booktitle}{\emph{What a {Pixel} {Can} {Tell}:
  {Text}-to-{Image} {Generation} and its {Disinformation} {Potential}?}}
\newblock \bibinfo{type}{{T}echnical {R}eport}. \bibinfo{institution}{Democracy
  Reporting International}.
\newblock


\bibitem[Brack et~al\mbox{.}(2022)]%
        {Brack_Schramowski_Friedrich_Hintersdorf_Kersting_2022}
\bibfield{author}{\bibinfo{person}{Manuel Brack}, \bibinfo{person}{Patrick
  Schramowski}, \bibinfo{person}{Felix Friedrich}, \bibinfo{person}{Dominik
  Hintersdorf}, {and} \bibinfo{person}{Kristian Kersting}.}
  \bibinfo{year}{2022}\natexlab{}.
\newblock \showarticletitle{The Stable Artist: Steering Semantics in Diffusion
  Latent Space}.
\newblock  (\bibinfo{year}{2022}).
\newblock
\urldef\tempurl%
\url{https://doi.org/10.48550/ARXIV.2212.06013}
\showDOI{\tempurl}


\bibitem[Brady(2020)]%
        {brady_deepfakes_2020}
\bibfield{author}{\bibinfo{person}{Madeline Brady}.}
  \bibinfo{year}{2020}\natexlab{}.
\newblock \bibinfo{booktitle}{\emph{Deepfakes: a new disinformation threat?}}
\newblock \bibinfo{type}{{T}echnical {R}eport}. \bibinfo{institution}{Democracy
  Reporting International}.
\newblock


\bibitem[Bremmer and Kupchan(2023)]%
        {bremmer_eurasia_2023}
\bibfield{author}{\bibinfo{person}{Ian Bremmer} {and} \bibinfo{person}{Cliff
  Kupchan}.} \bibinfo{year}{2023}\natexlab{}.
\newblock \bibinfo{booktitle}{\emph{Eurasia {Group} {Top} {Risks}}}.
\newblock \bibinfo{type}{{T}echnical {R}eport}.
\newblock


\bibitem[Brittain(2023a)]%
        {Brittain_2023b}
\bibfield{author}{\bibinfo{person}{Blake Brittain}.}
  \bibinfo{year}{2023}\natexlab{a}.
\newblock \showarticletitle{Getty Images lawsuit says Stability AI misused
  photos to train AI}.
\newblock \bibinfo{journal}{\emph{Reuters}} (\bibinfo{date}{Feb}
  \bibinfo{year}{2023}).
\newblock
\urldef\tempurl%
\url{https://www.reuters.com/legal/getty-images-lawsuit-says-stability-ai-misused-photos-train-ai-2023-02-06/}
\showURL{%
\tempurl}


\bibitem[Brittain(2023b)]%
        {Brittain_2023a}
\bibfield{author}{\bibinfo{person}{Blake Brittain}.}
  \bibinfo{year}{2023}\natexlab{b}.
\newblock \bibinfo{title}{Lawsuits accuse AI content creators of misusing
  copyrighted work | Reuters}.
\newblock
\newblock
\urldef\tempurl%
\url{https://www.reuters.com/legal/transactional/lawsuits-accuse-ai-content-creators-misusing-copyrighted-work-2023-01-17/}
\showURL{%
\tempurl}


\bibitem[Buolamwini and Gebru(2018)]%
        {buolamwini_gender_2018}
\bibfield{author}{\bibinfo{person}{Joy Buolamwini} {and}
  \bibinfo{person}{Timnit Gebru}.} \bibinfo{year}{2018}\natexlab{}.
\newblock \showarticletitle{Gender {Shades}: {Intersectional} {Accuracy}
  {Disparities} in {Commercial} {Gender} {Classification}}. In
  \bibinfo{booktitle}{\emph{Proceedings of the 1st {Conference} on {Fairness},
  {Accountability} and {Transparency}}}. \bibinfo{publisher}{PMLR},
  \bibinfo{pages}{77--91}.
\newblock
\urldef\tempurl%
\url{https://proceedings.mlr.press/v81/buolamwini18a.html}
\showURL{%
\tempurl}
\newblock
\shownote{ISSN: 2640-3498}.


\bibitem[Cabral(2020)]%
        {cabral2020liability}
\bibfield{author}{\bibinfo{person}{Tiago~S{\'e}rgio Cabral}.}
  \bibinfo{year}{2020}\natexlab{}.
\newblock \showarticletitle{Liability and artificial intelligence in the EU:
  Assessing the adequacy of the current Product Liability Directive}.
\newblock \bibinfo{journal}{\emph{Maastricht Journal of European and
  Comparative Law}} \bibinfo{volume}{27}, \bibinfo{number}{5}
  (\bibinfo{year}{2020}), \bibinfo{pages}{615--635}.
\newblock


\bibitem[Carlini et~al\mbox{.}(2023)]%
  {Carlini_Hayes_Nasr_Jagielski_Sehwag_Tramèr_Balle_Ippolito_Wallace_2023}
\bibfield{author}{\bibinfo{person}{Nicholas Carlini}, \bibinfo{person}{Jamie
  Hayes}, \bibinfo{person}{Milad Nasr}, \bibinfo{person}{Matthew Jagielski},
  \bibinfo{person}{Vikash Sehwag}, \bibinfo{person}{Florian Tramèr},
  \bibinfo{person}{Borja Balle}, \bibinfo{person}{Daphne Ippolito}, {and}
  \bibinfo{person}{Eric Wallace}.} \bibinfo{year}{2023}\natexlab{}.
\newblock \showarticletitle{Extracting Training Data from Diffusion Models}.
\newblock  \bibinfo{number}{arXiv:2301.13188} (\bibinfo{date}{Jan}
  \bibinfo{year}{2023}).
\newblock
\urldef\tempurl%
\url{http://arxiv.org/abs/2301.13188}
\showURL{%
\tempurl}
\newblock
\shownote{arXiv:2301.13188 [cs]}.


\bibitem[Cetinic and She(2022)]%
        {cetinic_understanding_2022}
\bibfield{author}{\bibinfo{person}{Eva Cetinic} {and} \bibinfo{person}{James
  She}.} \bibinfo{year}{2022}\natexlab{}.
\newblock \showarticletitle{Understanding and {Creating} {Art} with {AI}:
  {Review} and {Outlook}}.
\newblock \bibinfo{journal}{\emph{ACM Transactions on Multimedia Computing,
  Communications, and Applications}} \bibinfo{volume}{18}, \bibinfo{number}{2}
  (\bibinfo{date}{Feb.} \bibinfo{year}{2022}), \bibinfo{pages}{66:1--66:22}.
\newblock
\showISSN{1551-6857}
\urldef\tempurl%
\url{https://doi.org/10.1145/3475799}
\showDOI{\tempurl}


\bibitem[Cho et~al\mbox{.}(2022)]%
        {Cho_Zala_Bansal_2022}
\bibfield{author}{\bibinfo{person}{Jaemin Cho}, \bibinfo{person}{Abhay Zala},
  {and} \bibinfo{person}{Mohit Bansal}.} \bibinfo{year}{2022}\natexlab{}.
\newblock \showarticletitle{DALL-Eval: Probing the Reasoning Skills and Social
  Biases of Text-to-Image Generative Models}.
\newblock  \bibinfo{number}{arXiv:2202.04053} (\bibinfo{date}{Nov}
  \bibinfo{year}{2022}).
\newblock
\urldef\tempurl%
\url{http://arxiv.org/abs/2202.04053}
\showURL{%
\tempurl}
\newblock
\shownote{arXiv:2202.04053 [cs]}.


\bibitem[Coeckelbergh(2017)]%
        {coeckelbergh_can_2017}
\bibfield{author}{\bibinfo{person}{Mark Coeckelbergh}.}
  \bibinfo{year}{2017}\natexlab{}.
\newblock \showarticletitle{Can {Machines} {Create} {Art}?}
\newblock \bibinfo{journal}{\emph{Philosophy \& Technology}}
  \bibinfo{volume}{30}, \bibinfo{number}{3} (\bibinfo{date}{Sept.}
  \bibinfo{year}{2017}), \bibinfo{pages}{285--303}.
\newblock
\showISSN{2210-5433, 2210-5441}
\urldef\tempurl%
\url{https://doi.org/10.1007/s13347-016-0231-5}
\showDOI{\tempurl}


\bibitem[Coeckelbergh(2020)]%
        {coeckelbergh_ai_2020}
\bibfield{author}{\bibinfo{person}{Mark Coeckelbergh}.}
  \bibinfo{year}{2020}\natexlab{}.
\newblock \bibinfo{booktitle}{\emph{{AI} {Ethics}}}.
\newblock \bibinfo{publisher}{MIT Press}.
\newblock
\showISBNx{978-0-262-53819-0}
\newblock
\shownote{Google-Books-ID: Gs\_XDwAAQBAJ}.


\bibitem[Crawford and Smith(2023)]%
        {Crawford_Smith_2023}
\bibfield{author}{\bibinfo{person}{Angus Crawford} {and} \bibinfo{person}{Tony
  Smith}.} \bibinfo{year}{2023}\natexlab{}.
\newblock \bibinfo{title}{Illegal trade in AI child sex abuse images exposed}.
\newblock
\newblock
\urldef\tempurl%
\url{https://www.bbc.co.uk/news/uk-65932372}
\showURL{%
\tempurl}


\bibitem[Cremer et~al\mbox{.}(2023)]%
        {DeCremer2023}
\bibfield{author}{\bibinfo{person}{David~De Cremer},
  \bibinfo{person}{Nicola~Morini Bianzino}, {and} \bibinfo{person}{Ben Falk}.}
  \bibinfo{year}{2023}\natexlab{}.
\newblock \bibinfo{booktitle}{\emph{How Generative AI Could Disrupt Creative
  Work}}.
\newblock Harvard Business Review.
\newblock
\urldef\tempurl%
\url{https://hbr.org/2023/04/how-generative-ai-could-disrupt-creative-work}
\showURL{%
\tempurl}
\newblock
\shownote{AI And Machine Learning}.


\bibitem[Criddle and Murphy(2023)]%
        {criddle2023google}
\bibfield{author}{\bibinfo{person}{Cristina Criddle} {and}
  \bibinfo{person}{Hannah Murphy}.} \bibinfo{year}{2023}\natexlab{}.
\newblock \bibinfo{title}{Google to deploy generative AI to create
  sophisticated ad campaigns}.
\newblock
\newblock
\urldef\tempurl%
\url{https://www.ft.com/content/36d09d32-8735-466a-97a6-868dfa34bdd5}
\showURL{%
\tempurl}


\bibitem[Dang et~al\mbox{.}(2020)]%
        {dang_detection_2020}
\bibfield{author}{\bibinfo{person}{Hao Dang}, \bibinfo{person}{Feng Liu},
  \bibinfo{person}{Joel Stehouwer}, \bibinfo{person}{Xiaoming Liu}, {and}
  \bibinfo{person}{Anil Jain}.} \bibinfo{year}{2020}\natexlab{}.
\newblock \bibinfo{title}{On the {Detection} of {Digital} {Face}
  {Manipulation}}.
\newblock
\newblock
\urldef\tempurl%
\url{https://doi.org/10.48550/arXiv.1910.01717}
\showDOI{\tempurl}
\newblock
\shownote{arXiv:1910.01717 [cs]}.


\bibitem[Davenport(2023)]%
        {davenport2023cuebric}
\bibfield{author}{\bibinfo{person}{Tom Davenport}.}
  \bibinfo{year}{2023}\natexlab{}.
\newblock \showarticletitle{Cuebric: Generative AI Comes To Hollywood}.
\newblock \bibinfo{journal}{\emph{Forbes}} (\bibinfo{date}{Mar}
  \bibinfo{year}{2023}).
\newblock
\urldef\tempurl%
\url{https://www.forbes.com/sites/tomdavenport/2023/03/13/cuebric-generative-ai-comes-to-hollywood/?sh=340acd52174b}
\showURL{%
\tempurl}


\bibitem[Dev et~al\mbox{.}(2021)]%
        {Dev_Monajatipoor_Ovalle_Subramonian_Phillips_Chang_2021}
\bibfield{author}{\bibinfo{person}{Sunipa Dev}, \bibinfo{person}{Masoud
  Monajatipoor}, \bibinfo{person}{Anaelia Ovalle}, \bibinfo{person}{Arjun
  Subramonian}, \bibinfo{person}{Jeff Phillips}, {and} \bibinfo{person}{Kai-Wei
  Chang}.} \bibinfo{year}{2021}\natexlab{}.
\newblock \showarticletitle{Harms of Gender Exclusivity and Challenges in
  Non-Binary Representation in Language Technologies}. In
  \bibinfo{booktitle}{\emph{Proceedings of the 2021 Conference on Empirical
  Methods in Natural Language Processing}}. \bibinfo{publisher}{Association for
  Computational Linguistics}, \bibinfo{address}{Online and Punta Cana,
  Dominican Republic}, \bibinfo{pages}{1968–1994}.
\newblock
\urldef\tempurl%
\url{https://doi.org/10.18653/v1/2021.emnlp-main.150}
\showDOI{\tempurl}


\bibitem[Dev et~al\mbox{.}(2022)]%
        {Dev_Sheng_Zhao_Amstutz_Sun_Hou_Sanseverino_Kim_Nishi_Peng_2022}
\bibfield{author}{\bibinfo{person}{Sunipa Dev}, \bibinfo{person}{Emily Sheng},
  \bibinfo{person}{Jieyu Zhao}, \bibinfo{person}{Aubrie Amstutz},
  \bibinfo{person}{Jiao Sun}, \bibinfo{person}{Yu Hou}, \bibinfo{person}{Mattie
  Sanseverino}, \bibinfo{person}{Jiin Kim}, \bibinfo{person}{Akihiro Nishi},
  \bibinfo{person}{Nanyun Peng}, {and} \bibinfo{person}{Kai-Wei Chang}.}
  \bibinfo{year}{2022}\natexlab{}.
\newblock \showarticletitle{On Measures of Biases and Harms in NLP}. In
  \bibinfo{booktitle}{\emph{Findings of the Association for Computational
  Linguistics: AACL-IJCNLP 2022}}. \bibinfo{publisher}{Association for
  Computational Linguistics}, \bibinfo{address}{Online only},
  \bibinfo{pages}{246–267}.
\newblock
\urldef\tempurl%
\url{https://aclanthology.org/2022.findings-aacl.24}
\showURL{%
\tempurl}


\bibitem[Dhamala et~al\mbox{.}(2021)]%
        {Dhamala_Sun_Kumar_Krishna_Pruksachatkun_Chang_Gupta_2021}
\bibfield{author}{\bibinfo{person}{Jwala Dhamala}, \bibinfo{person}{Tony Sun},
  \bibinfo{person}{Varun Kumar}, \bibinfo{person}{Satyapriya Krishna},
  \bibinfo{person}{Yada Pruksachatkun}, \bibinfo{person}{Kai-Wei Chang}, {and}
  \bibinfo{person}{Rahul Gupta}.} \bibinfo{year}{2021}\natexlab{}.
\newblock \showarticletitle{BOLD: Dataset and Metrics for Measuring Biases in
  Open-Ended Language Generation}. In \bibinfo{booktitle}{\emph{Proceedings of
  the 2021 ACM Conference on Fairness, Accountability, and Transparency}}
  \emph{(\bibinfo{series}{FAccT ’21})}. \bibinfo{publisher}{Association for
  Computing Machinery}, \bibinfo{address}{New York, NY, USA},
  \bibinfo{pages}{862–872}.
\newblock
\showISBNx{978-1-4503-8309-7}
\urldef\tempurl%
\url{https://doi.org/10.1145/3442188.3445924}
\showDOI{\tempurl}


\bibitem[Dhariwal and Nichol(2022)]%
        {dhariwal_diffusion_2022}
\bibfield{author}{\bibinfo{person}{Prafulla Dhariwal} {and}
  \bibinfo{person}{Alexander~Quinn Nichol}.} \bibinfo{year}{2022}\natexlab{}.
\newblock \showarticletitle{Diffusion {Models} {Beat} {GANs} on {Image}
  {Synthesis}}.
\newblock
\urldef\tempurl%
\url{https://openreview.net/forum?id=AAWuCvzaVt}
\showURL{%
\tempurl}


\bibitem[Dinh et~al\mbox{.}(2017)]%
        {dinh_density_2017}
\bibfield{author}{\bibinfo{person}{Laurent Dinh}, \bibinfo{person}{Jascha
  Sohl-Dickstein}, {and} \bibinfo{person}{Samy Bengio}.}
  \bibinfo{year}{2017}\natexlab{}.
\newblock \bibinfo{title}{Density estimation using {Real} {NVP}}.
\newblock
\newblock
\urldef\tempurl%
\url{https://doi.org/10.48550/arXiv.1605.08803}
\showDOI{\tempurl}
\newblock
\shownote{arXiv:1605.08803 [cs, stat]}.


\bibitem[Dolhansky et~al\mbox{.}(2020)]%
        {dolhansky_deepfake_2020}
\bibfield{author}{\bibinfo{person}{Brian Dolhansky}, \bibinfo{person}{Joanna
  Bitton}, \bibinfo{person}{Ben Pflaum}, \bibinfo{person}{Jikuo Lu},
  \bibinfo{person}{Russ Howes}, \bibinfo{person}{Menglin Wang}, {and}
  \bibinfo{person}{Cristian~Canton Ferrer}.} \bibinfo{year}{2020}\natexlab{}.
\newblock \bibinfo{title}{The {DeepFake} {Detection} {Challenge} ({DFDC})
  {Dataset}}.
\newblock
\newblock
\urldef\tempurl%
\url{https://doi.org/10.48550/arXiv.2006.07397}
\showDOI{\tempurl}
\newblock
\shownote{arXiv:2006.07397 [cs]}.


\bibitem[Durmus et~al\mbox{.}(2023)]%
        {durmus2023towards}
\bibfield{author}{\bibinfo{person}{E. Durmus}, \bibinfo{person}{K. Nyugen},
  \bibinfo{person}{T.~I. Liao}, \bibinfo{person}{N. Schiefer},
  \bibinfo{person}{A. Askell}, \bibinfo{person}{A. Bakhtin},
  \bibinfo{person}{C. Chen}, \bibinfo{person}{Z. Hatfield-Dodds},
  \bibinfo{person}{D. Hernandez}, \bibinfo{person}{N. Joseph}, {and}
  \bibinfo{person}{L. Lovitt}.} \bibinfo{year}{2023}\natexlab{}.
\newblock \showarticletitle{Towards Measuring the Representation of Subjective
  Global Opinions in Language Models}.
\newblock \bibinfo{journal}{\emph{arXiv preprint arXiv:2306.16388}}
  (\bibinfo{year}{2023}).
\newblock


\bibitem[Edwards(2022)]%
        {BenjEdwards}
\bibfield{author}{\bibinfo{person}{Benj Edwards}.}
  \bibinfo{year}{2022}\natexlab{}.
\newblock \bibinfo{title}{Artist finds private medical record photos in popular
  AI Training Data Set}.
\newblock
\newblock
\urldef\tempurl%
\url{https://arstechnica.com/information-technology/2022/09/artist-finds-private-medical-record-photos-in-popular-ai-training-data-set/}
\showURL{%
\tempurl}


\bibitem[Edwards(2021)]%
        {edwards2021eu}
\bibfield{author}{\bibinfo{person}{Lilian Edwards}.}
  \bibinfo{year}{2021}\natexlab{}.
\newblock \showarticletitle{The EU AI Act: a summary of its significance and
  scope}.
\newblock \bibinfo{journal}{\emph{Artificial Intelligence (the EU AI Act)}}
  \bibinfo{volume}{1} (\bibinfo{year}{2021}).
\newblock


\bibitem[Eloundou et~al\mbox{.}(2023)]%
        {eloundou2023gpts}
\bibfield{author}{\bibinfo{person}{Tyna Eloundou}, \bibinfo{person}{Sam
  Manning}, \bibinfo{person}{Pamela Mishkin}, {and} \bibinfo{person}{Daniel
  Rock}.} \bibinfo{year}{2023}\natexlab{}.
\newblock \showarticletitle{Gpts are gpts: An early look at the labor market
  impact potential of large language models}.
\newblock \bibinfo{journal}{\emph{arXiv preprint arXiv:2303.10130}}
  (\bibinfo{year}{2023}).
\newblock


\bibitem[{Emad [@EMostaque]}(2023)]%
        {emad_emostaque_its_2023}
\bibfield{author}{\bibinfo{person}{{Emad [@EMostaque]}}.}
  \bibinfo{year}{2023}\natexlab{}.
\newblock \bibinfo{title}{It’s not that generative {AI} will replace
  {\textless}digital profession{\textgreater}. {\textless}members of digital
  profession{\textgreater} that use generative {AI} will replace
  {\textless}members of digital profession{\textgreater} that don’t.}
\newblock
\newblock
\urldef\tempurl%
\url{https://twitter.com/EMostaque/status/1633265477769199617}
\showURL{%
\tempurl}


\bibitem[Fatunde and Tse(2022)]%
        {fatunde_digital_2022}
\bibfield{author}{\bibinfo{person}{Mureji Fatunde} {and}
  \bibinfo{person}{Crystal Tse}.} \bibinfo{year}{2022}\natexlab{}.
\newblock \showarticletitle{Digital {Media} {Firm} {Stability} {AI} {Raises}
  {Funds} at \$1 {Billion} {Value}}.
\newblock \bibinfo{journal}{\emph{Bloomberg.com}} (\bibinfo{date}{Oct.}
  \bibinfo{year}{2022}).
\newblock
\urldef\tempurl%
\url{https://www.bloomberg.com/news/articles/2022-10-17/digital-media-firm-stability-ai-raises-funds-at-1-billion-value}
\showURL{%
\tempurl}


\bibitem[Franks and Waldman(2019)]%
        {franks_sex_2019}
\bibfield{author}{\bibinfo{person}{Mary~Anne Franks} {and}
  \bibinfo{person}{Ari~Ezra Waldman}.} \bibinfo{year}{2019}\natexlab{}.
\newblock \showarticletitle{Sex, {Lies}, and {Videotape}: {Deep} {Fakes} and
  {Free} {Speech} {Delusions}}.
\newblock  (\bibinfo{year}{2019}).
\newblock


\bibitem[Freeze et~al\mbox{.}(2021)]%
        {freeze_fake_2021}
\bibfield{author}{\bibinfo{person}{Melanie Freeze}, \bibinfo{person}{Mary
  Baumgartner}, \bibinfo{person}{Peter Bruno}, \bibinfo{person}{Jacob~R.
  Gunderson}, \bibinfo{person}{Joshua Olin}, \bibinfo{person}{Morgan~Quinn
  Ross}, {and} \bibinfo{person}{Justine Szafran}.}
  \bibinfo{year}{2021}\natexlab{}.
\newblock \showarticletitle{Fake {Claims} of {Fake} {News}: {Political}
  {Misinformation}, {Warnings}, and the {Tainted} {Truth} {Effect}}.
\newblock \bibinfo{journal}{\emph{Political Behavior}} \bibinfo{volume}{43},
  \bibinfo{number}{4} (\bibinfo{date}{Dec.} \bibinfo{year}{2021}),
  \bibinfo{pages}{1433--1465}.
\newblock
\showISSN{1573-6687}
\urldef\tempurl%
\url{https://doi.org/10.1007/s11109-020-09597-3}
\showDOI{\tempurl}


\bibitem[Frolov et~al\mbox{.}(2021)]%
        {frolov_adversarial_2021}
\bibfield{author}{\bibinfo{person}{Stanislav Frolov}, \bibinfo{person}{Tobias
  Hinz}, \bibinfo{person}{Federico Raue}, \bibinfo{person}{Jörn Hees}, {and}
  \bibinfo{person}{Andreas Dengel}.} \bibinfo{year}{2021}\natexlab{}.
\newblock \showarticletitle{Adversarial {Text}-to-{Image} {Synthesis}: {A}
  {Review}}.
\newblock \bibinfo{journal}{\emph{Neural Networks}}  \bibinfo{volume}{144}
  (\bibinfo{date}{Dec.} \bibinfo{year}{2021}), \bibinfo{pages}{187--209}.
\newblock
\showISSN{08936080}
\urldef\tempurl%
\url{https://doi.org/10.1016/j.neunet.2021.07.019}
\showDOI{\tempurl}
\newblock
\shownote{arXiv:2101.09983 [cs]}.


\bibitem[Fussell(2019)]%
        {fussell_2019}
\bibfield{author}{\bibinfo{person}{Sidney Fussell}.}
  \bibinfo{year}{2019}\natexlab{}.
\newblock \bibinfo{title}{How an attempt at correcting bias in tech goes
  wrong}.
\newblock
\newblock
\urldef\tempurl%
\url{https://www.theatlantic.com/technology/archive/2019/10/google-allegedly-used-homeless-train-pixel-phone/599668/}
\showURL{%
\tempurl}


\bibitem[Gamir-Ríos et~al\mbox{.}(2021)]%
        {gamir-rios_multimodal_2021}
\bibfield{author}{\bibinfo{person}{José Gamir-Ríos}, \bibinfo{person}{Raquel
  Tarullo}, {and} \bibinfo{person}{Miguel Ibáñez-Cuquerella}.}
  \bibinfo{year}{2021}\natexlab{}.
\newblock \showarticletitle{Multimodal disinformation about otherness on the
  internet . {The} spread of racist, xenophobic and {Islamophobic} fake news in
  2020}.
\newblock \bibinfo{journal}{\emph{Anàlisi}} (\bibinfo{date}{July}
  \bibinfo{year}{2021}), \bibinfo{pages}{49--64}.
\newblock
\urldef\tempurl%
\url{https://doi.org/10.5565/rev/analisi.3398>}
\showDOI{\tempurl}


\bibitem[Gebru et~al\mbox{.}(2021)]%
        {gebru2021datasheets}
\bibfield{author}{\bibinfo{person}{Timnit Gebru}, \bibinfo{person}{Jamie
  Morgenstern}, \bibinfo{person}{Briana Vecchione},
  \bibinfo{person}{Jennifer~Wortman Vaughan}, \bibinfo{person}{Hanna Wallach},
  \bibinfo{person}{Hal~Daum{\'e} Iii}, {and} \bibinfo{person}{Kate Crawford}.}
  \bibinfo{year}{2021}\natexlab{}.
\newblock \showarticletitle{Datasheets for datasets}.
\newblock \bibinfo{journal}{\emph{Commun. ACM}} \bibinfo{volume}{64},
  \bibinfo{number}{12} (\bibinfo{year}{2021}), \bibinfo{pages}{86--92}.
\newblock


\bibitem[Gehman et~al\mbox{.}(2020)]%
        {Gehman_Gururangan_Sap_Choi_Smith_2020}
\bibfield{author}{\bibinfo{person}{Samuel Gehman}, \bibinfo{person}{Suchin
  Gururangan}, \bibinfo{person}{Maarten Sap}, \bibinfo{person}{Yejin Choi},
  {and} \bibinfo{person}{Noah~A. Smith}.} \bibinfo{year}{2020}\natexlab{}.
\newblock \showarticletitle{RealToxicityPrompts: Evaluating Neural Toxic
  Degeneration in Language Models}. In \bibinfo{booktitle}{\emph{Findings of
  the Association for Computational Linguistics: EMNLP 2020}}.
  \bibinfo{publisher}{Association for Computational Linguistics},
  \bibinfo{address}{Online}, \bibinfo{pages}{3356–3369}.
\newblock
\urldef\tempurl%
\url{https://doi.org/10.18653/v1/2020.findings-emnlp.301}
\showDOI{\tempurl}


\bibitem[Ghosh and Fossas(2022)]%
        {Ghosh_Fossas_2022}
\bibfield{author}{\bibinfo{person}{Avijit Ghosh} {and}
  \bibinfo{person}{Genoveva Fossas}.} \bibinfo{year}{2022}\natexlab{}.
\newblock \showarticletitle{Can There be Art Without an Artist?}
\newblock  (\bibinfo{year}{2022}).
\newblock
\urldef\tempurl%
\url{https://doi.org/10.48550/ARXIV.2209.07667}
\showDOI{\tempurl}


\bibitem[Giomelakis et~al\mbox{.}(2021)]%
        {giomelakis_verification_2021}
\bibfield{author}{\bibinfo{person}{Dimitrios Giomelakis}, \bibinfo{person}{Olga
  Papadopoulou}, \bibinfo{person}{Symeon Papadopoulos}, {and}
  \bibinfo{person}{Andreas Veglis}.} \bibinfo{year}{2021}\natexlab{}.
\newblock \showarticletitle{Verification of {News} {Video} {Content}:
  {Findings} from a {Study} of {Journalism} {Students}}.
\newblock \bibinfo{journal}{\emph{Journalism Practice}} (\bibinfo{date}{Aug.}
  \bibinfo{year}{2021}), \bibinfo{pages}{1--30}.
\newblock
\showISSN{1751-2786, 1751-2794}
\urldef\tempurl%
\url{https://doi.org/10.1080/17512786.2021.1965905}
\showDOI{\tempurl}


\bibitem[Goldfarb-Tarrant et~al\mbox{.}(2023)]%
        {goldfarbtarrant2023prompt}
\bibfield{author}{\bibinfo{person}{Seraphina Goldfarb-Tarrant},
  \bibinfo{person}{Eddie Ungless}, \bibinfo{person}{Esma Balkir}, {and}
  \bibinfo{person}{Su~Lin Blodgett}.} \bibinfo{year}{2023}\natexlab{}.
\newblock \bibinfo{title}{This Prompt is Measuring <MASK>: Evaluating Bias
  Evaluation in Language Models}.
\newblock
\newblock
\showeprint[arxiv]{2305.12757}~[cs.CL]


\bibitem[Golladay and Holtfreter(2017)]%
        {golladay_consequences_2017}
\bibfield{author}{\bibinfo{person}{Katelyn Golladay} {and}
  \bibinfo{person}{Kristy Holtfreter}.} \bibinfo{year}{2017}\natexlab{}.
\newblock \showarticletitle{The {Consequences} of {Identity} {Theft}
  {Victimization}: {An} {Examination} of {Emotional} and {Physical} {Health}
  {Outcomes}}.
\newblock \bibinfo{journal}{\emph{Victims \& Offenders}} \bibinfo{volume}{12},
  \bibinfo{number}{5} (\bibinfo{date}{Sept.} \bibinfo{year}{2017}),
  \bibinfo{pages}{741--760}.
\newblock
\showISSN{1556-4886}
\urldef\tempurl%
\url{https://doi.org/10.1080/15564886.2016.1177766}
\showDOI{\tempurl}
\newblock
\shownote{Publisher: Routledge \_eprint:
  https://doi.org/10.1080/15564886.2016.1177766}.


\bibitem[Goodfellow et~al\mbox{.}(2014)]%
        {goodfellow_generative_2014}
\bibfield{author}{\bibinfo{person}{Ian~J. Goodfellow}, \bibinfo{person}{Jean
  Pouget-Abadie}, \bibinfo{person}{Mehdi Mirza}, \bibinfo{person}{Bing Xu},
  \bibinfo{person}{David Warde-Farley}, \bibinfo{person}{Sherjil Ozair},
  \bibinfo{person}{Aaron Courville}, {and} \bibinfo{person}{Yoshua Bengio}.}
  \bibinfo{year}{2014}\natexlab{}.
\newblock \bibinfo{title}{Generative {Adversarial} {Networks}}.
\newblock
\newblock
\urldef\tempurl%
\url{https://doi.org/10.48550/arXiv.1406.2661}
\showDOI{\tempurl}
\newblock
\shownote{arXiv:1406.2661 [cs, stat]}.


\bibitem[Hacker(2022)]%
        {hacker2022european}
\bibfield{author}{\bibinfo{person}{Philipp Hacker}.}
  \bibinfo{year}{2022}\natexlab{}.
\newblock \showarticletitle{The European AI Liability Directives--Critique of a
  Half-Hearted Approach and Lessons for the Future}.
\newblock \bibinfo{journal}{\emph{arXiv preprint arXiv:2211.13960}}
  (\bibinfo{year}{2022}).
\newblock


\bibitem[Hacker et~al\mbox{.}(2023)]%
        {hacker2023regulating}
\bibfield{author}{\bibinfo{person}{Philipp Hacker}, \bibinfo{person}{Andreas
  Engel}, {and} \bibinfo{person}{Marco Mauer}.}
  \bibinfo{year}{2023}\natexlab{}.
\newblock \showarticletitle{Regulating ChatGPT and other large generative AI
  models}. In \bibinfo{booktitle}{\emph{Proceedings of the 2023 ACM Conference
  on Fairness, Accountability, and Transparency}}. \bibinfo{pages}{1112--1123}.
\newblock


\bibitem[Hansen(2022)]%
        {hansen_ai_2022}
\bibfield{author}{\bibinfo{person}{Rune~Klingenberg Hansen}.}
  \bibinfo{year}{2022}\natexlab{}.
\newblock \bibinfo{title}{{AI} {Image} {Generator}: {This} {Is} {Someone}
  {Thinking} {About} {Data} {Ethics} · {Dataetisk} {Tænkehandletank}}.
\newblock
\newblock
\urldef\tempurl%
\url{https://dataethics.eu/ai-image-generator-this-is-someone-thinking-about-data-ethics/}
\showURL{%
\tempurl}


\bibitem[Hataya et~al\mbox{.}(2022)]%
        {Hataya_Bao_Arai_2022}
\bibfield{author}{\bibinfo{person}{Ryuichiro Hataya}, \bibinfo{person}{Han
  Bao}, {and} \bibinfo{person}{Hiromi Arai}.} \bibinfo{year}{2022}\natexlab{}.
\newblock \showarticletitle{Will Large-scale Generative Models Corrupt Future
  Datasets?}
\newblock  (\bibinfo{year}{2022}).
\newblock
\urldef\tempurl%
\url{https://doi.org/10.48550/ARXIV.2211.08095}
\showDOI{\tempurl}


\bibitem[Helberger and Diakopoulos(2023)]%
        {helberger2023chatgpt}
\bibfield{author}{\bibinfo{person}{Natali Helberger} {and}
  \bibinfo{person}{Nicholas Diakopoulos}.} \bibinfo{year}{2023}\natexlab{}.
\newblock \showarticletitle{ChatGPT and the AI Act}.
\newblock \bibinfo{journal}{\emph{Internet Policy Review}}
  \bibinfo{volume}{12}, \bibinfo{number}{1} (\bibinfo{year}{2023}).
\newblock


\bibitem[Helbing et~al\mbox{.}(2019)]%
        {helbing_will_2019}
\bibfield{author}{\bibinfo{person}{Dirk Helbing}, \bibinfo{person}{Bruno~S.
  Frey}, \bibinfo{person}{Gerd Gigerenzer}, \bibinfo{person}{Ernst Hafen},
  \bibinfo{person}{Michael Hagner}, \bibinfo{person}{Yvonne Hofstetter},
  \bibinfo{person}{Jeroen van~den Hoven}, \bibinfo{person}{Roberto~V. Zicari},
  {and} \bibinfo{person}{Andrej Zwitter}.} \bibinfo{year}{2019}\natexlab{}.
\newblock \showarticletitle{Will {Democracy} {Survive} {Big} {Data} and
  {Artificial} {Intelligence}?}
\newblock In \bibinfo{booktitle}{\emph{Towards {Digital} {Enlightenment}:
  {Essays} on the {Dark} and {Light} {Sides} of the {Digital} {Revolution}}},
  \bibfield{editor}{\bibinfo{person}{Dirk Helbing}} (Ed.).
  \bibinfo{publisher}{Springer International Publishing},
  \bibinfo{address}{Cham}, \bibinfo{pages}{73--98}.
\newblock
\showISBNx{978-3-319-90869-4}
\urldef\tempurl%
\url{https://doi.org/10.1007/978-3-319-90869-4_7}
\showDOI{\tempurl}


\bibitem[Hern(2019)]%
        {hern_2019}
\bibfield{author}{\bibinfo{person}{Alex Hern}.}
  \bibinfo{year}{2019}\natexlab{}.
\newblock \bibinfo{title}{Revealed: Catastrophic effects of working as a
  facebook moderator}.
\newblock
\newblock
\urldef\tempurl%
\url{https://www.theguardian.com/technology/2019/sep/17/revealed-catastrophic-effects-working-facebook-moderator}
\showURL{%
\tempurl}


\bibitem[Ho et~al\mbox{.}(2020)]%
        {ho_denoising_2020}
\bibfield{author}{\bibinfo{person}{Jonathan Ho}, \bibinfo{person}{Ajay Jain},
  {and} \bibinfo{person}{Pieter Abbeel}.} \bibinfo{year}{2020}\natexlab{}.
\newblock \showarticletitle{Denoising {Diffusion} {Probabilistic} {Models}}.
\newblock  (\bibinfo{date}{June} \bibinfo{year}{2020}).
\newblock
\urldef\tempurl%
\url{https://doi.org/10.48550/arXiv.2006.11239}
\showDOI{\tempurl}


\bibitem[Horton(2023)]%
        {Horton_2023}
\bibfield{author}{\bibinfo{person}{Adrian Horton}.}
  \bibinfo{year}{2023}\natexlab{}.
\newblock \bibinfo{title}{Marvel faces backlash over ai-generated opening
  credits}.
\newblock
\newblock
\urldef\tempurl%
\url{https://www.theguardian.com/tv-and-radio/2023/jun/21/marvel-ai-generated-credits-backlash}
\showURL{%
\tempurl}


\bibitem[Houde et~al\mbox{.}(2020)]%
        {houde_business_2020}
\bibfield{author}{\bibinfo{person}{Stephanie Houde}, \bibinfo{person}{Vera
  Liao}, \bibinfo{person}{Jacquelyn Martino}, \bibinfo{person}{Michael Muller},
  \bibinfo{person}{David Piorkowski}, \bibinfo{person}{John Richards},
  \bibinfo{person}{Justin Weisz}, {and} \bibinfo{person}{Yunfeng Zhang}.}
  \bibinfo{year}{2020}\natexlab{}.
\newblock \bibinfo{title}{Business (mis){Use} {Cases} of {Generative} {AI}}.
\newblock
\newblock
\urldef\tempurl%
\url{http://arxiv.org/abs/2003.07679}
\showURL{%
\tempurl}
\newblock
\shownote{arXiv:2003.07679 [cs]}.


\bibitem[Hutchinson et~al\mbox{.}(2022)]%
        {Hutchinson_Baldridge_Prabhakaran_2022}
\bibfield{author}{\bibinfo{person}{Ben Hutchinson}, \bibinfo{person}{Jason
  Baldridge}, {and} \bibinfo{person}{Vinodkumar Prabhakaran}.}
  \bibinfo{year}{2022}\natexlab{}.
\newblock \showarticletitle{Underspecification in Scene
  Description-to-Depiction Tasks}. \bibinfo{publisher}{arXiv}.
\newblock
\urldef\tempurl%
\url{https://doi.org/10.48550/ARXIV.2210.05815}
\showDOI{\tempurl}


\bibitem[Interesse(2022)]%
        {interesse2022china}
\bibfield{author}{\bibinfo{person}{Giulia Interesse}.}
  \bibinfo{year}{2022}\natexlab{}.
\newblock \bibinfo{booktitle}{\emph{China to Regulate Deep Synthesis (Deepfake)
  Technology Starting 2023}}.
\newblock
\urldef\tempurl%
\url{https://www.china-briefing.com/news/china-to-regulate-deep-synthesis-deep-fake-technology-starting-january-2023/}
\showURL{%
\tempurl}
\newblock
\shownote{Accessed: 2023-07-08}.


\bibitem[Jankowicz et~al\mbox{.}(2021)]%
        {jankowicz_addressing_2021}
\bibfield{author}{\bibinfo{person}{Nina Jankowicz}, \bibinfo{person}{Sandra
  Pepera}, {and} \bibinfo{person}{Molly Middlehurst}.}
  \bibinfo{year}{2021}\natexlab{}.
\newblock \bibinfo{booktitle}{\emph{Addressing {Online} {Misogyny} and
  {Gendered} {Disinformation}: {A} {How}-{To} {Guide}}}.
\newblock \bibinfo{type}{{T}echnical {R}eport}. \bibinfo{institution}{National
  Democracy Institution}.
\newblock


\bibitem[Jursenas et~al\mbox{.}(2021)]%
        {jursenas_double-edged_2021}
\bibfield{author}{\bibinfo{person}{Alfonsas Jursenas},
  \bibinfo{person}{Kasparas Karlauskas}, \bibinfo{person}{Gediminas
  Maskeliunas}, {and} \bibinfo{person}{Julius Ruseckas}.}
  \bibinfo{year}{2021}\natexlab{}.
\newblock \bibinfo{booktitle}{\emph{The {Double}-{Edged} {Sword} of {AI}:
  {Enabler} of {Disinformation}}}.
\newblock \bibinfo{type}{{T}echnical {R}eport}. \bibinfo{institution}{Nato
  Strategic Communications}.
\newblock


\bibitem[Kapelyukh et~al\mbox{.}(2022)]%
        {Kapelyukh_Vosylius_Johns_2022}
\bibfield{author}{\bibinfo{person}{Ivan Kapelyukh}, \bibinfo{person}{Vitalis
  Vosylius}, {and} \bibinfo{person}{Edward Johns}.}
  \bibinfo{year}{2022}\natexlab{}.
\newblock \showarticletitle{DALL-E-Bot: Introducing Web-Scale Diffusion Models
  to Robotics}.
\newblock  \bibinfo{number}{arXiv:2210.02438} (\bibinfo{date}{Nov}
  \bibinfo{year}{2022}).
\newblock
\urldef\tempurl%
\url{http://arxiv.org/abs/2210.02438}
\showURL{%
\tempurl}
\newblock
\shownote{arXiv:2210.02438 [cs]}.


\bibitem[Kasirzadeh(2022)]%
        {Kasirzadeh_2022}
\bibfield{author}{\bibinfo{person}{Atoosa Kasirzadeh}.}
  \bibinfo{year}{2022}\natexlab{}.
\newblock \showarticletitle{Algorithmic Fairness and Structural Injustice:
  Insights from Feminist Political Philosophy}. In
  \bibinfo{booktitle}{\emph{AIES '22: Proceedings of the 2022 AAAI/ACM
  Conference on AI, Ethics, and Society}}. \bibinfo{pages}{349--356}.
\newblock
\urldef\tempurl%
\url{https://doi.org/10.1145/3514094.3534188}
\showDOI{\tempurl}


\bibitem[Kasirzadeh and Gabriel(2023)]%
        {kasirzadeh_2023_conversation}
\bibfield{author}{\bibinfo{person}{Atoosa Kasirzadeh} {and}
  \bibinfo{person}{Iason Gabriel}.} \bibinfo{year}{2023}\natexlab{}.
\newblock \showarticletitle{In conversation with Artificial Intelligence:
  aligning language models with human values}.
\newblock \bibinfo{journal}{\emph{Philosophy \& Technology}}
  \bibinfo{volume}{36}, \bibinfo{number}{2} (\bibinfo{year}{2023}),
  \bibinfo{pages}{1--24}.
\newblock
\urldef\tempurl%
\url{https://doi.org/10.1007/s13347-023-00606-x}
\showDOI{\tempurl}


\bibitem[Kasirzadeh and Klein(2021)]%
        {kasirzadeh2021ethical}
\bibfield{author}{\bibinfo{person}{Atoosa Kasirzadeh} {and}
  \bibinfo{person}{Colin Klein}.} \bibinfo{year}{2021}\natexlab{}.
\newblock \showarticletitle{The ethical gravity thesis: Marrian levels and the
  persistence of bias in automated decision-making systems}. In
  \bibinfo{booktitle}{\emph{Proceedings of the 2021 AAAI/ACM Conference on AI,
  Ethics, and Society}}. \bibinfo{pages}{618--626}.
\newblock


\bibitem[Kazim et~al\mbox{.}(2022)]%
        {kazim2022proposed}
\bibfield{author}{\bibinfo{person}{Emre Kazim}, \bibinfo{person}{Osman
  G{\"u}{\c{c}}l{\"u}t{\"u}rk}, \bibinfo{person}{Denise Almeida},
  \bibinfo{person}{Charles Kerrigan}, \bibinfo{person}{Elizabeth Lomas},
  \bibinfo{person}{Adriano Koshiyama}, \bibinfo{person}{Airlie Hilliard}, {and}
  \bibinfo{person}{Markus Trengove}.} \bibinfo{year}{2022}\natexlab{}.
\newblock \showarticletitle{Proposed EU AI Act—Presidency compromise text:
  select overview and comment on the changes to the proposed regulation}.
\newblock \bibinfo{journal}{\emph{AI and Ethics}} (\bibinfo{year}{2022}),
  \bibinfo{pages}{1--7}.
\newblock


\bibitem[Keküllüoğlu et~al\mbox{.}(2016)]%
        {Keküllüoğlu_Kökciyan_Yolum_2016}
\bibfield{author}{\bibinfo{person}{Dilara Keküllüoğlu},
  \bibinfo{person}{Nadin Kökciyan}, {and} \bibinfo{person}{Pınar Yolum}.}
  \bibinfo{year}{2016}\natexlab{}.
\newblock \showarticletitle{Strategies for Privacy Negotiation in Online Social
  Networks}. In \bibinfo{booktitle}{\emph{Proceedings of the 1st International
  Workshop on AI for Privacy and Security}}. \bibinfo{publisher}{ACM},
  \bibinfo{address}{The Hague Netherlands}, \bibinfo{pages}{1–8}.
\newblock
\showISBNx{978-1-4503-4304-6}
\urldef\tempurl%
\url{https://doi.org/10.1145/2970030.2970035}
\showDOI{\tempurl}


\bibitem[Khoo et~al\mbox{.}(2022)]%
        {khoo_deepfake_2022}
\bibfield{author}{\bibinfo{person}{Brandon Khoo}, \bibinfo{person}{Raphaël
  C.-W. Phan}, {and} \bibinfo{person}{Chern-Hong Lim}.}
  \bibinfo{year}{2022}\natexlab{}.
\newblock \showarticletitle{Deepfake attribution: {On} the source
  identification of artificially generated images}.
\newblock \bibinfo{journal}{\emph{WIREs Data Mining and Knowledge Discovery}}
  \bibinfo{volume}{12}, \bibinfo{number}{3} (\bibinfo{year}{2022}),
  \bibinfo{pages}{e1438}.
\newblock
\showISSN{1942-4795}
\urldef\tempurl%
\url{https://doi.org/10.1002/widm.1438}
\showDOI{\tempurl}
\newblock
\shownote{\_eprint: https://onlinelibrary.wiley.com/doi/pdf/10.1002/widm.1438}.


\bibitem[Kingma and Welling(2022)]%
        {kingma_auto-encoding_2022}
\bibfield{author}{\bibinfo{person}{Diederik~P. Kingma} {and}
  \bibinfo{person}{Max Welling}.} \bibinfo{year}{2022}\natexlab{}.
\newblock \bibinfo{title}{Auto-{Encoding} {Variational} {Bayes}}.
\newblock
\newblock
\urldef\tempurl%
\url{https://doi.org/10.48550/arXiv.1312.6114}
\showDOI{\tempurl}
\newblock
\shownote{arXiv:1312.6114 [cs, stat]}.


\bibitem[Kuran and McCaffery(2004)]%
        {Kuran_McCaffery_2004}
\bibfield{author}{\bibinfo{person}{Timur Kuran} {and}
  \bibinfo{person}{Edward~J. McCaffery}.} \bibinfo{year}{2004}\natexlab{}.
\newblock \showarticletitle{Expanding Discrimination Research: Beyond Ethnicity
  and to the Web*}.
\newblock \bibinfo{journal}{\emph{Social Science Quarterly}}
  \bibinfo{volume}{85}, \bibinfo{number}{3} (\bibinfo{year}{2004}),
  \bibinfo{pages}{713–730}.
\newblock
\showISSN{1540-6237}
\urldef\tempurl%
\url{https://doi.org/10.1111/j.0038-4941.2004.00241.x}
\showDOI{\tempurl}


\bibitem[Kökciyan et~al\mbox{.}(2017)]%
        {Kökciyan_Yaglikci_Yolum_2017}
\bibfield{author}{\bibinfo{person}{Nadin Kökciyan}, \bibinfo{person}{Nefise
  Yaglikci}, {and} \bibinfo{person}{Pinar Yolum}.}
  \bibinfo{year}{2017}\natexlab{}.
\newblock \showarticletitle{An Argumentation Approach for Resolving Privacy
  Disputes in Online Social Networks}.
\newblock \bibinfo{journal}{\emph{ACM Transactions on Internet Technology}}
  \bibinfo{volume}{17}, \bibinfo{number}{3} (\bibinfo{date}{Aug}
  \bibinfo{year}{2017}), \bibinfo{pages}{1–22}.
\newblock
\showISSN{1533-5399, 1557-6051}
\urldef\tempurl%
\url{https://doi.org/10.1145/3003434}
\showDOI{\tempurl}


\bibitem[Langer et~al\mbox{.}(2021)]%
        {Langer_Oster_Speith_Hermanns_Kästner_Schmidt_Sesing_Baum_2021}
\bibfield{author}{\bibinfo{person}{Markus Langer}, \bibinfo{person}{Daniel
  Oster}, \bibinfo{person}{Timo Speith}, \bibinfo{person}{Holger Hermanns},
  \bibinfo{person}{Lena Kästner}, \bibinfo{person}{Eva Schmidt},
  \bibinfo{person}{Andreas Sesing}, {and} \bibinfo{person}{Kevin Baum}.}
  \bibinfo{year}{2021}\natexlab{}.
\newblock \showarticletitle{What do we want from Explainable Artificial
  Intelligence (XAI)? – A stakeholder perspective on XAI and a conceptual
  model guiding interdisciplinary XAI research}.
\newblock \bibinfo{journal}{\emph{Artificial Intelligence}}
  \bibinfo{volume}{296} (\bibinfo{date}{Jul} \bibinfo{year}{2021}),
  \bibinfo{pages}{103473}.
\newblock
\showISSN{0004-3702}
\urldef\tempurl%
\url{https://doi.org/10.1016/j.artint.2021.103473}
\showDOI{\tempurl}


\bibitem[Lawson et~al\mbox{.}(2023)]%
        {lawson_tribalism_2023}
\bibfield{author}{\bibinfo{person}{M.~Asher Lawson}, \bibinfo{person}{Shikhar
  Anand}, {and} \bibinfo{person}{Hemant Kakkar}.}
  \bibinfo{year}{2023}\natexlab{}.
\newblock \showarticletitle{Tribalism and tribulations: {The} social costs of
  not sharing fake news}.
\newblock \bibinfo{journal}{\emph{Journal of Experimental Psychology: General}}
  \bibinfo{volume}{152}, \bibinfo{number}{3} (\bibinfo{year}{2023}),
  \bibinfo{pages}{611--631}.
\newblock
\showISSN{1939-2222}
\urldef\tempurl%
\url{https://doi.org/10.1037/xge0001374}
\showDOI{\tempurl}
\newblock
\shownote{Place: US Publisher: American Psychological Association}.


\bibitem[Lazer et~al\mbox{.}(2018)]%
        {lazer_science_2018}
\bibfield{author}{\bibinfo{person}{David M.~J. Lazer},
  \bibinfo{person}{Matthew~A. Baum}, \bibinfo{person}{Yochai Benkler},
  \bibinfo{person}{Adam~J. Berinsky}, \bibinfo{person}{Kelly~M. Greenhill},
  \bibinfo{person}{Filippo Menczer}, \bibinfo{person}{Miriam~J. Metzger},
  \bibinfo{person}{Brendan Nyhan}, \bibinfo{person}{Gordon Pennycook},
  \bibinfo{person}{David Rothschild}, \bibinfo{person}{Michael Schudson},
  \bibinfo{person}{Steven~A. Sloman}, \bibinfo{person}{Cass~R. Sunstein},
  \bibinfo{person}{Emily~A. Thorson}, \bibinfo{person}{Duncan~J. Watts}, {and}
  \bibinfo{person}{Jonathan~L. Zittrain}.} \bibinfo{year}{2018}\natexlab{}.
\newblock \showarticletitle{The science of fake news}.
\newblock \bibinfo{journal}{\emph{Science}} \bibinfo{volume}{359},
  \bibinfo{number}{6380} (\bibinfo{date}{March} \bibinfo{year}{2018}),
  \bibinfo{pages}{1094--1096}.
\newblock
\urldef\tempurl%
\url{https://doi.org/10.1126/science.aao2998}
\showDOI{\tempurl}
\newblock
\shownote{Publisher: American Association for the Advancement of Science}.


\bibitem[Leibowicz et~al\mbox{.}(2021)]%
        {leibowicz_deepfake_2021}
\bibfield{author}{\bibinfo{person}{Claire~R. Leibowicz}, \bibinfo{person}{Sean
  McGregor}, {and} \bibinfo{person}{Aviv Ovadya}.}
  \bibinfo{year}{2021}\natexlab{}.
\newblock \showarticletitle{The {Deepfake} {Detection} {Dilemma}: {A}
  {Multistakeholder} {Exploration} of {Adversarial} {Dynamics} in {Synthetic}
  {Media}}. In \bibinfo{booktitle}{\emph{Proceedings of the 2021 {AAAI}/{ACM}
  {Conference} on {AI}, {Ethics}, and {Society}}}
  \emph{(\bibinfo{series}{{AIES} '21})}. \bibinfo{publisher}{Association for
  Computing Machinery}, \bibinfo{address}{New York, NY, USA},
  \bibinfo{pages}{736--744}.
\newblock
\showISBNx{978-1-4503-8473-5}
\urldef\tempurl%
\url{https://doi.org/10.1145/3461702.3462584}
\showDOI{\tempurl}


\bibitem[Leonhardt(2022)]%
        {leonhardt_crisis_2022}
\bibfield{author}{\bibinfo{person}{David Leonhardt}.}
  \bibinfo{year}{2022}\natexlab{}.
\newblock \showarticletitle{‘{A} {Crisis} {Coming}’: {The} {Twin} {Threats}
  to {American} {Democracy}}.
\newblock \bibinfo{journal}{\emph{The New York Times}} (\bibinfo{date}{Sept.}
  \bibinfo{year}{2022}).
\newblock
\showISSN{0362-4331}
\urldef\tempurl%
\url{https://www.nytimes.com/2022/09/17/us/american-democracy-threats.html}
\showURL{%
\tempurl}


\bibitem[Liang et~al\mbox{.}(2022)]%
        {liang2022holistic}
\bibfield{author}{\bibinfo{person}{Percy Liang}, \bibinfo{person}{Rishi
  Bommasani}, \bibinfo{person}{Tony Lee}, \bibinfo{person}{Dimitris Tsipras},
  \bibinfo{person}{Dilara Soylu}, \bibinfo{person}{Michihiro Yasunaga},
  \bibinfo{person}{Yian Zhang}, \bibinfo{person}{Deepak Narayanan},
  \bibinfo{person}{Yuhuai Wu}, \bibinfo{person}{Ananya Kumar}, {et~al\mbox{.}}}
  \bibinfo{year}{2022}\natexlab{}.
\newblock \showarticletitle{Holistic evaluation of language models}.
\newblock \bibinfo{journal}{\emph{arXiv preprint arXiv:2211.09110}}
  (\bibinfo{year}{2022}).
\newblock


\bibitem[Lomas(2022)]%
        {Lomas_2022}
\bibfield{author}{\bibinfo{person}{Natasha Lomas}.}
  \bibinfo{year}{2022}\natexlab{}.
\newblock \bibinfo{title}{Shutterstock to integrate OpenAI’s DALL-E 2 and
  launch fund for contributor artists | TechCrunch}.
\newblock
\newblock
\urldef\tempurl%
\url{https://techcrunch.com/2022/10/25/shutterstock-openai-dall-e-2/}
\showURL{%
\tempurl}


\bibitem[Lovato et~al\mbox{.}(2022)]%
        {lovato_diverse_2022}
\bibfield{author}{\bibinfo{person}{Juniper Lovato}, \bibinfo{person}{Laurent
  Hébert-Dufresne}, \bibinfo{person}{Jonathan St-Onge},
  \bibinfo{person}{Randall Harp}, \bibinfo{person}{Gabriela~Salazar Lopez},
  \bibinfo{person}{Sean~P. Rogers}, \bibinfo{person}{Ijaz~Ul Haq}, {and}
  \bibinfo{person}{Jeremiah Onaolapo}.} \bibinfo{year}{2022}\natexlab{}.
\newblock \showarticletitle{Diverse {Misinformation}: {Impacts} of {Human}
  {Biases} on {Detection} of {Deepfakes} on {Networks}}.
\newblock  (\bibinfo{year}{2022}).
\newblock
\urldef\tempurl%
\url{https://doi.org/10.48550/ARXIV.2210.10026}
\showDOI{\tempurl}
\newblock
\shownote{Publisher: arXiv Version Number: 2}.


\bibitem[Madiega and Chahri(2023)]%
        {Madiega2023}
\bibfield{author}{\bibinfo{person}{Tambiama Madiega} {and}
  \bibinfo{person}{Samy Chahri}.} \bibinfo{year}{2023}\natexlab{}.
\newblock \bibinfo{booktitle}{\emph{Artificial intelligence act}}.
\newblock EU Legislation in Progress.
\newblock
\urldef\tempurl%
\url{https://www.europarl.europa.eu/RegData/etudes/BRIE/2021/698792/EPRS_BRI(2021)698792_EN.pdf}
\showURL{%
\tempurl}
\newblock
\shownote{BRIEFING}.


\bibitem[Markl(2022)]%
        {Markl_2022}
\bibfield{author}{\bibinfo{person}{Nina Markl}.}
  \bibinfo{year}{2022}\natexlab{}.
\newblock \showarticletitle{Mind the data gap(s): Investigating power in speech
  and language datasets}. In \bibinfo{booktitle}{\emph{Proceedings of the
  Second Workshop on Language Technology for Equality, Diversity and
  Inclusion}}. \bibinfo{publisher}{Association for Computational Linguistics},
  \bibinfo{address}{Dublin, Ireland}, \bibinfo{pages}{1–12}.
\newblock
\urldef\tempurl%
\url{https://doi.org/10.18653/v1/2022.ltedi-1.1}
\showDOI{\tempurl}


\bibitem[Mathews et~al\mbox{.}(2023)]%
        {mathews_explainable_2023}
\bibfield{author}{\bibinfo{person}{Sherin Mathews}, \bibinfo{person}{Shivangee
  Trivedi}, \bibinfo{person}{Amanda House}, \bibinfo{person}{Steve Povolny},
  {and} \bibinfo{person}{Celeste Fralick}.} \bibinfo{year}{2023}\natexlab{}.
\newblock \showarticletitle{An explainable deepfake detection framework on a
  novel unconstrained dataset}.
\newblock \bibinfo{journal}{\emph{Complex \& Intelligent Systems}}
  (\bibinfo{date}{Jan.} \bibinfo{year}{2023}).
\newblock
\showISSN{2199-4536, 2198-6053}
\urldef\tempurl%
\url{https://doi.org/10.1007/s40747-022-00956-7}
\showDOI{\tempurl}


\bibitem[Mauro and Schellmann(2023)]%
        {Mauro_Schellmann_2023}
\bibfield{author}{\bibinfo{person}{Gianluca Mauro} {and} \bibinfo{person}{Hilke
  Schellmann}.} \bibinfo{year}{2023}\natexlab{}.
\newblock \showarticletitle{‘There is no standard’: investigation finds AI
  algorithms objectify women’s bodies}.
\newblock \bibinfo{journal}{\emph{The Guardian}} (\bibinfo{date}{Feb}
  \bibinfo{year}{2023}).
\newblock
\showISSN{0261-3077}
\urldef\tempurl%
\url{https://www.theguardian.com/technology/2023/feb/08/biased-ai-algorithms-racy-women-bodies}
\showURL{%
\tempurl}


\bibitem[Millière(2022a)]%
        {Milliere_2022}
\bibfield{author}{\bibinfo{person}{Raphaël Millière}.}
  \bibinfo{year}{2022}\natexlab{a}.
\newblock \showarticletitle{Adversarial Attacks on Image Generation With
  Made-Up Words}.
\newblock \bibinfo{journal}{\emph{ArXiv}} (\bibinfo{year}{2022}).
\newblock
\urldef\tempurl%
\url{https://doi.org/10.48550/arXiv.2208.04135}
\showDOI{\tempurl}


\bibitem[Millière(2022b)]%
        {milliere_deep_2022}
\bibfield{author}{\bibinfo{person}{Raphaël Millière}.}
  \bibinfo{year}{2022}\natexlab{b}.
\newblock \showarticletitle{Deep learning and synthetic media}.
\newblock \bibinfo{journal}{\emph{Synthese}} \bibinfo{volume}{200},
  \bibinfo{number}{3} (\bibinfo{date}{May} \bibinfo{year}{2022}),
  \bibinfo{pages}{231}.
\newblock
\showISSN{1573-0964}
\urldef\tempurl%
\url{https://doi.org/10.1007/s11229-022-03739-2}
\showDOI{\tempurl}


\bibitem[Mishkin et~al\mbox{.}(2022)]%
        {mishkin_dalle_2022}
\bibfield{author}{\bibinfo{person}{Pamela Mishkin}, \bibinfo{person}{Lama
  Ahmad}, \bibinfo{person}{Miles Brundage}, \bibinfo{person}{Gretchen Krueger},
  {and} \bibinfo{person}{Girish Sastry}.} \bibinfo{year}{2022}\natexlab{}.
\newblock \showarticletitle{{DALL}·{E} 2 {Preview} - {Risks} and
  {Limitations}}.
\newblock  (\bibinfo{year}{2022}).
\newblock
\urldef\tempurl%
\url{[https://github.com/openai/dalle-2-preview/blob/main/system-card.md](https://github.com/openai/dalle-2-preview/blob/main/system-card.md)}
\showURL{%
\tempurl}


\bibitem[M{\"o}kander et~al\mbox{.}(2022)]%
        {mokander2022us}
\bibfield{author}{\bibinfo{person}{Jakob M{\"o}kander}, \bibinfo{person}{Prathm
  Juneja}, \bibinfo{person}{David~S Watson}, {and} \bibinfo{person}{Luciano
  Floridi}.} \bibinfo{year}{2022}\natexlab{}.
\newblock \showarticletitle{The US Algorithmic Accountability Act of 2022 vs.
  The EU Artificial Intelligence Act: what can they learn from each other?}
\newblock \bibinfo{journal}{\emph{Minds and Machines}} \bibinfo{volume}{32},
  \bibinfo{number}{4} (\bibinfo{year}{2022}), \bibinfo{pages}{751--758}.
\newblock


\bibitem[Moreno(2022)]%
        {Moreno_2022}
\bibfield{author}{\bibinfo{person}{Johan Moreno}.}
  \bibinfo{year}{2022}\natexlab{}.
\newblock \bibinfo{title}{Shutterstock Will Soon Offer Licensed DALL-E 2
  Images, Showing What The Future Of Generative AI Might Look Like}.
\newblock
\newblock
\urldef\tempurl%
\url{https://www.forbes.com/sites/johanmoreno/2022/10/26/shutterstock-will-soon-offer-ai-generated-images-showing-what-the-future-of-dall-e-might-look-like/}
\showURL{%
\tempurl}


\bibitem[Nadimpalli and Rattani(2022)]%
        {nadimpalli_gbdf_2022}
\bibfield{author}{\bibinfo{person}{Aakash~Varma Nadimpalli} {and}
  \bibinfo{person}{Ajita Rattani}.} \bibinfo{year}{2022}\natexlab{}.
\newblock \bibinfo{booktitle}{\emph{{GBDF}: {Gender} {Balanced} {DeepFake}
  {Dataset} {Towards} {Fair} {DeepFake} {Detection}}}.
\newblock


\bibitem[Nemitz(2018)]%
        {nemitz_constitutional_2018}
\bibfield{author}{\bibinfo{person}{Paul Nemitz}.}
  \bibinfo{year}{2018}\natexlab{}.
\newblock \showarticletitle{Constitutional democracy and technology in the age
  of artificial intelligence}.
\newblock \bibinfo{journal}{\emph{Philosophical Transactions of the Royal
  Society A: Mathematical, Physical and Engineering Sciences}}
  \bibinfo{volume}{376}, \bibinfo{number}{2133} (\bibinfo{date}{Oct.}
  \bibinfo{year}{2018}), \bibinfo{pages}{20180089}.
\newblock
\urldef\tempurl%
\url{https://doi.org/10.1098/rsta.2018.0089}
\showDOI{\tempurl}
\newblock
\shownote{Publisher: Royal Society}.


\bibitem[Newton and Dhole(2023)]%
        {Newton_Dhole_2023}
\bibfield{author}{\bibinfo{person}{Alexis Newton} {and}
  \bibinfo{person}{Kaustubh Dhole}.} \bibinfo{year}{2023}\natexlab{}.
\newblock \showarticletitle{Is AI Art Another Industrial Revolution in the
  Making?}
\newblock  (\bibinfo{year}{2023}).
\newblock
\urldef\tempurl%
\url{https://doi.org/10.48550/ARXIV.2301.05133}
\showDOI{\tempurl}


\bibitem[Nguyen et~al\mbox{.}(2022)]%
        {nguyen_deep_2022}
\bibfield{author}{\bibinfo{person}{Thanh~Thi Nguyen}, \bibinfo{person}{Quoc
  Viet~Hung Nguyen}, \bibinfo{person}{Dung~Tien Nguyen},
  \bibinfo{person}{Duc~Thanh Nguyen}, \bibinfo{person}{Thien Huynh-The},
  \bibinfo{person}{Saeid Nahavandi}, \bibinfo{person}{Thanh~Tam Nguyen},
  \bibinfo{person}{Quoc-Viet Pham}, {and} \bibinfo{person}{Cuong~M. Nguyen}.}
  \bibinfo{year}{2022}\natexlab{}.
\newblock \showarticletitle{Deep learning for deepfakes creation and detection:
  {A} survey}.
\newblock \bibinfo{journal}{\emph{Computer Vision and Image Understanding}}
  \bibinfo{volume}{223} (\bibinfo{date}{Oct.} \bibinfo{year}{2022}),
  \bibinfo{pages}{103525}.
\newblock
\showISSN{10773142}
\urldef\tempurl%
\url{https://doi.org/10.1016/j.cviu.2022.103525}
\showDOI{\tempurl}


\bibitem[Nichol(2022)]%
        {nichol_2022}
\bibfield{author}{\bibinfo{person}{Alex Nichol}.}
  \bibinfo{year}{2022}\natexlab{}.
\newblock \bibinfo{title}{Dall·E 2 pre-training mitigations}.
\newblock
\newblock
\urldef\tempurl%
\url{https://openai.com/research/dall-e-2-pre-training-mitigations}
\showURL{%
\tempurl}


\bibitem[Nichol et~al\mbox{.}(2022)]%
        {nichol_glide_2022}
\bibfield{author}{\bibinfo{person}{Alex Nichol}, \bibinfo{person}{Prafulla
  Dhariwal}, \bibinfo{person}{Aditya Ramesh}, \bibinfo{person}{Pranav Shyam},
  \bibinfo{person}{Pamela Mishkin}, \bibinfo{person}{Bob McGrew},
  \bibinfo{person}{Ilya Sutskever}, {and} \bibinfo{person}{Mark Chen}.}
  \bibinfo{year}{2022}\natexlab{}.
\newblock \bibinfo{title}{{GLIDE}: {Towards} {Photorealistic} {Image}
  {Generation} and {Editing} with {Text}-{Guided} {Diffusion} {Models}}.
\newblock
\newblock
\urldef\tempurl%
\url{https://doi.org/10.48550/arXiv.2112.10741}
\showDOI{\tempurl}
\newblock
\shownote{arXiv:2112.10741 [cs]}.


\bibitem[Nour and Gelfand(2021)]%
        {nour_deepfakes_2021}
\bibfield{author}{\bibinfo{person}{Nika Nour} {and} \bibinfo{person}{Julia
  Gelfand}.} \bibinfo{year}{2021}\natexlab{}.
\newblock \showarticletitle{Deepfakes: {A} {Digital} {Transformation} {Leads}
  to {Misinformation}}.
\newblock  (\bibinfo{year}{2021}).
\newblock


\bibitem[Obedkov(2023)]%
        {Obedkov2023}
\bibfield{author}{\bibinfo{person}{Evgeny Obedkov}.}
  \bibinfo{year}{2023}\natexlab{}.
\newblock \bibinfo{booktitle}{\emph{Game illustrator jobs in China down 70\%
  due to rapid AI adoption}}.
\newblock Game World Observer.
\newblock
\urldef\tempurl%
\url{https://gameworldobserver.com/2023/04/12/game-artist-jobs-china-down-70-percent-gen-ai-adoption}
\showURL{%
\tempurl}


\bibitem[Oord et~al\mbox{.}(2016)]%
        {oord_conditional_2016}
\bibfield{author}{\bibinfo{person}{Aaron van~den Oord}, \bibinfo{person}{Nal
  Kalchbrenner}, \bibinfo{person}{Oriol Vinyals}, \bibinfo{person}{Lasse
  Espeholt}, \bibinfo{person}{Alex Graves}, {and} \bibinfo{person}{Koray
  Kavukcuoglu}.} \bibinfo{year}{2016}\natexlab{}.
\newblock \bibinfo{title}{Conditional {Image} {Generation} with {PixelCNN}
  {Decoders}}.
\newblock
\newblock
\urldef\tempurl%
\url{https://doi.org/10.48550/arXiv.1606.05328}
\showDOI{\tempurl}
\newblock
\shownote{arXiv:1606.05328 [cs]}.


\bibitem[Oppenlaender(2022)]%
        {Oppenlaender_2022a}
\bibfield{author}{\bibinfo{person}{Jonas Oppenlaender}.}
  \bibinfo{year}{2022}\natexlab{}.
\newblock \showarticletitle{The Creativity of Text-to-Image Generation}. In
  \bibinfo{booktitle}{\emph{Proceedings of the 25th International Academic
  Mindtrek Conference}} \emph{(\bibinfo{series}{Academic Mindtrek ’22})}.
  \bibinfo{publisher}{Association for Computing Machinery},
  \bibinfo{address}{New York, NY, USA}, \bibinfo{pages}{192–202}.
\newblock
\showISBNx{978-1-4503-9955-5}
\urldef\tempurl%
\url{https://doi.org/10.1145/3569219.3569352}
\showDOI{\tempurl}


\bibitem[O'Sullivan and Passantino(2023)]%
        {osullivan2023}
\bibfield{author}{\bibinfo{person}{Donie O'Sullivan} {and} \bibinfo{person}{Jon
  Passantino}.} \bibinfo{year}{2023}\natexlab{}.
\newblock \bibinfo{booktitle}{\emph{‘Verified’ Twitter accounts share fake
  image of ‘explosion’ near Pentagon, causing confusion}}.
\newblock CNN.
\newblock
\urldef\tempurl%
\url{https://edition.cnn.com/2023/05/22/tech/twitter-fake-image-pentagon-explosion/index.html}
\showURL{%
\tempurl}


\bibitem[Paris and Donovan(2019)]%
        {paris_deepfakes_2019}
\bibfield{author}{\bibinfo{person}{Britt Paris} {and} \bibinfo{person}{Joan
  Donovan}.} \bibinfo{year}{2019}\natexlab{}.
\newblock \showarticletitle{{DEEPFAKES} {AND} {CHEAP} {FAKES}}.
\newblock  (\bibinfo{year}{2019}).
\newblock


\bibitem[Pariser(2011)]%
        {pariser_filter_2011}
\bibfield{author}{\bibinfo{person}{Eli Pariser}.}
  \bibinfo{year}{2011}\natexlab{}.
\newblock \bibinfo{booktitle}{\emph{The {Filter} {Bubble}: {How} the {New}
  {Personalized} {Web} {Is} {Changing} {What} {We} {Read} and {How} {We}
  {Think}}}.
\newblock \bibinfo{publisher}{Penguin}.
\newblock
\showISBNx{978-1-101-51512-9}
\newblock
\shownote{Google-Books-ID: wcalrOI1YbQC}.


\bibitem[Park et~al\mbox{.}(2022)]%
        {park_judge_2022}
\bibfield{author}{\bibinfo{person}{Seongbeom Park}, \bibinfo{person}{Suhong
  Moon}, {and} \bibinfo{person}{Jinkyu Kim}.} \bibinfo{year}{2022}\natexlab{}.
\newblock \bibinfo{title}{Judge, {Localize}, and {Edit}: {Ensuring} {Visual}
  {Commonsense} {Morality} for {Text}-to-{Image} {Generation}}.
\newblock
\newblock
\urldef\tempurl%
\url{https://doi.org/10.48550/arXiv.2212.03507}
\showDOI{\tempurl}
\newblock
\shownote{arXiv:2212.03507 [cs]}.


\bibitem[Perrigo(2023)]%
        {perrigo_2023}
\bibfield{author}{\bibinfo{person}{Billy Perrigo}.}
  \bibinfo{year}{2023}\natexlab{}.
\newblock \bibinfo{title}{OpenAI used Kenyan workers on less than 2 per hour:
  Exclusive}.
\newblock
\newblock
\urldef\tempurl%
\url{https://time.com/6247678/openai-chatgpt-kenya-workers/}
\showURL{%
\tempurl}


\bibitem[Pu et~al\mbox{.}(2022)]%
        {pu_fairness_2022}
\bibfield{author}{\bibinfo{person}{Muxin Pu}, \bibinfo{person}{Meng~Yi Kuan},
  \bibinfo{person}{Nyee~Thoang Lim}, \bibinfo{person}{Chun~Yong Chong}, {and}
  \bibinfo{person}{Mei~Kuan Lim}.} \bibinfo{year}{2022}\natexlab{}.
\newblock \showarticletitle{Fairness {Evaluation} in {Deepfake} {Detection}
  {Models} using {Metamorphic} {Testing}}. In \bibinfo{booktitle}{\emph{2022
  {IEEE}/{ACM} 7th {International} {Workshop} on {Metamorphic} {Testing}
  ({MET})}}. \bibinfo{pages}{7--14}.
\newblock
\urldef\tempurl%
\url{https://doi.org/10.1145/3524846.3527337}
\showDOI{\tempurl}


\bibitem[Qiu et~al\mbox{.}(2022)]%
        {Qiu_Zhu_Shi_Wenzel_Tang_Zhao_Li_Li_2022}
\bibfield{author}{\bibinfo{person}{Jielin Qiu}, \bibinfo{person}{Yi Zhu},
  \bibinfo{person}{Xingjian Shi}, \bibinfo{person}{Florian Wenzel},
  \bibinfo{person}{Zhiqiang Tang}, \bibinfo{person}{Ding Zhao},
  \bibinfo{person}{Bo Li}, {and} \bibinfo{person}{Mu Li}.}
  \bibinfo{year}{2022}\natexlab{}.
\newblock \showarticletitle{Are Multimodal Models Robust to Image and Text
  Perturbations?}
\newblock  (\bibinfo{year}{2022}).
\newblock
\urldef\tempurl%
\url{https://doi.org/10.48550/ARXIV.2212.08044}
\showDOI{\tempurl}


\bibitem[Queerinai et~al\mbox{.}(2023)]%
        {queerinai_queer_2023}
\bibfield{author}{\bibinfo{person}{Organizers~Of Queerinai},
  \bibinfo{person}{Anaelia Ovalle}, \bibinfo{person}{Arjun Subramonian},
  \bibinfo{person}{Ashwin Singh}, \bibinfo{person}{Claas Voelcker},
  \bibinfo{person}{Danica~J. Sutherland}, \bibinfo{person}{Davide Locatelli},
  \bibinfo{person}{Eva Breznik}, \bibinfo{person}{Filip Klubicka},
  \bibinfo{person}{Hang Yuan}, \bibinfo{person}{Hetvi J}, \bibinfo{person}{Huan
  Zhang}, \bibinfo{person}{Jaidev Shriram}, \bibinfo{person}{Kruno Lehman},
  \bibinfo{person}{Luca Soldaini}, \bibinfo{person}{Maarten Sap},
  \bibinfo{person}{Marc~Peter Deisenroth}, \bibinfo{person}{Maria~Leonor
  Pacheco}, \bibinfo{person}{Maria Ryskina}, \bibinfo{person}{Martin Mundt},
  \bibinfo{person}{Milind Agarwal}, \bibinfo{person}{Nyx Mclean},
  \bibinfo{person}{Pan Xu}, \bibinfo{person}{A Pranav}, \bibinfo{person}{Raj
  Korpan}, \bibinfo{person}{Ruchira Ray}, \bibinfo{person}{Sarah Mathew},
  \bibinfo{person}{Sarthak Arora}, \bibinfo{person}{St John},
  \bibinfo{person}{Tanvi Anand}, \bibinfo{person}{Vishakha Agrawal},
  \bibinfo{person}{William Agnew}, \bibinfo{person}{Yanan Long},
  \bibinfo{person}{Zijie~J. Wang}, \bibinfo{person}{Zeerak Talat},
  \bibinfo{person}{Avijit Ghosh}, \bibinfo{person}{Nathaniel Dennler},
  \bibinfo{person}{Michael Noseworthy}, \bibinfo{person}{Sharvani Jha},
  \bibinfo{person}{Emi Baylor}, \bibinfo{person}{Aditya Joshi},
  \bibinfo{person}{Natalia~Y. Bilenko}, \bibinfo{person}{Andrew Mcnamara},
  \bibinfo{person}{Raphael Gontijo-Lopes}, \bibinfo{person}{Alex Markham},
  \bibinfo{person}{Evyn Dong}, \bibinfo{person}{Jackie Kay},
  \bibinfo{person}{Manu Saraswat}, \bibinfo{person}{Nikhil Vytla}, {and}
  \bibinfo{person}{Luke Stark}.} \bibinfo{year}{2023}\natexlab{}.
\newblock \showarticletitle{Queer {In} {AI}: {A} {Case} {Study} in
  {Community}-{Led} {Participatory} {AI}}. In
  \bibinfo{booktitle}{\emph{Proceedings of the 2023 {ACM} {Conference} on
  {Fairness}, {Accountability}, and {Transparency}}}
  \emph{(\bibinfo{series}{{FAccT} '23})}. \bibinfo{publisher}{Association for
  Computing Machinery}, \bibinfo{address}{New York, NY, USA},
  \bibinfo{pages}{1882--1895}.
\newblock
\showISBNx{9798400701924}
\urldef\tempurl%
\url{https://doi.org/10.1145/3593013.3594134}
\showDOI{\tempurl}


\bibitem[Radford et~al\mbox{.}(2021)]%
        {radford_learning_2021}
\bibfield{author}{\bibinfo{person}{Alec Radford}, \bibinfo{person}{Jong~Wook
  Kim}, \bibinfo{person}{Chris Hallacy}, \bibinfo{person}{Aditya Ramesh},
  \bibinfo{person}{Gabriel Goh}, \bibinfo{person}{Sandhini Agarwal},
  \bibinfo{person}{Girish Sastry}, \bibinfo{person}{Amanda Askell},
  \bibinfo{person}{Pamela Mishkin}, \bibinfo{person}{Jack Clark},
  \bibinfo{person}{Gretchen Krueger}, {and} \bibinfo{person}{Ilya Sutskever}.}
  \bibinfo{year}{2021}\natexlab{}.
\newblock \bibinfo{title}{Learning {Transferable} {Visual} {Models} {From}
  {Natural} {Language} {Supervision}}.
\newblock
\newblock
\urldef\tempurl%
\url{https://doi.org/10.48550/arXiv.2103.00020}
\showDOI{\tempurl}
\newblock
\shownote{arXiv:2103.00020 [cs]}.


\bibitem[Ramesh et~al\mbox{.}(2022)]%
        {ramesh_hierarchical_2022}
\bibfield{author}{\bibinfo{person}{Aditya Ramesh}, \bibinfo{person}{Prafulla
  Dhariwal}, \bibinfo{person}{Alex Nichol}, \bibinfo{person}{Casey Chu}, {and}
  \bibinfo{person}{Mark Chen}.} \bibinfo{year}{2022}\natexlab{}.
\newblock \bibinfo{title}{Hierarchical {Text}-{Conditional} {Image}
  {Generation} with {CLIP} {Latents}}.
\newblock
\newblock
\urldef\tempurl%
\url{https://doi.org/10.48550/arXiv.2204.06125}
\showDOI{\tempurl}
\newblock
\shownote{arXiv:2204.06125 [cs]}.


\bibitem[Raval et~al\mbox{.}(2022)]%
        {raval_survey_2022}
\bibfield{author}{\bibinfo{person}{Mehul~S Raval}, \bibinfo{person}{Mohendra
  Roy}, {and} \bibinfo{person}{Minoru Kuribayashi}.}
  \bibinfo{year}{2022}\natexlab{}.
\newblock \showarticletitle{Survey on {Vision} based {Fake} {News} {Detection}
  and its {Impact} {Analysis}}.
\newblock \bibinfo{journal}{\emph{2022 Asia-Pacific Signal and Information
  Processing Association Annual Summit and Conference (APSIPA ASC)}}
  (\bibinfo{date}{Nov.} \bibinfo{year}{2022}), \bibinfo{pages}{1837--1841}.
\newblock
\urldef\tempurl%
\url{https://doi.org/10.23919/APSIPAASC55919.2022.9980089}
\showDOI{\tempurl}
\newblock
\shownote{Conference Name: 2022 Asia Pacific Signal and Information Processing
  Association Annual Summit and Conference (APSIPA ASC) ISBN: 9786165904773
  Place: Chiang Mai, Thailand Publisher: IEEE}.


\bibitem[Reed et~al\mbox{.}(2016)]%
        {reed_generative_2016}
\bibfield{author}{\bibinfo{person}{Scott~E. Reed}, \bibinfo{person}{Zeynep
  Akata}, \bibinfo{person}{Xinchen Yan}, \bibinfo{person}{L. Logeswaran},
  \bibinfo{person}{B. Schiele}, {and} \bibinfo{person}{Honglak Lee}.}
  \bibinfo{year}{2016}\natexlab{}.
\newblock \showarticletitle{Generative {Adversarial} {Text} to {Image}
  {Synthesis}}.
\newblock \bibinfo{journal}{\emph{ArXiv}} (\bibinfo{date}{May}
  \bibinfo{year}{2016}).
\newblock
\urldef\tempurl%
\url{https://www.semanticscholar.org/paper/6c7f040a150abf21dbcefe1f22e0f98fa184f41a}
\showURL{%
\tempurl}


\bibitem[Renaud et~al\mbox{.}(2023)]%
        {renaud2023chatgpt}
\bibfield{author}{\bibinfo{person}{Karen Renaud}, \bibinfo{person}{Merrill
  Warkentin}, {and} \bibinfo{person}{George Westerman}.}
  \bibinfo{year}{2023}\natexlab{}.
\newblock \bibinfo{booktitle}{\emph{From ChatGPT to HackGPT: Meeting the
  Cybersecurity Threat of Generative AI}}.
\newblock \bibinfo{publisher}{MIT Sloan Management Review}.
\newblock


\bibitem[Rocha et~al\mbox{.}(2021)]%
        {rocha_impact_2021}
\bibfield{author}{\bibinfo{person}{Yasmim~Mendes Rocha},
  \bibinfo{person}{Gabriel~Acácio de Moura}, \bibinfo{person}{Gabriel~Alves
  Desidério}, \bibinfo{person}{Carlos~Henrique de Oliveira},
  \bibinfo{person}{Francisco~Dantas Lourenço}, {and}
  \bibinfo{person}{Larissa~Deadame de Figueiredo~Nicolete}.}
  \bibinfo{year}{2021}\natexlab{}.
\newblock \showarticletitle{The impact of fake news on social media and its
  influence on health during the {COVID}-19 pandemic: a systematic review}.
\newblock \bibinfo{journal}{\emph{Zeitschrift Fur Gesundheitswissenschaften}}
  (\bibinfo{date}{Oct.} \bibinfo{year}{2021}), \bibinfo{pages}{1--10}.
\newblock
\showISSN{2198-1833}
\urldef\tempurl%
\url{https://doi.org/10.1007/s10389-021-01658-z}
\showDOI{\tempurl}


\bibitem[Rombach et~al\mbox{.}(2022)]%
        {Rombach_2022_CVPR}
\bibfield{author}{\bibinfo{person}{Robin Rombach}, \bibinfo{person}{Andreas
  Blattmann}, \bibinfo{person}{Dominik Lorenz}, \bibinfo{person}{Patrick
  Esser}, {and} \bibinfo{person}{Bj\"orn Ommer}.}
  \bibinfo{year}{2022}\natexlab{}.
\newblock \showarticletitle{High-Resolution Image Synthesis With Latent
  Diffusion Models}. In \bibinfo{booktitle}{\emph{Proceedings of the IEEE/CVF
  Conference on Computer Vision and Pattern Recognition (CVPR)}}.
  \bibinfo{pages}{10684--10695}.
\newblock


\bibitem[Saharia et~al\mbox{.}(2022)]%
        {saharia_photorealistic_2022}
\bibfield{author}{\bibinfo{person}{Chitwan Saharia}, \bibinfo{person}{William
  Chan}, \bibinfo{person}{Saurabh Saxena}, \bibinfo{person}{Lala Li},
  \bibinfo{person}{Jay Whang}, \bibinfo{person}{Emily Denton},
  \bibinfo{person}{Seyed Kamyar~Seyed Ghasemipour},
  \bibinfo{person}{Burcu~Karagol Ayan}, \bibinfo{person}{S.~Sara Mahdavi},
  \bibinfo{person}{Rapha~Gontijo Lopes}, \bibinfo{person}{Tim Salimans},
  \bibinfo{person}{Jonathan Ho}, \bibinfo{person}{David~J. Fleet}, {and}
  \bibinfo{person}{Mohammad Norouzi}.} \bibinfo{year}{2022}\natexlab{}.
\newblock \bibinfo{title}{Photorealistic {Text}-to-{Image} {Diffusion} {Models}
  with {Deep} {Language} {Understanding}}.
\newblock
\newblock
\urldef\tempurl%
\url{http://arxiv.org/abs/2205.11487}
\showURL{%
\tempurl}
\newblock
\shownote{arXiv:2205.11487 [cs]}.


\bibitem[Salminen et~al\mbox{.}(2020)]%
        {Salminen_Jung_Chowdhury_Jansen_2020}
\bibfield{author}{\bibinfo{person}{Joni Salminen}, \bibinfo{person}{Soon-gyo
  Jung}, \bibinfo{person}{Shammur Chowdhury}, {and} \bibinfo{person}{Bernard~J.
  Jansen}.} \bibinfo{year}{2020}\natexlab{}.
\newblock \showarticletitle{Analyzing Demographic Bias in Artificially
  Generated Facial Pictures}. In \bibinfo{booktitle}{\emph{Extended Abstracts
  of the 2020 CHI Conference on Human Factors in Computing Systems}}
  \emph{(\bibinfo{series}{CHI EA ’20})}. \bibinfo{publisher}{Association for
  Computing Machinery}, \bibinfo{address}{New York, NY, USA},
  \bibinfo{pages}{1–8}.
\newblock
\showISBNx{978-1-4503-6819-3}
\urldef\tempurl%
\url{https://doi.org/10.1145/3334480.3382791}
\showDOI{\tempurl}


\bibitem[Saltz et~al\mbox{.}(2021)]%
        {saltz_encounters_2021}
\bibfield{author}{\bibinfo{person}{Emily Saltz}, \bibinfo{person}{Claire~R
  Leibowicz}, {and} \bibinfo{person}{Claire Wardle}.}
  \bibinfo{year}{2021}\natexlab{}.
\newblock \showarticletitle{Encounters with {Visual} {Misinformation} and
  {Labels} {Across} {Platforms}: {An} {Interview} and {Diary} {Study} to
  {Inform} {Ecosystem} {Approaches} to {Misinformation} {Interventions}}. In
  \bibinfo{booktitle}{\emph{Extended {Abstracts} of the 2021 {CHI} {Conference}
  on {Human} {Factors} in {Computing} {Systems}}} \emph{(\bibinfo{series}{{CHI}
  {EA} '21})}. \bibinfo{publisher}{Association for Computing Machinery},
  \bibinfo{address}{New York, NY, USA}, \bibinfo{pages}{1--6}.
\newblock
\showISBNx{978-1-4503-8095-9}
\urldef\tempurl%
\url{https://doi.org/10.1145/3411763.3451807}
\showDOI{\tempurl}


\bibitem[Samuelson(2023)]%
        {samuelson2023legal}
\bibfield{author}{\bibinfo{person}{Pamela Samuelson}.}
  \bibinfo{year}{2023}\natexlab{}.
\newblock \showarticletitle{Legal Challenges to Generative AI, Part I}.
\newblock \bibinfo{journal}{\emph{Commun. ACM}} \bibinfo{volume}{66},
  \bibinfo{number}{7} (\bibinfo{year}{2023}), \bibinfo{pages}{20--23}.
\newblock


\bibitem[Sebastian(2023)]%
        {sebastian2023chatgpt}
\bibfield{author}{\bibinfo{person}{Glorin Sebastian}.}
  \bibinfo{year}{2023}\natexlab{}.
\newblock \showarticletitle{Do ChatGPT and other AI chatbots pose a
  cybersecurity risk?: An exploratory study}.
\newblock \bibinfo{journal}{\emph{International Journal of Security and Privacy
  in Pervasive Computing (IJSPPC)}} \bibinfo{volume}{15}, \bibinfo{number}{1}
  (\bibinfo{year}{2023}), \bibinfo{pages}{1--11}.
\newblock


\bibitem[Seneviratne(2022)]%
        {seneviratne_blockchain_2022}
\bibfield{author}{\bibinfo{person}{Oshani Seneviratne}.}
  \bibinfo{year}{2022}\natexlab{}.
\newblock \showarticletitle{Blockchain for {Social} {Good}: {Combating}
  {Misinformation} on the {Web} with {AI} and {Blockchain}}.
\newblock \bibinfo{journal}{\emph{14th ACM Web Science Conference 2022}}
  (\bibinfo{date}{June} \bibinfo{year}{2022}), \bibinfo{pages}{435--442}.
\newblock
\urldef\tempurl%
\url{https://doi.org/10.1145/3501247.3539016}
\showDOI{\tempurl}
\newblock
\shownote{Conference Name: WebSci '22: 14th ACM Web Science Conference 2022
  ISBN: 9781450391917 Place: Barcelona Spain Publisher: ACM}.


\bibitem[Seneviratne et~al\mbox{.}(2022)]%
        {Seneviratne_Senanayake_Rasnayaka_Vidanaarachchi_Thompson_2022}
\bibfield{author}{\bibinfo{person}{Sachith Seneviratne},
  \bibinfo{person}{Damith Senanayake}, \bibinfo{person}{Sanka Rasnayaka},
  \bibinfo{person}{Rajith Vidanaarachchi}, {and} \bibinfo{person}{Jason
  Thompson}.} \bibinfo{year}{2022}\natexlab{}.
\newblock \showarticletitle{DALLE-URBAN: Capturing the urban design expertise
  of large text to image transformers}.
\newblock  \bibinfo{number}{arXiv:2208.04139} (\bibinfo{date}{Oct}
  \bibinfo{year}{2022}).
\newblock
\urldef\tempurl%
\url{http://arxiv.org/abs/2208.04139}
\showURL{%
\tempurl}
\newblock
\shownote{arXiv:2208.04139 [cs]}.


\bibitem[Seow et~al\mbox{.}(2022)]%
        {seow_comprehensive_2022}
\bibfield{author}{\bibinfo{person}{Jia~Wen Seow}, \bibinfo{person}{Mei~Kuan
  Lim}, \bibinfo{person}{Raphaël~C.W. Phan}, {and} \bibinfo{person}{Joseph~K.
  Liu}.} \bibinfo{year}{2022}\natexlab{}.
\newblock \showarticletitle{A comprehensive overview of {Deepfake}:
  {Generation}, detection, datasets, and opportunities}.
\newblock \bibinfo{journal}{\emph{Neurocomputing}}  \bibinfo{volume}{513}
  (\bibinfo{date}{Nov.} \bibinfo{year}{2022}), \bibinfo{pages}{351--371}.
\newblock
\showISSN{09252312}
\urldef\tempurl%
\url{https://doi.org/10.1016/j.neucom.2022.09.135}
\showDOI{\tempurl}


\bibitem[Shmueli et~al\mbox{.}(2021)]%
        {Shmueli_Fell_Ray_Ku_2021}
\bibfield{author}{\bibinfo{person}{Boaz Shmueli}, \bibinfo{person}{Jan Fell},
  \bibinfo{person}{Soumya Ray}, {and} \bibinfo{person}{Lun-Wei Ku}.}
  \bibinfo{year}{2021}\natexlab{}.
\newblock \showarticletitle{Beyond Fair Pay: Ethical Implications of NLP
  Crowdsourcing}. In \bibinfo{booktitle}{\emph{Proceedings of the 2021
  Conference of the North American Chapter of the Association for Computational
  Linguistics: Human Language Technologies}}. \bibinfo{publisher}{Association
  for Computational Linguistics}, \bibinfo{address}{Online},
  \bibinfo{pages}{3758–3769}.
\newblock
\urldef\tempurl%
\url{https://doi.org/10.18653/v1/2021.naacl-main.295}
\showDOI{\tempurl}


\bibitem[Sloane et~al\mbox{.}(2022)]%
        {Sloane_Moss_Awomolo_Forlano_2022}
\bibfield{author}{\bibinfo{person}{Mona Sloane}, \bibinfo{person}{Emanuel
  Moss}, \bibinfo{person}{Olaitan Awomolo}, {and} \bibinfo{person}{Laura
  Forlano}.} \bibinfo{year}{2022}\natexlab{}.
\newblock \showarticletitle{Participation Is not a Design Fix for Machine
  Learning}. In \bibinfo{booktitle}{\emph{Equity and Access in Algorithms,
  Mechanisms, and Optimization}}. \bibinfo{publisher}{ACM},
  \bibinfo{address}{Arlington VA USA}, \bibinfo{pages}{1–6}.
\newblock
\showISBNx{978-1-4503-9477-2}
\urldef\tempurl%
\url{https://doi.org/10.1145/3551624.3555285}
\showDOI{\tempurl}


\bibitem[Sohl-Dickstein et~al\mbox{.}(2015)]%
        {sohl-dickstein_deep_2015}
\bibfield{author}{\bibinfo{person}{Jascha Sohl-Dickstein},
  \bibinfo{person}{Eric~A. Weiss}, \bibinfo{person}{Niru Maheswaranathan},
  {and} \bibinfo{person}{Surya Ganguli}.} \bibinfo{year}{2015}\natexlab{}.
\newblock \bibinfo{title}{Deep {Unsupervised} {Learning} using {Nonequilibrium}
  {Thermodynamics}}.
\newblock
\newblock
\urldef\tempurl%
\url{https://doi.org/10.48550/arXiv.1503.03585}
\showDOI{\tempurl}
\newblock
\shownote{arXiv:1503.03585 [cond-mat, q-bio, stat]}.


\bibitem[Solaiman(2023)]%
        {solaiman2023gradient}
\bibfield{author}{\bibinfo{person}{Irene Solaiman}.}
  \bibinfo{year}{2023}\natexlab{}.
\newblock \showarticletitle{The gradient of generative AI release: Methods and
  considerations}. In \bibinfo{booktitle}{\emph{Proceedings of the 2023 ACM
  Conference on Fairness, Accountability, and Transparency}}.
  \bibinfo{pages}{111--122}.
\newblock


\bibitem[Somepalli et~al\mbox{.}(2022)]%
        {somepalli_diffusion_2022}
\bibfield{author}{\bibinfo{person}{Gowthami Somepalli}, \bibinfo{person}{Vasu
  Singla}, \bibinfo{person}{Micah Goldblum}, \bibinfo{person}{Jonas Geiping},
  {and} \bibinfo{person}{Tom Goldstein}.} \bibinfo{year}{2022}\natexlab{}.
\newblock \bibinfo{title}{Diffusion {Art} or {Digital} {Forgery}?
  {Investigating} {Data} {Replication} in {Diffusion} {Models}}.
\newblock
\newblock
\urldef\tempurl%
\url{https://doi.org/10.48550/arXiv.2212.03860}
\showDOI{\tempurl}
\newblock
\shownote{arXiv:2212.03860 [cs]}.


\bibitem[Song et~al\mbox{.}(2020)]%
        {song_score-based_2020}
\bibfield{author}{\bibinfo{person}{Yang Song}, \bibinfo{person}{Jascha
  Sohl-Dickstein}, \bibinfo{person}{Diederik~P. Kingma},
  \bibinfo{person}{Abhishek Kumar}, \bibinfo{person}{Stefano Ermon}, {and}
  \bibinfo{person}{Ben Poole}.} \bibinfo{year}{2020}\natexlab{}.
\newblock \showarticletitle{Score-{Based} {Generative} {Modeling} through
  {Stochastic} {Differential} {Equations}}.
\newblock  (\bibinfo{date}{Nov.} \bibinfo{year}{2020}).
\newblock
\urldef\tempurl%
\url{https://doi.org/10.48550/arXiv.2011.13456}
\showDOI{\tempurl}


\bibitem[Struppek et~al\mbox{.}(2022)]%
        {Struppek_Hintersdorf_Kersting_2022}
\bibfield{author}{\bibinfo{person}{Lukas Struppek}, \bibinfo{person}{Dominik
  Hintersdorf}, {and} \bibinfo{person}{Kristian Kersting}.}
  \bibinfo{year}{2022}\natexlab{}.
\newblock \showarticletitle{Rickrolling the Artist: Injecting Invisible
  Backdoors into Text-Guided Image Generation Models}.
\newblock  (\bibinfo{year}{2022}).
\newblock
\urldef\tempurl%
\url{https://doi.org/10.48550/ARXIV.2211.02408}
\showDOI{\tempurl}


\bibitem[Sweeney(2009)]%
        {Sweeney_2009}
\bibfield{author}{\bibinfo{person}{Naofse~Mac Sweeney}.}
  \bibinfo{year}{2009}\natexlab{}.
\newblock \showarticletitle{Beyond Ethnicity: The Overlooked Diversity of Group
  Identities}.
\newblock \bibinfo{journal}{\emph{Journal of Mediterranean Archaeology}}
  \bibinfo{volume}{22}, \bibinfo{number}{1} (\bibinfo{date}{Jun}
  \bibinfo{year}{2009}), \bibinfo{pages}{101–126}.
\newblock
\showISSN{17431700, 09527648}
\urldef\tempurl%
\url{https://doi.org/10.1558/jmea.v22i1.101}
\showDOI{\tempurl}


\bibitem[Takahashi(2023)]%
        {takahashi2023ai}
\bibfield{author}{\bibinfo{person}{Dean Takahashi}.}
  \bibinfo{year}{2023}\natexlab{}.
\newblock \bibinfo{title}{AI Games and AI Film Festival will highlight how
  generative AI is taking root}.
\newblock
\newblock
\urldef\tempurl%
\url{https://venturebeat.com/games/ai-games-and-ai-film-festival-will-highlight-how-generative-ai-is-taking-root/}
\showURL{%
\tempurl}


\bibitem[Tolosana et~al\mbox{.}(2020)]%
        {tolosana_deepfakes_2020}
\bibfield{author}{\bibinfo{person}{Ruben Tolosana}, \bibinfo{person}{Ruben
  Vera-Rodriguez}, \bibinfo{person}{Julian Fierrez}, \bibinfo{person}{Aythami
  Morales}, {and} \bibinfo{person}{Javier Ortega-Garcia}.}
  \bibinfo{year}{2020}\natexlab{}.
\newblock \showarticletitle{Deepfakes and beyond: {A} {Survey} of face
  manipulation and fake detection}.
\newblock \bibinfo{journal}{\emph{Information Fusion}}  \bibinfo{volume}{64}
  (\bibinfo{date}{Dec.} \bibinfo{year}{2020}), \bibinfo{pages}{131--148}.
\newblock
\showISSN{1566-2535}
\urldef\tempurl%
\url{https://doi.org/10.1016/j.inffus.2020.06.014}
\showDOI{\tempurl}


\bibitem[Tomasev et~al\mbox{.}(2022)]%
        {tomasev_manifestations_2022}
\bibfield{author}{\bibinfo{person}{Nenad Tomasev},
  \bibinfo{person}{Jonathan~Leader Maynard}, {and} \bibinfo{person}{Iason
  Gabriel}.} \bibinfo{year}{2022}\natexlab{}.
\newblock \showarticletitle{Manifestations of {Xenophobia} in {AI} {Systems}}.
\newblock  (\bibinfo{year}{2022}).
\newblock
\urldef\tempurl%
\url{https://doi.org/10.48550/ARXIV.2212.07877}
\showDOI{\tempurl}
\newblock
\shownote{Publisher: arXiv Version Number: 1}.


\bibitem[Topaz et~al\mbox{.}(2022)]%
        {Topaz_Higdon_Epps-Darling_Siau_Kerkhoff_Mendiratta_Young_2022}
\bibfield{author}{\bibinfo{person}{Chad~M. Topaz}, \bibinfo{person}{Jude
  Higdon}, \bibinfo{person}{Avriel Epps-Darling}, \bibinfo{person}{Ethan Siau},
  \bibinfo{person}{Harper Kerkhoff}, \bibinfo{person}{Shivani Mendiratta},
  {and} \bibinfo{person}{Eric Young}.} \bibinfo{year}{2022}\natexlab{}.
\newblock \showarticletitle{Race- and gender-based under-representation of
  creative contributors: art, fashion, film, and music}.
\newblock \bibinfo{journal}{\emph{Humanities and Social Sciences
  Communications}} \bibinfo{volume}{9}, \bibinfo{number}{11}
  (\bibinfo{date}{Jun} \bibinfo{year}{2022}), \bibinfo{pages}{1–11}.
\newblock
\showISSN{2662-9992}
\urldef\tempurl%
\url{https://doi.org/10.1057/s41599-022-01239-9}
\showDOI{\tempurl}


\bibitem[Trauthig(2022)]%
        {trauthig_whatsapp_2022}
\bibfield{author}{\bibinfo{person}{Inga Trauthig}.}
  \bibinfo{year}{2022}\natexlab{}.
\newblock \bibinfo{title}{{WhatsApp}, {Misinformation}, and {Latino}
  {Political} {Discourse} in the {U}.{S}.}
\newblock
\newblock
\urldef\tempurl%
\url{https://techpolicy.press/whatsapp-misinformation-and-latino-political-discourse-in-the-u-s/}
\showURL{%
\tempurl}


\bibitem[Trotta and Pierson(2023)]%
        {Trotta_Pierson_2023}
\bibfield{author}{\bibinfo{person}{Daniel Trotta} {and}
  \bibinfo{person}{Brendan Pierson}.} \bibinfo{year}{2023}\natexlab{}.
\newblock \bibinfo{title}{US judges halt healthcare bans for Transgender
  Youth}.
\newblock
\newblock
\urldef\tempurl%
\url{https://www.reuters.com/legal/us-judges-halt-healthcare-bans-transgender-youth-2023-07-03/}
\showURL{%
\tempurl}


\bibitem[Ungless et~al\mbox{.}(2023)]%
        {Ungless_Ross_Lauscher_2023}
\bibfield{author}{\bibinfo{person}{Eddie~L. Ungless}, \bibinfo{person}{Björn
  Ross}, {and} \bibinfo{person}{Anne Lauscher}.}
  \bibinfo{year}{2023}\natexlab{}.
\newblock \showarticletitle{Stereotypes and Smut: The (Mis)representation of
  Non-cisgender Identities by Text-to-Image Models}.
\newblock  \bibinfo{number}{arXiv:2305.17072} (\bibinfo{date}{May}
  \bibinfo{year}{2023}).
\newblock
\urldef\tempurl%
\url{http://arxiv.org/abs/2305.17072}
\showURL{%
\tempurl}
\newblock
\shownote{arXiv:2305.17072 [cs]}.


\bibitem[Vaccari and Chadwick(2020)]%
        {vaccari_deepfakes_2020}
\bibfield{author}{\bibinfo{person}{Cristian Vaccari} {and}
  \bibinfo{person}{Andrew Chadwick}.} \bibinfo{year}{2020}\natexlab{}.
\newblock \showarticletitle{Deepfakes and {Disinformation}: {Exploring} the
  {Impact} of {Synthetic} {Political} {Video} on {Deception}, {Uncertainty},
  and {Trust} in {News}}.
\newblock \bibinfo{journal}{\emph{Social Media + Society}} \bibinfo{volume}{6},
  \bibinfo{number}{1} (\bibinfo{date}{Jan.} \bibinfo{year}{2020}),
  \bibinfo{pages}{2056305120903408}.
\newblock
\showISSN{2056-3051}
\urldef\tempurl%
\url{https://doi.org/10.1177/2056305120903408}
\showDOI{\tempurl}
\newblock
\shownote{Publisher: SAGE Publications Ltd}.


\bibitem[Vartiainen and Tedre(2023)]%
        {vartiainen2023using}
\bibfield{author}{\bibinfo{person}{Henriikka Vartiainen} {and}
  \bibinfo{person}{Matti Tedre}.} \bibinfo{year}{2023}\natexlab{}.
\newblock \showarticletitle{Using artificial intelligence in craft education:
  crafting with text-to-image generative models}.
\newblock \bibinfo{journal}{\emph{Digital Creativity}} (\bibinfo{year}{2023}),
  \bibinfo{pages}{1--21}.
\newblock


\bibitem[Vaswani et~al\mbox{.}(2017)]%
        {vaswani_attention_2017}
\bibfield{author}{\bibinfo{person}{Ashish Vaswani}, \bibinfo{person}{Noam
  Shazeer}, \bibinfo{person}{Niki Parmar}, \bibinfo{person}{Jakob Uszkoreit},
  \bibinfo{person}{Llion Jones}, \bibinfo{person}{Aidan~N. Gomez},
  \bibinfo{person}{Lukasz Kaiser}, {and} \bibinfo{person}{Illia Polosukhin}.}
  \bibinfo{year}{2017}\natexlab{}.
\newblock \showarticletitle{Attention {Is} {All} {You} {Need}}.
\newblock  (\bibinfo{date}{June} \bibinfo{year}{2017}).
\newblock
\urldef\tempurl%
\url{https://doi.org/10.48550/arXiv.1706.03762}
\showDOI{\tempurl}


\bibitem[Veale et~al\mbox{.}(2023)]%
        {veale2023ai}
\bibfield{author}{\bibinfo{person}{Michael Veale}, \bibinfo{person}{Kira
  Matus}, {and} \bibinfo{person}{Robert Gorwa}.}
  \bibinfo{year}{2023}\natexlab{}.
\newblock \showarticletitle{AI and Global Governance: Modalities, Rationales,
  Tensions}.
\newblock \bibinfo{journal}{\emph{Annual Review of Law and Social Science}}
  \bibinfo{volume}{19} (\bibinfo{year}{2023}).
\newblock
\urldef\tempurl%
\url{https://doi.org/10.1146/annurev-lawsocsci-020223-040749}
\showURL{%
\tempurl}
\newblock
\shownote{Review in Advance first posted online on June 28, 2023. (Changes may
  still occur before final publication.)}.


\bibitem[Verdoliva(2020)]%
        {verdoliva_media_2020}
\bibfield{author}{\bibinfo{person}{Luisa Verdoliva}.}
  \bibinfo{year}{2020}\natexlab{}.
\newblock \bibinfo{title}{Media {Forensics} and {DeepFakes}: an overview}.
\newblock
\newblock
\urldef\tempurl%
\url{http://arxiv.org/abs/2001.06564}
\showURL{%
\tempurl}
\newblock
\shownote{arXiv:2001.06564 [cs]}.


\bibitem[Vyas et~al\mbox{.}(2023)]%
        {Vyas_Kakade_Barak_2023}
\bibfield{author}{\bibinfo{person}{Nikhil Vyas}, \bibinfo{person}{Sham Kakade},
  {and} \bibinfo{person}{Boaz Barak}.} \bibinfo{year}{2023}\natexlab{}.
\newblock \showarticletitle{Provable Copyright Protection for Generative
  Models}.
\newblock  \bibinfo{number}{arXiv:2302.10870} (\bibinfo{date}{Feb}
  \bibinfo{year}{2023}).
\newblock
\urldef\tempurl%
\url{http://arxiv.org/abs/2302.10870}
\showURL{%
\tempurl}
\newblock
\shownote{arXiv:2302.10870 [cs, stat]}.


\bibitem[Wang et~al\mbox{.}(2022a)]%
        {Wang_Liu_Zhang_Kleiman_Kim_Zhao_Shirai_Narayanan_Russakovsky_2022}
\bibfield{author}{\bibinfo{person}{Angelina Wang}, \bibinfo{person}{Alexander
  Liu}, \bibinfo{person}{Ryan Zhang}, \bibinfo{person}{Anat Kleiman},
  \bibinfo{person}{Leslie Kim}, \bibinfo{person}{Dora Zhao},
  \bibinfo{person}{Iroha Shirai}, \bibinfo{person}{Arvind Narayanan}, {and}
  \bibinfo{person}{Olga Russakovsky}.} \bibinfo{year}{2022}\natexlab{a}.
\newblock \showarticletitle{REVISE: A Tool for Measuring and Mitigating Bias in
  Visual Datasets}.
\newblock \bibinfo{journal}{\emph{International Journal of Computer Vision}}
  \bibinfo{volume}{130}, \bibinfo{number}{7} (\bibinfo{date}{Jul}
  \bibinfo{year}{2022}), \bibinfo{pages}{1790–1810}.
\newblock
\showISSN{0920-5691}
\urldef\tempurl%
\url{https://doi.org/10.1007/s11263-022-01625-5}
\showDOI{\tempurl}


\bibitem[Wang et~al\mbox{.}(2022b)]%
        {wang_imagen_2022}
\bibfield{author}{\bibinfo{person}{Su Wang}, \bibinfo{person}{Chitwan Saharia},
  \bibinfo{person}{Ceslee Montgomery}, \bibinfo{person}{Jordi Pont-Tuset},
  \bibinfo{person}{Shai Noy}, \bibinfo{person}{Stefano Pellegrini},
  \bibinfo{person}{Yasumasa Onoe}, \bibinfo{person}{Sarah Laszlo},
  \bibinfo{person}{David~J. Fleet}, \bibinfo{person}{Radu Soricut},
  \bibinfo{person}{Jason Baldridge}, \bibinfo{person}{Mohammad Norouzi},
  \bibinfo{person}{Peter Anderson}, {and} \bibinfo{person}{William Chan}.}
  \bibinfo{year}{2022}\natexlab{b}.
\newblock \bibinfo{title}{Imagen {Editor} and {EditBench}: {Advancing} and
  {Evaluating} {Text}-{Guided} {Image} {Inpainting}}.
\newblock
\newblock
\urldef\tempurl%
\url{https://doi.org/10.48550/arXiv.2212.06909}
\showDOI{\tempurl}
\newblock
\shownote{arXiv:2212.06909 [cs]}.


\bibitem[Weatherbed(2023)]%
        {Weatherbed_2023}
\bibfield{author}{\bibinfo{person}{Jess Weatherbed}.}
  \bibinfo{year}{2023}\natexlab{}.
\newblock \bibinfo{title}{Levi’s will test AI-generated clothing models to
  “increase diversity”}.
\newblock
\newblock
\urldef\tempurl%
\url{https://www.theverge.com/2023/3/27/23658385/levis-ai-generated-clothing-model-diversity-denim}
\showURL{%
\tempurl}


\bibitem[Weidinger et~al\mbox{.}(2021)]%
        {weidinger_ethical_2021}
\bibfield{author}{\bibinfo{person}{Laura Weidinger}, \bibinfo{person}{John
  Mellor}, \bibinfo{person}{Maribeth Rauh}, \bibinfo{person}{Conor Griffin},
  \bibinfo{person}{Jonathan Uesato}, \bibinfo{person}{Po-Sen Huang},
  \bibinfo{person}{Myra Cheng}, \bibinfo{person}{Mia Glaese},
  \bibinfo{person}{Borja Balle}, \bibinfo{person}{Atoosa Kasirzadeh},
  \bibinfo{person}{Zac Kenton}, \bibinfo{person}{Sasha Brown},
  \bibinfo{person}{Will Hawkins}, \bibinfo{person}{Tom Stepleton},
  \bibinfo{person}{Courtney Biles}, \bibinfo{person}{Abeba Birhane},
  \bibinfo{person}{Julia Haas}, \bibinfo{person}{Laura Rimell},
  \bibinfo{person}{Lisa~Anne Hendricks}, \bibinfo{person}{William Isaac},
  \bibinfo{person}{Sean Legassick}, \bibinfo{person}{Geoffrey Irving}, {and}
  \bibinfo{person}{Iason Gabriel}.} \bibinfo{year}{2021}\natexlab{}.
\newblock \bibinfo{title}{Ethical and social risks of harm from {Language}
  {Models}}.
\newblock
\newblock
\urldef\tempurl%
\url{https://doi.org/10.48550/arXiv.2112.04359}
\showDOI{\tempurl}
\newblock
\shownote{arXiv:2112.04359 [cs]}.


\bibitem[Weidinger et~al\mbox{.}(2022)]%
        {weidinger_taxonomy_2022}
\bibfield{author}{\bibinfo{person}{Laura Weidinger}, \bibinfo{person}{Jonathan
  Uesato}, \bibinfo{person}{Maribeth Rauh}, \bibinfo{person}{Conor Griffin},
  \bibinfo{person}{Po-Sen Huang}, \bibinfo{person}{John Mellor},
  \bibinfo{person}{Amelia Glaese}, \bibinfo{person}{Myra Cheng},
  \bibinfo{person}{Borja Balle}, \bibinfo{person}{Atoosa Kasirzadeh},
  \bibinfo{person}{Courtney Biles}, \bibinfo{person}{Sasha Brown},
  \bibinfo{person}{Zac Kenton}, \bibinfo{person}{Will Hawkins},
  \bibinfo{person}{Tom Stepleton}, \bibinfo{person}{Abeba Birhane},
  \bibinfo{person}{Lisa~Anne Hendricks}, \bibinfo{person}{Laura Rimell},
  \bibinfo{person}{William Isaac}, \bibinfo{person}{Julia Haas},
  \bibinfo{person}{Sean Legassick}, \bibinfo{person}{Geoffrey Irving}, {and}
  \bibinfo{person}{Iason Gabriel}.} \bibinfo{year}{2022}\natexlab{}.
\newblock \showarticletitle{Taxonomy of {Risks} posed by {Language} {Models}}.
  In \bibinfo{booktitle}{\emph{2022 {ACM} {Conference} on {Fairness},
  {Accountability}, and {Transparency}}}. \bibinfo{publisher}{ACM},
  \bibinfo{address}{Seoul Republic of Korea}, \bibinfo{pages}{214--229}.
\newblock
\showISBNx{978-1-4503-9352-2}
\urldef\tempurl%
\url{https://doi.org/10.1145/3531146.3533088}
\showDOI{\tempurl}


\bibitem[Weisz et~al\mbox{.}(2023)]%
        {weisz_toward_2023}
\bibfield{author}{\bibinfo{person}{Justin~D. Weisz}, \bibinfo{person}{Michael
  Muller}, \bibinfo{person}{Jessica He}, {and} \bibinfo{person}{Stephanie
  Houde}.} \bibinfo{year}{2023}\natexlab{}.
\newblock \showarticletitle{Toward {General} {Design} {Principles} for
  {Generative} {AI} {Applications}}.
\newblock  (\bibinfo{year}{2023}).
\newblock
\urldef\tempurl%
\url{https://doi.org/10.48550/ARXIV.2301.05578}
\showDOI{\tempurl}
\newblock
\shownote{Publisher: arXiv Version Number: 1}.


\bibitem[West et~al\mbox{.}(2019)]%
        {West_Whittaker_Crawford_2019}
\bibfield{author}{\bibinfo{person}{Sarah~Myers West}, \bibinfo{person}{Meredith
  Whittaker}, {and} \bibinfo{person}{Kate Crawford}.}
  \bibinfo{year}{2019}\natexlab{}.
\newblock \bibinfo{booktitle}{\emph{Discriminating Systems: Gender, Race and
  Power in AI.}}
\newblock
\urldef\tempurl%
\url{Retrieved from https://ainowinstitute.org/ discriminatingsystems.html.}
\showURL{%
\tempurl}


\bibitem[Westerlund(2019)]%
        {westerlund_emergence_2019}
\bibfield{author}{\bibinfo{person}{Mika Westerlund}.}
  \bibinfo{year}{2019}\natexlab{}.
\newblock \showarticletitle{The {Emergence} of {Deepfake} {Technology}: {A}
  {Review}}.
\newblock \bibinfo{journal}{\emph{Technology Innovation Management Review}}
  \bibinfo{volume}{9}, \bibinfo{number}{11} (\bibinfo{year}{2019}),
  \bibinfo{pages}{40--53}.
\newblock
\showISSN{1927-0321}
\urldef\tempurl%
\url{https://doi.org/10.22215/timreview/1282}
\showDOI{\tempurl}
\newblock
\shownote{Place: Ottawa Publisher: Talent First Network}.


\bibitem[Widder et~al\mbox{.}(2022)]%
        {widder_limits_2022}
\bibfield{author}{\bibinfo{person}{David~Gray Widder}, \bibinfo{person}{Dawn
  Nafus}, \bibinfo{person}{Laura Dabbish}, {and} \bibinfo{person}{James
  Herbsleb}.} \bibinfo{year}{2022}\natexlab{}.
\newblock \showarticletitle{Limits and {Possibilities} for \&\#x201c;{Ethical}
  {AI}\&\#x201d; in {Open} {Source}: {A} {Study} of {Deepfakes}}. In
  \bibinfo{booktitle}{\emph{2022 {ACM} {Conference} on {Fairness},
  {Accountability}, and {Transparency}}} \emph{(\bibinfo{series}{{FAccT}
  '22})}. \bibinfo{publisher}{Association for Computing Machinery},
  \bibinfo{address}{New York, NY, USA}, \bibinfo{pages}{2035--2046}.
\newblock
\showISBNx{978-1-4503-9352-2}
\urldef\tempurl%
\url{https://doi.org/10.1145/3531146.3533779}
\showDOI{\tempurl}


\bibitem[Wiggers(2023)]%
        {Wiggers_2023}
\bibfield{author}{\bibinfo{person}{Kyle Wiggers}.}
  \bibinfo{year}{2023}\natexlab{}.
\newblock \bibinfo{title}{The current legal cases against generative AI are
  just the beginning}.
\newblock
\newblock
\urldef\tempurl%
\url{https://techcrunch.com/2023/01/27/the-current-legal-cases-against-generative-ai-are-just-the-beginning/}
\showURL{%
\tempurl}


\bibitem[Wolfe et~al\mbox{.}(2022)]%
        {Wolfe_Yang_Howe_Caliskan_2022}
\bibfield{author}{\bibinfo{person}{Robert Wolfe}, \bibinfo{person}{Yiwei Yang},
  \bibinfo{person}{Bill Howe}, {and} \bibinfo{person}{Aylin Caliskan}.}
  \bibinfo{year}{2022}\natexlab{}.
\newblock \showarticletitle{Contrastive Language-Vision AI Models Pretrained on
  Web-Scraped Multimodal Data Exhibit Sexual Objectification Bias}.
\newblock  \bibinfo{number}{arXiv:2212.11261} (\bibinfo{date}{Dec}
  \bibinfo{year}{2022}).
\newblock
\urldef\tempurl%
\url{http://arxiv.org/abs/2212.11261}
\showURL{%
\tempurl}
\newblock
\shownote{arXiv:2212.11261 [cs]}.


\bibitem[Xu et~al\mbox{.}(2022)]%
        {xu_comprehensive_2022}
\bibfield{author}{\bibinfo{person}{Ying Xu}, \bibinfo{person}{Philipp
  Terhörst}, \bibinfo{person}{Kiran Raja}, {and} \bibinfo{person}{Marius
  Pedersen}.} \bibinfo{year}{2022}\natexlab{}.
\newblock \showarticletitle{A {Comprehensive} {Analysis} of {AI} {Biases} in
  {DeepFake} {Detection} {With} {Massively} {Annotated} {Databases}}.
\newblock  (\bibinfo{year}{2022}).
\newblock
\urldef\tempurl%
\url{https://doi.org/10.48550/ARXIV.2208.05845}
\showDOI{\tempurl}
\newblock
\shownote{Publisher: arXiv Version Number: 1}.


\bibitem[Yang et~al\mbox{.}(2022)]%
        {https://doi.org/10.48550/arxiv.2209.00796}
\bibfield{author}{\bibinfo{person}{Ling Yang}, \bibinfo{person}{Zhilong Zhang},
  \bibinfo{person}{Yang Song}, \bibinfo{person}{Shenda Hong},
  \bibinfo{person}{Runsheng Xu}, \bibinfo{person}{Yue Zhao},
  \bibinfo{person}{Yingxia Shao}, \bibinfo{person}{Wentao Zhang},
  \bibinfo{person}{Bin Cui}, {and} \bibinfo{person}{Ming-Hsuan Yang}.}
  \bibinfo{year}{2022}\natexlab{}.
\newblock \bibinfo{title}{Diffusion Models: A Comprehensive Survey of Methods
  and Applications}.
\newblock
\newblock
\urldef\tempurl%
\url{https://doi.org/10.48550/ARXIV.2209.00796}
\showDOI{\tempurl}


\bibitem[Yu et~al\mbox{.}(2022)]%
        {yu_scaling_2022}
\bibfield{author}{\bibinfo{person}{Jiahui Yu}, \bibinfo{person}{Yuanzhong Xu},
  \bibinfo{person}{Jing~Yu Koh}, \bibinfo{person}{Thang Luong},
  \bibinfo{person}{Gunjan Baid}, \bibinfo{person}{Zirui Wang},
  \bibinfo{person}{Vijay Vasudevan}, \bibinfo{person}{Alexander Ku},
  \bibinfo{person}{Yinfei Yang}, \bibinfo{person}{Burcu~Karagol Ayan},
  \bibinfo{person}{Ben Hutchinson}, \bibinfo{person}{Wei Han},
  \bibinfo{person}{Zarana Parekh}, \bibinfo{person}{Xin Li},
  \bibinfo{person}{Han Zhang}, \bibinfo{person}{Jason Baldridge}, {and}
  \bibinfo{person}{Yonghui Wu}.} \bibinfo{year}{2022}\natexlab{}.
\newblock \showarticletitle{Scaling {Autoregressive} {Models} for
  {Content}-{Rich} {Text}-to-{Image} {Generation}}.
\newblock  (\bibinfo{year}{2022}).
\newblock
\urldef\tempurl%
\url{https://doi.org/10.48550/ARXIV.2206.10789}
\showDOI{\tempurl}
\newblock
\shownote{Publisher: arXiv Version Number: 1}.


\bibitem[Yu et~al\mbox{.}(2020)]%
        {yu_artificial_2020}
\bibfield{author}{\bibinfo{person}{Ning Yu}, \bibinfo{person}{Vladislav
  Skripniuk}, \bibinfo{person}{Sahar Abdelnabi}, {and} \bibinfo{person}{Mario
  Fritz}.} \bibinfo{year}{2020}\natexlab{}.
\newblock \showarticletitle{Artificial {Fingerprinting} for {Generative}
  {Models}: {Rooting} {Deepfake} {Attribution} in {Training} {Data}}.
\newblock  (\bibinfo{date}{July} \bibinfo{year}{2020}).
\newblock
\urldef\tempurl%
\url{https://doi.org/10.48550/arXiv.2007.08457}
\showDOI{\tempurl}


\bibitem[Zhao et~al\mbox{.}(2017)]%
        {Zhao_Wang_Yatskar_Ordonez_Chang_2017}
\bibfield{author}{\bibinfo{person}{Jieyu Zhao}, \bibinfo{person}{Tianlu Wang},
  \bibinfo{person}{Mark Yatskar}, \bibinfo{person}{Vicente Ordonez}, {and}
  \bibinfo{person}{Kai-Wei Chang}.} \bibinfo{year}{2017}\natexlab{}.
\newblock \showarticletitle{Men Also Like Shopping: Reducing Gender Bias
  Amplification using Corpus-level Constraints}. In
  \bibinfo{booktitle}{\emph{Proceedings of the 2017 Conference on Empirical
  Methods in Natural Language Processing}}. \bibinfo{publisher}{Association for
  Computational Linguistics}, \bibinfo{address}{Copenhagen, Denmark},
  \bibinfo{pages}{2979–2989}.
\newblock
\urldef\tempurl%
\url{https://doi.org/10.18653/v1/D17-1323}
\showDOI{\tempurl}


\bibitem[Zhou et~al\mbox{.}(2023)]%
        {zhou_synthetic_2023}
\bibfield{author}{\bibinfo{person}{Jiawei Zhou}, \bibinfo{person}{Yixuan
  Zhang}, \bibinfo{person}{Qianni Luo}, \bibinfo{person}{Andrea~G Parker},
  {and} \bibinfo{person}{Munmun De~Choudhury}.}
  \bibinfo{year}{2023}\natexlab{}.
\newblock \showarticletitle{Synthetic {Lies}: {Understanding} {AI}-{Generated}
  {Misinformation} and {Evaluating} {Algorithmic} and {Human} {Solutions}}. In
  \bibinfo{booktitle}{\emph{Proceedings of the 2023 {CHI} {Conference} on
  {Human} {Factors} in {Computing} {Systems}}} \emph{(\bibinfo{series}{{CHI}
  '23})}. \bibinfo{publisher}{Association for Computing Machinery},
  \bibinfo{address}{New York, NY, USA}, \bibinfo{pages}{1--20}.
\newblock
\showISBNx{978-1-4503-9421-5}
\urldef\tempurl%
\url{https://doi.org/10.1145/3544548.3581318}
\showDOI{\tempurl}


\end{thebibliography}

\appendix

\section{Taxonomy Methodology}
We conducted our searches utilising the Semantic Scholar API. Semantic Scholar index over 200 million academic papers. To capture relevant papers we selected five seed papers covering biased training data, biased image generation and bias in text-to-image models \cite{bianchi_easily_2022,Salminen_Jung_Chowdhury_Jansen_2020,Cho_Zala_Bansal_2022,Bansal_Yin_Monajatipoor_Chang_2022,birhane_multimodal_2021}. To capture papers relevant to misinformation harms, we selected three papers relevant to either deep fakes or synthetic media \cite{tolosana_deepfakes_2020, westerlund_emergence_2019} or diffusion technology and evaluation \cite{https://doi.org/10.48550/arxiv.2209.00796}. Our search returned over 300 papers. 43 of these papers provided substantial and useful discussions of text-to-image technologies. Through extensive manual searches we identified a further 40 papers, most of which were technical papers. Collected papers were then analysed for stakeholders, risks, empirical investigations and open research questions. 

Our taxonomy of risks initially adopted an inductive-deductive approach, in that we preempted the existence of three broad categories (discrimination and exclusion, harmful misuse, misinformation) and derived subcategories from analysis of the papers. We then retroactively identified potential ``gaps'' in the literature, based in part on analogous research into the harms of other technologies, plus identifying key stakeholders that have not been addressed. These gaps are clearly identified in the table.

\end{document}